\documentclass[useAMS,usenatbib]{mn2e}
   \usepackage{graphicx}
   \usepackage{amsmath}
   \usepackage{color}
   \usepackage{epstopdf}
   \usepackage{pdfsync}
   \usepackage{indentfirst}
   \usepackage{gensymb}
   \usepackage{float}
\font\fiverm=cmr5             \font\sevenrm=cmr7

          \font\sixrm=cmr6       

\def\dover#1#2{\hbox{${{\displaystyle#1 \vphantom{(} }\over{
   \displaystyle #2 \vphantom{(} }}$}}
\def\hskneg{\hspace{-6pt}}
\def\hsknegsm{\hspace{-3pt}}
{\catcode`\@=11                                                  
\gdef\SchlangeUnter#1#2{\lower2pt\vbox{\baselineskip 0pt\lineskip0pt    
\ialign{$\m@th#1\hfil##\hfil$\crcr#2\crcr\sim\crcr}}}}           
\def\gtrsim{\mathrel{\mathpalette\SchlangeUnter>}}               
\def\lesssim{\mathrel{\mathpalette\SchlangeUnter<}}    

\def\fsc{\alpha_{\hbox{\sevenrm f}}}                                
                                
\def\lambar{\lambda\llap {--}}

\def\teq#1{$\, #1\,$}                         
\def\split{\gamma\to\gamma\gamma}
\def\pairs{\gamma\to e^+e^-}
\def\calFErber{{\cal F}_{\hbox{\sixrm Erber}}}
\def\calFTH{{\cal F}_{\hbox{\sixrm TH}}}
\def\rns{R_{\hbox{\sixrm NS}}}

\def\erg{\varepsilon}
\def\eesc{\varepsilon_{\rm esc}}
\def\thetakB{\Theta_{\hbox{\sixrm kB}}}

\def\rE{r_{\hbox{\fiverm E}}}

\def\gammae{\gamma_{\hbox{\sixrm e}}}
\def\deltaE{\delta_{\hbox{\fiverm E}}}
\def\omegaE{\omega_{\hbox{\fiverm E}}}
\def\zetae{\zeta_{\hbox{\fiverm E}}}
\def\Thetae{\Theta_{\hbox{\sixrm e}}}
\def\thetaE{\theta_{\hbox{\fiverm E}}}
\def\rEvec{\hat{\boldsymbol{r}}_{\hbox{\fiverm E}}}
\def\thetaEvec{\hat{\boldsymbol{\theta}}_{\hbox{\fiverm E}}}
\def\phiEvec{\hat{\boldsymbol{\phi}}_{\hbox{\fiverm E}}}
\def\Bvec{\boldsymbol{B}}
\def\BEvec{\hat{\boldsymbol{B}}_{\hbox{\fiverm E}}}
\def\BEmag{B_{\hbox{\fiverm E}}}
\def\kvechat{\boldsymbol{k}}
\def\xmax{x_{\hbox{\sevenrm max}}}
\def\rmax{r_{\hbox{\sevenrm max}}}
\def\emax{\erg_{\hbox{\sevenrm max}}}
\def\PsiE{\Psi_{\hbox{\sixrm E}}}
\def\PsiL{\Psi_{\hbox{\sixrm L}}}
\def\kGR{\boldsymbol{k}_{\hbox{\fiverm GR}}}
\def\BGR{\boldsymbol{B}_{\hbox{\fiverm GR}}}

\def\Omegavec{\boldsymbol{\hat{\Omega}}}
\def\muvec{\boldsymbol{\hat{\mu}}}
\def\nvec{\boldsymbol{\hat{n} }_v}
\def\BEtotvec{{\boldsymbol{B}}_{\hbox{\sixrm E}}}
\def\eescsp{\varepsilon^{sp}_{\rm esc}}
\def\efmax{\varepsilon_f^{\rm max}}
\def\thetamin{\theta_{\rm min}}
\def\thetamax{\theta_{\rm max}}

\title[Opacities for Photon Splitting and Pair Creation in Neutron Star Magnetospheres]
{Opacities for Photon Splitting and Pair Creation in Neutron Star Magnetospheres}
\author[K. Hu, M.~G. Baring, Z. Wadiasingh \& A.~K. Harding]{
Kun Hu,$^{1}$\thanks{E-mail: kh38@rice.edu (KH); baring@rice.edu (MGB); zwadiasingh@gmail.com (ZW); ahardingx@yahoo.com (AKH)\vskip -50pt} 
Matthew G. Baring,$^{1}$
Zorawar Wadiasingh$^{2,3}$ and Alice K. Harding$^{2}$\\
$^{1}$Department of Physics and Astronomy - MS 108, Rice University,
6100 Main Street, Houston, Texas 77251-1892, USA
\\
$^{2}$Gravitational Astrophysics Laboratory, Code 663,
NASA's Goddard Space Flight Center, Greenbelt, MD 20770, USA
\\
$^{3}$Centre for Space Research, North-West University, Potchefstroom, South Africa
\vspace{-10pt}}

\begin{document}

\date{Accepted 2019 April 5. Received 2019 April 3; in original form 2019 February 12.}

\newcommand{\vol}[2]{$\,$\rm #1\rm , #2}                 

\pagerange{\pageref{firstpage}--24} \pubyear{2019}

\maketitle

\label{firstpage}

\vspace{-50pt}
\begin{abstract}
Over the last four decades, persistent and flaring emission of magnetars observed by various telescopes
has provided us with a suite of light curves and spectra in soft and hard X-rays, with no emission yet detected above around 1 MeV.
Attenuation of such high-energy photons by magnetic pair creation and photon splitting is expected to 
be active in the magnetospheres of magnetars, possibly accounting for the paucity of gamma-rays in their signals.
This paper explores polarization-dependent opacities for these two QED processes 
in static vacuum dipole magnetospheres of highly-magnetized neutron stars, calculating
attenuation lengths and determining escape energies, which are the maximum photon energies for transparency out to infinity. 
The numerical trajectory integral analysis in flat and curved spacetimes 
provides upper bounds of a few MeV or less to the visible energies for magnetars for locales proximate to the stellar surface.
Photon splitting opacity alone puts constraints on the possible emission locales in their magnetospheres:
regions within field loops of maximum altitudes \teq{\rmax\sim 2-4} stellar radii are not commensurate with
maximum detected energies of around 250 keV. 
These constraints apply not only to magnetar flares but also to their quiescent hard X-ray tail emission. 
An exploration of photon splitting attenuation in the context of a resonant inverse Compton scattering model for the hard X-ray tails
derives distinctive phase-resolved spectroscopic and polarimetric signatures, of significant interest
for future MeV-band missions such as AMEGO and e-ASTROGAM.
\end{abstract}


\section{Introduction}

Magnetars are isolated neutron stars characterized by strong magnetic fields 
\teq{\sim 10^{13} - 10^{15}} G inferred 
from their rotational spin-down evolution, high persistent X-ray luminosities, 
and transient flaring activity in hard X-rays.  Historically, magnetars have been categorized in 
two groups: soft gamma-ray repeaters (SGRs) and anomalous X-ray pulsars (AXPs). 
The magnetar scenario was first introduced  by \cite{DT92} for SGRs and later extended 
to AXPs by \cite{TD96}, wherein it was proposed that they are powered
by the decay of magnetic fields.
At present, around 30 magnetars have been observed -- for a list of properties, 
see the McGill Magnetar Catalog \citep{OK14} and its contemporaneous
online version\footnote{http://www.physics.mcgill.ca/~pulsar/magnetar/main.html}.

Magnetars evince bright, persistent pulsed X-ray emission, with luminosities often 
exceeding the rotational energy loss rate by several orders of magnitude; this disparity 
promotes the hypothesis that their strong magnetic fields power their emission. The soft X-ray
spectra of magnetars can generally be fit by blackbody models with temperatures around 0.3--1 keV,
connected to a steep power-law tail between 1 keV and 10 keV with an index around 2 to 4
\citep[e.g.,][]{Mereghetti08,Vigano13,Turolla15}. The thermal component of the soft X-ray emission is
believed to come from the surface of the star, while a leading model for the production of 
the non-thermal part is repeated resonant cyclotron scattering in a magnetospheric corona
\citep[e.g.,][]{TLK02,Lyutikov06}. 

A third component of the persistent emission appears above 10 keV, fit 
with power-law models of an index between 1 to 2 \citep{Kuiper06,Gotz06,Hartog08a,Hartog08b,Younes17}, 
i.e., much flatter than the soft X-ray spectra. These hard tails cannot extend much beyond a few hundred
keV due to constraining upper bounds obtained by COMPTEL at MeV energies and {\it Fermi}-LAT above 100 MeV
\citep[e.g.,][]{Fermi_mag_bounds}.   Thus, a spectral break or turnover is indicated at around 200-700 keV,
and a possible suggestion of this appears in {\it Fermi}-GBM spectra of 4U 0142+61 
and 1E 1841-045 \citep{terBeek12}.
For this non-thermal hard X-ray component, resonant inverse Compton scattering 
is believed to be the most efficient radiative mechanism for its production
\citep{BH07,FT07,Beloborodov13,Wadiasingh18}. This process is resonant at the cyclotron
frequency, increasing the cross section over the familiar non-magnetic Thomson value by 
around 2--3 orders of magnitude when \teq{B\lesssim 4.4\times 10^{13}} Gauss \citep[e.g., see][]{gonthier00}.

In addition to these quiescent signals, recurrent bursts of typical luminosities  \teq{10^{38}-10^{42}} erg/sec
have been observed for many magnetars.  Some episodes of such bursts
last hours to days, over which tens to hundreds of individual short bursts can occur. The spectra
of short bursts can be fit by a thermal bremsstrahlung model with temperatures around 30 keV
\citep{Gogus99,Gogus00,Feroci04}. In some sources a statistically better 
fit for burst spectra is obtained with a two-blackbody
model \citep{Israel08,Lin12,vdH12,Younes14}, thereby promoting the possibility of thermal gradients 
existing in the magnetosphere.

Magnetars also exhibit giant flares of luminosities \teq{10^{44} - 10^{47}} erg/sec at hard X-ray energies 
extending up to around 1 MeV.  They have been observed for only three
SGRs \citep[e.g.,][]{Mazets79,Hurley99,Hurley05}, and for each of these, only once over roughly a 40 year period.
They are characterized by a short, intense spike lasting for about 0.2 s. This peak count rate is then followed by
an energetic pulsating tail lasting several minutes. 
The ``fireball scenario'' is the leading candidate in explaining the radiative dissipation and  cooling phases of giant flares
\citep[e.g.,][]{DT92,TD95}.  In this picture, the energy released by the reconfiguration of crustal
magnetic fields generates numerous electron-positron pairs in the magnetosphere.  While 
much plasma flows out along the open field lines, an optically-thick
pair fireball is trapped by the closed field lines for many rotation periods, both zones generating
pulsating, cooling tails of giant flares.  \cite{vanPutten16} determined that fireball-driven,
high opacity outflows possess sufficient relativistic beaming to approximately yield the observed 
pulse fractions.

The upper bounds to energies of both quiescent and flare signals may be due to the intrinsic nature of 
plasma in the emission region, for example the effective temperature or mean Lorentz factor of the energetic 
pairs when heated by magnetic dissipation.  Yet it is also suggestive of hard X-ray and
gamma-ray attenuation in the magnetosphere, a prospect that is explored in this paper.
Two exotic physical processes, magnetic pair creation and magnetic photon splitting, are
expected to be effective in attenuating hard X-rays and gamma-rays in the magnetospheres of
magnetars. One photon pair creation \teq{\gamma\to e^{\pm}} is permitted in strong-field environment because
the external magnetic field can absorb momentum perpendicular to the field \citep[e.g.,][]{Erber66}. This process operates
effectively for photons with energies above the absolute pair creation threshold of
\teq{2m_ec^2}. Below the pair threshold, magnetic photon splitting, a higher-order QED process \citep[e.g.,][]{Adler71}, can
efficiently attenuate high energy radiation. When the magnetic field strength exceeds around \teq{10^{13}}Gauss, 
these two processes generate
opacity for high energy photons in the magnetospheres, perhaps helping explain the 
observed maximum energies and upper
limits imposed by COMPTEL and {\it Fermi}-LAT.

Opacity calculations have been presented in a number of papers over the years.
\cite{Baring95} discussed the spectral softening of SGR flares caused
by repeated photon splitting. \cite{HBG96} computed attenuation lengths and escape energies for both pair
creation and photon splitting for photons emergent from quasi-polar locales of neutron stars.
\cite{HBG97} and \cite{BH01} carried out cascade simulations incorporating
splitting and pair creation in pulsar magnetospheres, 
discussing a possible ``radio death line" for high-field pulsars \citep{BH98} caused by the suppression of pair
creation. More recently, a detailed discussion on the pair creation opacity in neutron star
magnetospheres was conducted by \cite{SB14}. Their work, which focused mostly on emission from near
the polar surface zones of neutron stars, used transparency criteria to constrain emission altitudes for 
gamma-ray pulsars in the context of {\it Fermi}-LAT and 
\v{C}erenkov telescope observations.

In this paper, we calculate photon splitting and pair creation opacities in inner magnetospheres 
of neutron stars, applicable to arbitrary colatitudes and a substantial range of altitudes 
in closed field line zones above the stellar surface. 
The focus on these regions makes for a more pertinent connection with physical models 
of magnetar emission than was afforded with previous expositions.  We present the attenuation
lengths and escape energies as functions of emission position for photons emitted parallel or at
small angles to the local magnetic fields.  
Starting from flat spacetime, we deliver opacity calculations both analytically and numerically.
The numerical results here are in broad agreement with previous papers
\citep{HBG97,BH01,SB14} that focus mainly on the polar cap region of neutron stars.
Analytic approximations are deduced for photon splitting 
both to guide validation of the full numerical evaluations, and to supplement similar results offered 
in \cite{SB14} for pair creation.  

Special emphasis is given to the case where photons are emitted along specific field loops, 
which is broadly applicable to models where high degrees of Doppler beaming are operable.
In particular, comparatively simple empirical approximations
for photon splitting and pair creation escape energies are provided for implementation in 
emission models explored elsewhere.
As a case illustration, the opacity calculation is incorporated into the resonant inverse Compton
scattering model of the hard X-ray tails to illustrate the phase-resolved spectroscopic and 
polarimetric characteristics for a spinning magnetar.  This demonstrates the substantial variability 
of the maximum energy with pulse phase, helping to identify observational diagnostics that
are possible with sensitive hard X-ray and Compton telescopes, 
both present and future.  The influences of general relativity (GR) 
are discussed in Section~\ref{sec:GR}, which are found to increase opacity in most parameter
regimes, especially for emission directed inward. 
Therein, a ``photosphere" plot is presented, where the opaque region in the
magnetosphere for specific photon energy is depicted, and the main impact 
of GR in expanding the volume of opaque regions is clearly evident.

\section{Opacity Geometry and Physics}
 \label{sec:geometry_physics}

To assess the importance of the photon splitting and pair creation in magnetars, 
the path-integrated opacity for these two processes is calculated.  From these
are derived two central quantities, the attenuation length and the escape energy. 
Before developing the opacity calculations, the essentials of the geometry 
and the quantum physics are outlined in this Section.

\subsection{Radiative Transfer Geometry}
 \label{sec:geometry}

\begin{figure*}
 \begin{minipage}{17.5cm}
\vspace*{-10pt}
\centerline{\hskip 10pt \includegraphics[height=7.8cm]{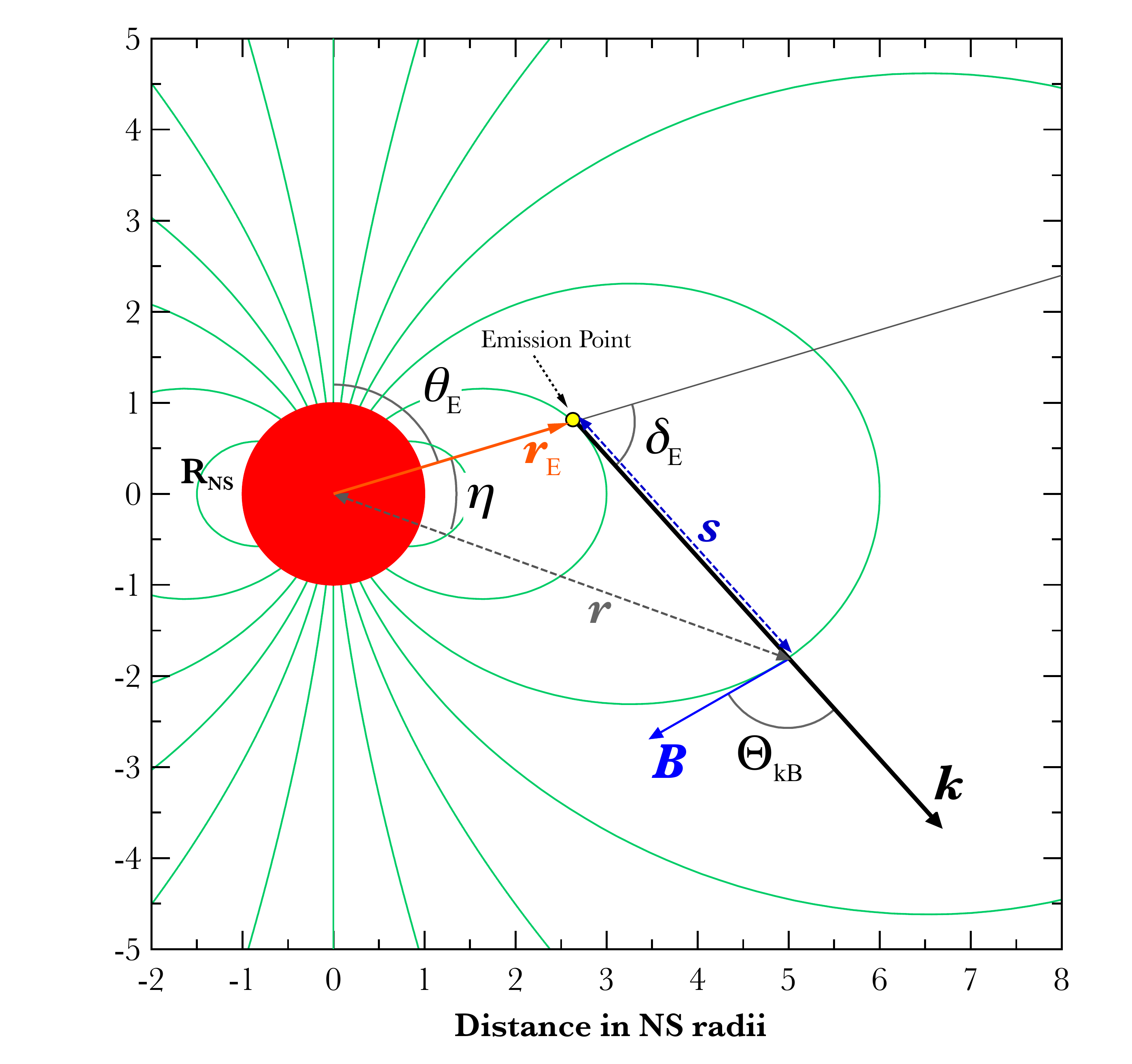}
   \hskip -10pt\includegraphics[height=7.8cm]{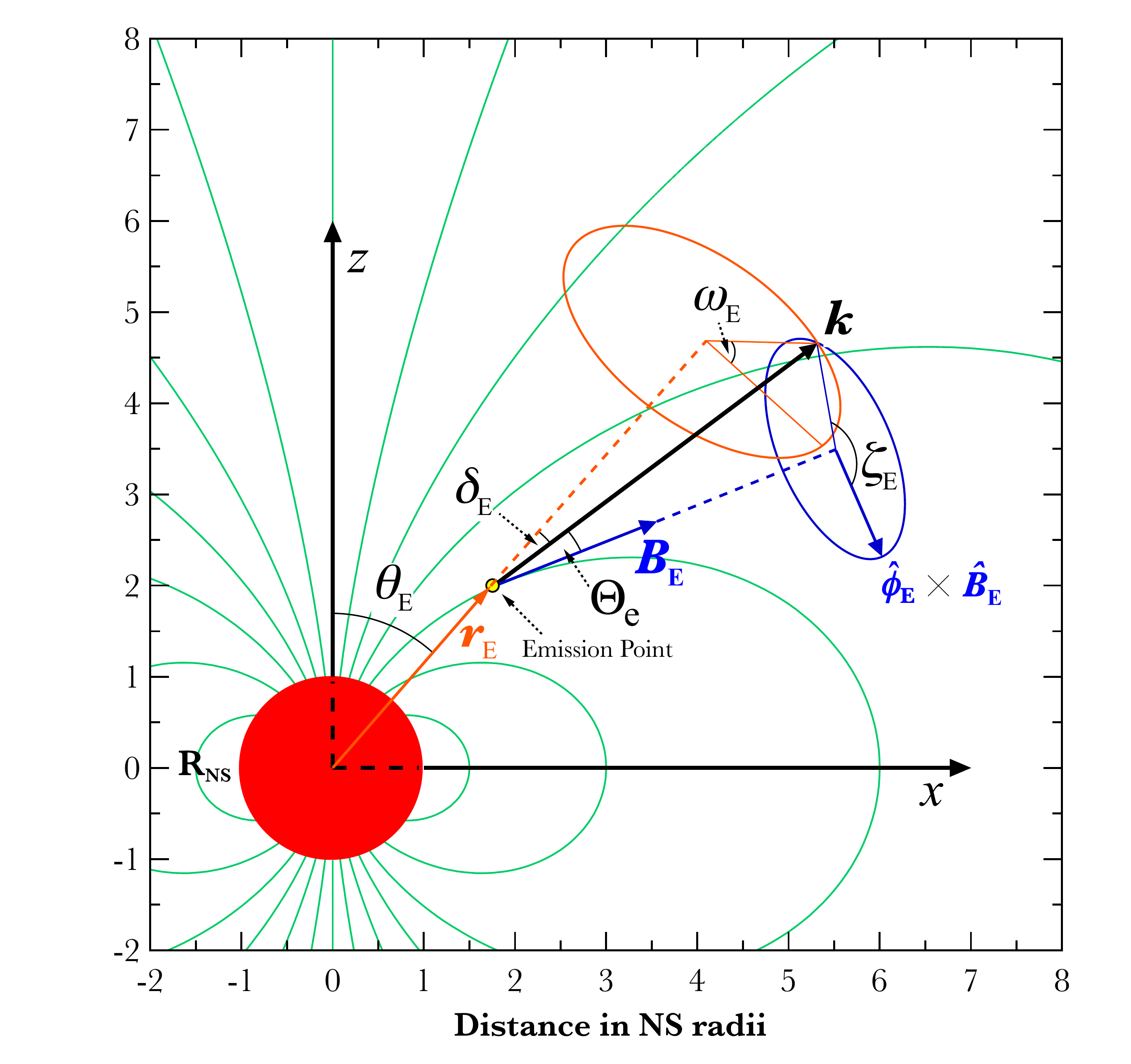}}
\vspace*{-5pt}
\caption{{\it Left Panel}: Photon propagation geometry in a dipole magnetic field, 
with green curves representing field lines, specifically for meridional cases 
in the \teq{x-z} plane where the light path is coplanar with the field loops.   
The photon emission point is at an altitude \teq{\rE = h \rns} and colatitude \teq{\thetaE}.  
The photon trajectory, represented by the black line, is a straight line 
for Euclidean geometry; a curved path (not shown here) applies for general relativistic 
considerations.  At any location along the photon path, \teq{\kvechat}
is the photon momentum vector, and \teq{\Bvec} is the local
magnetic field vector; the angle between these two vectors is
\teq{\thetakB}, given in Eq.~(\ref{eq:cos_thetakB}).  All such locations
are defined by the propagation angle \teq{\eta}, with the radial
position \teq{r} relative to the center of the neutron star, and the
distance \teq{s} from the point of emission being related via Eq.~(\ref{eq:r_path}).
{\it Right Panel}: Geometry pertinent to the trajectory at the point of photon emission, for 
cases where \teq{\Bvec} and \teq{\kvechat} do not necessarily lie in the meridional \teq{x-z} plane,
which is also the (\teq{\BEvec , \phiEvec \times \BEvec}) plane.  
The magnetic field vector \teq{{\boldsymbol{B}}_{\hbox{\fiverm E}}} at the emission point is also displayed.
Here, \teq{\deltaE} is the angle between \teq{\kvechat} and  \teq{\boldsymbol{\rE}}, the radial vector at the point
of emission, and \teq{\omegaE} is the azimuth angle with respect to the \teq{\thetaEvec} vector at
this point, given in Eq.~(\ref{eq:omegaE_def}).  In addition, \teq{\Thetae} is the angle between \teq{\kvechat} and
\teq{\BEvec}, and \teq{\zetae} is the azimuth angle around \teq{\BEvec} with respect to the 
\teq{\phiEvec \times \BEvec} vector at the emission location. 
Either \teq{(\deltaE, \omegaE)} or \teq{(\Thetae, \zetae)} determine the photon propagation direction, 
and relations between them are given in Eqs.~(\ref{eq:cos_deltaE_omegaE}) 
and~(\ref{eq:sin_deltaE_omegaE}).
 \label{fig:geometry}}
\end{minipage}
\end{figure*}

For opacity determination, a test photon is presumed to be emitted from a 
specific location with an altitude \teq{\rE} and a polar colatitude \teq{\thetaE} in the 
magnetosphere of a neutron star. The photon then propagates in the direction of 
its momentum vector which makes an angle \teq{\deltaE} with respect to its 
emission location vector \teq{\boldsymbol{\rE}}.  The optical depth for the 
splitting of photons and for pair creation takes the form of a trajectory integral:
\begin{equation}
   \tau(l)  \;=\; \int_{0}^{l} {\cal{R}} \,ds\quad ,   
 \label{eq:opt_depth}
\end{equation}
where \(\cal{R}\) is the attenuation coefficient in units of cm$^{-1}$  and 
\teq{l} is the cumulative path length along the photon trajectory out to some 
observation point, \teq{ds} being the differential path length.
The emission and propagation geometry is depicted in Fig.~\ref{fig:geometry}. 
The {\it attenuation length} \teq{L} is defined to be the path length over which the 
optical depth equals unity:
\begin{equation}
   \tau(L)  \;=\; \int_{0}^{L} {\cal{R}} \,ds \; =\; 1 \quad .
 \label{eq:atten_length_def}
\end{equation}
Since \teq{{\cal{R}}} is generally a strongly increasing function of photon energy \teq{\varepsilon}, 
\teq{\tau(\infty) <1} domains can be realized when \teq{\varepsilon} is below some critical value. 
This criterion defines the {\it escape energy} \teq{\varepsilon_{esc}}: it is the energy 
at which the attenuation length becomes infinity, so that photons with lower energies 
escape the magnetosphere of the star.  Thus the escape energy demarcates the 
boundary between transparency and opacity of the magnetosphere, a signature 
role that is identified in \cite{HBG97,BH01} and references therein.

In Sections 3 and 4, only flat spacetime propagation is treated in our considerations of 
photon attenuation.  This approximation is generally applicable for emission at
moderate or high altitudes.  Yet for emission locales within 1-2 stellar radii of the surface, 
general relativistic (GR) influences become important.  The main ones are 
curved photon trajectories, gravitational redshift, the deformation of the dipole field and 
concomitant changes in field strengths, all of which will be addressed in Section 5,
wherein \teq{ds} will represent coordinate length along a geodesic. GR effects will perturb the results 
obtained in Sections 3 and 4.  Yet, Minkowski spacetime considerations provide important
insights into the characteristics of photon attenuation, and facilitate the 
derivation of useful analytic approximations.
 
The flat spacetime geometry of the photon trajectory is illustrated in the left panel of Fig. \ref{fig:geometry}. 
Throughout, a vacuum dipole magnetic field is assumed, with a polar coordinate representation
\begin{equation}
   \boldsymbol{B} \; =\; \dover{B_p \rns^3}{2r^3}
   \left( 2\cos{\theta}\,\hat{\boldsymbol{r}}+\sin{\theta}\,\hat{\boldsymbol{\theta}} \right)\quad,
 \label{eq:B_dipole}
\end{equation}
where the angle \teq{\theta} represents the magnetic colatitude. 
Here \teq{B_p} is the surface polar (when \teq{\theta =0}) magnetic field strength 
and \teq{\rns} is the radius of the star. Throughout this paper, we assume \teq{\rns=} 10$^6$ cm.
The dipole assumption is a convenient choice, suited to attenuation processes that take place 
at low altitude: since magnetars are generally slow rotators, rotational aberration 
influences are small.  Departures from dipolar morphology as imbued
in twisted magnetospheres \citep[e.g.,][]{TLK02,cb17} will alter the computed photon opacities
somewhat; the extent to which they modify the overall character of the results presented below,
if at all, is deferred to future work.

\subsubsection{Coordinate Bases and the Photon Momentum \teq{\kvechat}}

We now identify three groups of coordinate bases to describe the trajectory of photons: 
(i) \teq{\{\hat{\boldsymbol{x}},\hat{\boldsymbol{y}},\hat{\boldsymbol{z}}\}} 
is a right-handed coordinate system with an origin at the star's center, its z-axis 
coinciding with the magnetic axis of the star and the emission point of the photon 
(without loss of generality) lying in the \teq{x-z} plane -- it applies to any point on the photon path; 
(ii) \teq{\{\hat{\boldsymbol{r}},\hat{\boldsymbol{\theta}},\hat{\boldsymbol{\phi}}\}} 
is the corresponding spherical polar coordinate system at any location in the path, and
(iii) \teq{\{\rEvec,\thetaEvec,\phiEvec\}} 
is the particular spherical polar coordinate system that applies specifically to the emission point. 
Obviously, the polar coordinate system (ii) is most suited to describing the magnetic 
field structure, while the emission point polar system (iii) is advantageous when specifying 
the direction of the photon trajectory.
The relations between the Cartesian and emission point polar coordinates are
\begin{eqnarray}
   \hat{\boldsymbol{x}} & = & \sin{\thetaE}\, \rEvec + \cos{\thetaE}\, \thetaEvec \quad ,\nonumber\\
   \hat{\boldsymbol{y}} & = & \phiEvec \quad ,
 \label{Cartesian_polar_relns}\\
   \hat{\boldsymbol{z}}& = & \cos{\thetaE}\, \rEvec - \sin{\thetaE}\, \thetaEvec \quad , \nonumber
\end{eqnarray}
with an analogous system of identities for general positions along the path.
One can then express the unit momentum vector \teq{\kvechat} 
of the emitted photon in terms of coordinates at the emission point:
\begin{equation}
   \kvechat \; =\; \cos\deltaE\, \rEvec
      + \sin\deltaE \cos\omegaE\,  \thetaEvec 
      + \sin\deltaE \sin\omegaE\, \phiEvec \quad .
 \label{eq:kvec_polar_reln}
\end{equation}
In the absence of GR modifications, the vector \teq{\kvechat} is a constant 
along the photon path.
In this expression, \teq{\deltaE} is the angle between {\bf k} and the 
radial direction at the emission point (see the left panel of Fig.~\ref{fig:geometry}), 
and \teq{\omegaE} is an azimuth angle at the emission point relative 
to the \teq{\hat{\boldsymbol{x}} - \hat{\boldsymbol{z}}} plane,
as illustrated in the right panel of Fig.~\ref{fig:geometry}.  Specifically, 
\begin{equation}
   \cos{\omegaE} \; =\; \dover{\kvechat\cdot\thetaEvec}{\sin{\deltaE}} \quad .
 \label{eq:omegaE_def}
\end{equation}
For flat spacetime propagation, the position vector {\bf r} along the path 
is given by \teq{\mathbf{r} = \rE \rEvec + s \kvechat},
where \teq{s} is the path length of the photon trajectory.

A sometimes advantageous rearrangement of the emission locale 
coordinates can be realized by rotating in the \teq{\rEvec - \thetaEvec} plane to align one 
basis axis with the magnetic field direction.  This can be encapsulated in the identities
\begin{eqnarray}
   \rEvec \hskneg & = & \hskneg \dover{1}{\sqrt{3\cos^2\thetaE+1}}
   \left[ 2\cos\thetaE \,\BEvec - \sin\thetaE \bigl(\phiEvec \times \BEvec \bigr) \right] \; , \nonumber\\[-5.5pt]
 \label{eq:rE_thetaE_rotate}\\[-5.5pt]
   \thetaEvec \hskneg & = & \hskneg \dover{1}{\sqrt{3\cos^2\thetaE+1}}
   \left[ \sin\thetaE\, \BEvec + 2\cos\thetaE \bigl(\phiEvec \times \BEvec \bigr) \right] \; .\nonumber
\end{eqnarray}
The vectors \teq{\BEvec} and \teq{\phiEvec  \times \BEvec} are marked on the right panel of 
Fig.~\ref{fig:geometry}.  Using this choice, one can specify the photon momentum 
vector \teq{\kvechat} at the 
emission point in terms of a \teq{\{\BEvec , \phiEvec , \phiEvec \times \BEvec\}} basis.
This can be completed by inserting Eq.~(\ref{eq:rE_thetaE_rotate}) into 
Eq.~(\ref{eq:kvec_polar_reln}), though the explicit result is not displayed here.  
An alternative is to employ the angle \teq{\Thetae} 
between \teq{\kvechat} and \teq{\BEvec} and the azimuthal angle \teq{\zetae} with 
respect to \teq{\phiEvec \times \BEvec} in a spherical triangle construction.  These 
two angles are depicted in Fig.~\ref{fig:geometry}.  Then one has
\begin{equation}
   \kvechat \, =\, \cos\Thetae \, \BEvec 
      + \sin\Thetae \cos\zetae \bigl(\phiEvec \times \BEvec \bigr)
      + \sin\Thetae \sin\zetae \,\phiEvec  \; .
 \label{eq:kvec_BEvec_form}
\end{equation}
Equating the coefficients of this expression with those of the form obtained from 
Eq.~(\ref{eq:kvec_polar_reln}) using the substitution captured in 
Eq.~(\ref{eq:rE_thetaE_rotate}) yields
\begin{eqnarray}
   \cos\deltaE \hskneg & = & \hskneg \kvechat \cdot \rEvec \; =\; 
   \dover{2\cos{\Thetae}\cos\thetaE - \sin\Thetae \cos{\zetae}\sin{\thetaE}}{\sqrt{3\cos^2\thetaE+1}} \; ,\nonumber\\[-5.5pt]
 \label{eq:cos_deltaE_omegaE} \\[-5.5pt]
   \cos\omegaE \hskneg & = & \hskneg \dover{\kvechat \cdot \thetaEvec}{\sin\deltaE} \; =\;
   \dover{2\cos{\Thetae}\sin\thetaE + \sin \Thetae\cos{\zetae}\cos{\thetaE}}{\sin\deltaE \sqrt{3\cos^2\thetaE+1}} \nonumber
\end{eqnarray}
from the basis vectors in the meridional ($ \BEvec , \phiEvec \times \BEvec $) plane.  
To these can be added a third identity pertaining to the \teq{\phiEvec} direction:
\begin{equation}
    \sin\deltaE \sin\omegaE \; =\; \sin\Thetae \sin\zetae\quad .
 \label{eq:sin_deltaE_omegaE}
\end{equation}
These three relations can prove useful in expressing key results just below.

\subsubsection{Relative direction of the Photon \teq{\kvechat} and the Field \teq{\Bvec}}

In Section~\ref{sec:split_pair_phys} it will become evident that the attenuation coefficients for the 
quantum processes considered here are critically sensitive to
the angle \teq{\thetakB} between the momentum of the photon at any point 
on its trajectory and the local magnetic field.  This angle is given by
\begin{equation}
   \cos\thetakB \; =\;  \dover{\kvechat\cdot\boldsymbol{B}}{\vert \boldsymbol{B}\vert} 
   \quad \hbox{or}\quad
   \sin\thetakB \; =\;  \dover{\vert \kvechat\times\boldsymbol{B}\vert}{\vert \boldsymbol{B}\vert} \quad ,
 \label{eq:thetakB_def}
\end{equation}
noting that \teq{\vert \kvechat \vert =1}.
We seek an expression for this angle in terms of the polar angles \teq{\theta ,\phi} at points along 
the path, the pathlength \teq{s}, and the emission locale parameters 
\teq{(\rE, \thetaE)} and direction parameters \teq{(\deltaE, \omegaE )}.  Since \teq{\kvechat} is fixed,
the direction of the field at each locale along the 
trajectory determines the value of \teq{\thetakB}. This angle can be expressed using 
local polar coordinates \teq{(r,\theta,\phi)} along its trajectory.  The distance from 
the star center is
\begin{equation}
   r \;\equiv\; \vert \boldsymbol{r} \vert
   \; =\; \bigl\vert \boldsymbol{\rE}+s\kvechat \bigr\vert
   \; =\;\sqrt{\rE^2+s^2+2 s \rE\cos{\deltaE}}\quad ,
 \label{eq:r_path}
\end{equation}
which is quickly gleaned from Fig.~\ref{fig:geometry}.
Similarly, one can express the {\bf propagation angle} \teq{\eta} 
depicted in the left panel of Fig.~\ref{fig:geometry} using the 
cosine or sine rules: 
\begin{equation}
   \cos\eta \; =\; \dover{r^2+\rE^2-s^2}{2r\rE}
   \quad ,\quad
   \sin\eta \; =\; \dover{s}{r}\, \sin\deltaE \quad .
 \label{eq:eta_def}
\end{equation}
These definitions apply regardless of whether or not the propagation 
vector \teq{\kvechat} is co-planar with magnetic field loops.
This angle will prove useful for the development of analytic 
asymptotic approximations in due course, serving as an
alternative to the pathlength variable \teq{s}.  At the point 
of emission, \teq{\eta=0}.  As \teq{r\to\infty} out to an observer, \teq{\boldsymbol{k}} becomes 
essentially parallel  to the local radial vector \teq{\boldsymbol{r}}, and \teq{\eta\to\deltaE}:
see the left panel of Fig.~\ref{fig:geometry} or Eq.~(\ref{eq:kvec_polar_reln}).

Expressions for the two polar coordinate angles contain more involved 
geometrical algebra.  Given that \teq{\mathbf{r} = \rE \rEvec + s \kvechat},
using Eqs.~(\ref{Cartesian_polar_relns})
and~(\ref{eq:kvec_polar_reln}) the polar angle (magnetic colatitude)
can be expressed via
\begin{eqnarray}
   \cos\theta & = & \dover{\boldsymbol{r}\cdot\hat{\boldsymbol{z}}}{r} \nonumber\\[-5.5pt]
 \label{eq:cos_theta_path}\\[-5.5pt]
   & = & \dover{(\rE + s\cos{\deltaE})\cos{\thetaE} - s\sin{\deltaE}\cos{\omegaE}\sin{\thetaE}}{
              \sqrt{\rE^2+s^2+2 s \rE\cos{\deltaE}}} \; .\nonumber
\end{eqnarray}
One could form a result using \teq{\sin\theta = \vert \boldsymbol{r} \times \hat{\boldsymbol{z}}\vert /r}.
Yet, in practice, only a form for \teq{r^2\sin^2\theta} is needed, and this is 
routinely obtained from Eq.~(\ref{eq:cos_theta_path}).  The azimuthal angle or 
magnetic longitude is encapsulated in the relation
\begin{eqnarray}
   r_x & \equiv & \boldsymbol{r}\cdot\hat{\boldsymbol{x}} \; =\; r\, \sin\theta \cos\phi \nonumber\\[-5.5pt]
 \label{eq:cos_phi_path}\\[-5.5pt]
   & = & (\rE+s\cos{\deltaE})\sin{\thetaE}+s\sin\deltaE \cos\omegaE \cos \thetaE \quad .\nonumber
\end{eqnarray}
The alternative relation for \teq{\phi} is
\begin{equation}
   r_y \; \equiv \; \boldsymbol{r}\cdot\hat{\boldsymbol{y}} \; =\; r\, \sin\theta \sin\phi 
   \; =\; s\sin{\deltaE}\sin{\omegaE} \quad ;
 \label{eq:sin_phi_path}
\end{equation}
both forms can be employed in Eq.~(\ref{eq:cos_thetakB}) below
to eliminate the explicit dependence on \teq{\phi}.
Observe that forming the dot product using the perpendicular axes 
\teq{\hat{\boldsymbol{x}}} and \teq{\hat{\boldsymbol{y}}} is the most convenient path to 
isolating these identities for the projections of \teq{\boldsymbol{r}} onto the 
\teq{(x,y)}-plane.  Summing the squares of Eqs.~(\ref{eq:cos_phi_path})
and~(\ref{eq:sin_phi_path}) yields the form for \teq{r_x^2+r_y^2=r^2\sin^2\theta}.
Then adding to this \teq{r^2\cos^2\theta} using Eq.~(\ref{eq:cos_theta_path}) 
and equating to the value of \teq{r^2} obtained from Eq.~(\ref{eq:r_path}) generates
spherical trigonometric relations between \teq{\deltaE}, \teq{\thetaE} and \teq{\omegaE}
to close the group of geometric identities.

To express the magnetic field in Eq.~(\ref{eq:B_dipole}) we now have 
forms for two basis vectors, namely the radial direction
\begin{equation}
   \hat{\boldsymbol{r}} \; =\; \dover{\rE \rEvec + s \kvechat}{\sqrt{\rE^2+s^2+2 s \rE\cos{\deltaE}}} \quad ,
 \label{eq:rvechat_form}
\end{equation}
where \teq{\kvechat} is written in Eq.~(\ref{eq:kvec_polar_reln}), 
and the polar angle direction
\begin{eqnarray}
   \hat{\boldsymbol{\theta}} & = & (\sin{\thetaE}\cos{\theta}\cos{\phi}-\cos{\thetaE}\sin{\theta}) \,\rEvec \nonumber\\[2.0pt]
   && \quad +(\cos{\thetaE}\cos{\theta}\cos{\phi}+\sin{\thetaE\sin{\theta}}) \,\thetaEvec 
 \label{eq:thetavechat_form}\\[2.0pt]
   && \quad +\cos{\theta}\sin{\phi} \,\phiEvec \quad .\nonumber
\end{eqnarray}
This second vector is obtained from the \teq{\{\rEvec,\thetaEvec,\phiEvec\}} triad 
via a rotation through \teq{\theta-\thetaE} about the \teq{y} axis 
followed by a rotation through \teq{\phi} about the \teq{z}-axis.
Then, selecting the dot product form in Eq.~(\ref{eq:thetakB_def}), one can 
evaluate \teq{\cos\thetakB} as
\begin{eqnarray}
   \cos\thetakB & = & \dover{1}{\sqrt{3\cos^2\theta+1}}\bigg[2\cos{\theta}\left(\frac{\rE\cos{\deltaE}+s}{r}\right)\bigg.\nonumber\\[3.0pt]
   & + & \hspace{-6pt} \sin{\theta}\cos{\deltaE}(\sin{\thetaE}\cos{\theta}\cos{\phi}-\cos{\thetaE}\sin{\theta})\nonumber\\[-4.5pt]
 \label{eq:cos_thetakB}\\[-3.5pt]
   & + & \hspace{-6pt} \sin{\theta}\sin{\deltaE}\cos{\omegaE}(\cos{\thetaE}\cos{\theta}\cos{\phi}+\sin{\thetaE\sin{\theta}}) \nonumber\\
   & + & \hspace{-6pt} \sin{\theta}\sin{\deltaE}\sin{\omegaE}\cos{\theta}\sin{\phi}\bigg] \quad . \nonumber
\end{eqnarray}
Converting Eq.~(\ref{eq:cos_thetakB}) into a sine, one can then compute the attenuation opacity
at any point of the photon trajectory, using the relativistic quantum forms for the 
attenuation rates expounded in Section~\ref{sec:split_pair_phys}.

A case of particular interest is that where photons propagate in the plane of a set of field loops,
termed the {\bf meridional plane},
a situation that is depicted in the left panel of Fig.~\ref{fig:geometry}.
This is fairly representative of general expectations because in most models of 
pulsar and magnetar hard X-ray and gamma-ray emission, 
the photons are created by ultra-relativistic electrons moving essentially along 
field lines, at least at low magnetospheric altitudes.  One can therefore set \teq{\phi=0}.  Doppler aberration for  
\teq{\gammae\gg1} circumstances generates very small angles \teq{\Thetae \lesssim 1/\gammae\ll1}
between the photon momentum vector and the magnetic field at the point of emission.  
Thus, \teq{\thetakB} normally starts at very small values, and 
increases as the photon propagates.   In cases where photons are
emitted exactly parallel to the local field line, i.e., \teq{\Thetae=0}, 
one can set \teq{\omegaE = 0}, and it then follows 
that \teq{2\tan\deltaE = \tan\thetaE}. Then the 
algebra for the angle \teq{\thetakB} simplifies dramatically, yielding the forms
\begin{eqnarray}
   \cos\thetakB\hsknegsm  & =  & \hsknegsm 
   \dover{\sin\theta\sin(\deltaE - \eta)+2\cos{\theta}\cos(\deltaE - \eta)}{\sqrt{3\cos^2\theta+1}} \nonumber\\[-1.5pt]
 \label{eq:thetakB_plane}\\[-1.5pt]
   \sin\thetakB\hsknegsm & = & \hsknegsm 
   \dover{\sin\theta\cos(\deltaE - \eta)-2\cos{\theta}\sin(\deltaE - \eta)}{\sqrt{3\cos^2\theta+1}} \quad ,\nonumber
\end{eqnarray}
remembering that \teq{\eta = \theta - \thetaE} in this special meridional case.
This result reproduces Eq.~(14) of \citet{SB14}.   Observe that here
\teq{\thetakB} is independent of the \teq{r} coordinate because of the 
self similarity of the dipole field configuration that is sampled for 
propagation in planes containing field loops; in general, \teq{\thetakB} is 
dependent on \teq{r} --
see Eq.~(\ref{eq:cos_thetakB}).

\subsection{Photon Splitting and Pair Creation Physics}
 \label{sec:split_pair_phys}

Having outlined the geometrical construction pertinent to the 
propagation calculation, now the key results for 
the attenuation coefficients \teq{{\cal R}} for the two quantum 
processes under consideration, photon splitting 
\teq{\split} and pair creation \teq{\pairs}, are summarized.
Extensive discussion of the two processes is provided in 
\cite{HBG97,BH01,HL06}, and references therein, so here 
a briefer exposition suffices.

Magnetic photon splitting is a third-order QED process 
in which a single photon splits into two lower-energy photons \cite{Adler71,BH97}.  
It is forbidden in the absence of an external field due to a 
charge conjugation symmetry (Furry's theorem) concerning propagators in the triangular 
Feynman diagram, and becomes possible as the 
field breaks this symmetry.  It operates efficiently and competes effectively with
one photon (magnetic) pair production in magnetic fields above \teq{\sim
10^{13}}Gauss \citep{HBG97} because it has no mass/energy threshold, and 
so can attenuate hard X-rays, typically in the 50-500 keV band.
The magnetized vacuum is dispersive and birefringent, so that the
refractive index deviates slightly from unity, and the phase velocity of light
is less than \teq{c} and depends on the polarization state of photons.
The physics of this vacuum dispersion is addressed extensively in \cite{Adler71} and \cite{Shabad75},
and we note that the directions of the resultant photons are almost but not quite co-linear with 
that of the attenuated photon.
In the limit of very weak dispersion, there are three polarization modes 
of splitting that are permitted by the CP (charge-parity) invariance in QED: 
\teq{\perp \rightarrow \parallel \parallel}, \teq{\parallel \rightarrow \perp \parallel} 
and \teq{\perp \rightarrow \perp \perp}.  Here, as in many previous 
papers, we adopt the linear polarization convention for light where
\teq{\parallel} (ordinary or O-mode) refers to the state with the photon's {\it electric}
field vector parallel to the plane containing the magnetic field \teq{\boldsymbol{B}}
and the photon's momentum vector (i.e., \teq{\boldsymbol{k}}), while \teq{\perp} 
(extraordinary or X-mode) denotes the photon's electric field vector being normal to this plane. 

\cite{Adler71} observed that in order to conserve both energy and momentum, 
the influence of vacuum dispersion was to restrict \teq{\split} to the 
\teq{\perp \rightarrow \parallel \parallel} mode below pair production
threshold, and prohibit the other two 
CP-permitted channels, {\it kinematic selection rules} that can 
potentially seed strong polarization of X-ray spectra.  This result was derived 
in the weak, linear domain assuming that magnetized vacuum contributions dominate 
the dispersion relation, wherein pair creation \teq{\pairs} is the dominant 
absorptive process.  In highly super-critical fields, quadratic and higher order resonant
contributions to the generalized vacuum polarizability tensor 
may modify this result.  In parallel, when plasma densities are high, or very near 
the cyclotron resonance, plasma dispersion competes with the magnetic vacuum 
contribution, thereby modifying both the propagation characteristics and the 
splitting selection rules.  The competition between vacuum and plasma dispersion 
spawns the so-called vacuum resonance \citep[e.g., see][ and references therein]{Meszaros1992}, 
at which polarization 
states for photons are modified in magnetized media with strong density gradients.
Examples for sites sampling the vacuum resonance include accretion columns 
in X-ray pulsars \citep[e.g.,][]{MV78,GPS78}
and the surface layers of magnetars \citep[e.g.,][]{Ozel2001,Lai2003}.
While we provide here forms for all three CP-permitted splitting rates, 
throughout much of the ensuing exposition, it will be assumed that 
Adler's kinematic selection rules apply, yielding an emphasis on results 
for \teq{\perp \rightarrow \parallel \parallel} splittings.

Fully general expressions for photon splitting rates are mathematically
complicated, and usually not required.  S-matrix determinations 
of the rates provided in \cite{Baring00} involve triple summations of combinations of 
elementary functions over the Landau level quantum numbers of virtual electrons.  
Euler-Heisenberg formulations are offered in \cite{Adler71}, and these incorporate 
triple integrals.  In the low-energy limit well below the pair creation 
threshold of \teq{\erg\sin\thetakB =2} (hereafter, all photon energies 
are expressed dimensionlessly, scaled by \teq{m_ec^2}), comparatively 
simple, single-integral expressions are obtainable because the energy and
magnetic field dependences of the splitting amplitudes become separable.  
These are the forms employed here.  To express them, we introduce 
two functions that serve as reaction amplitude coefficients, 
\teq{{\cal M}_1}, \teq{{\cal M}_2} defined by 
\begin{equation}
   {\cal M}_{\sigma} \; =\; \dover{1}{B^4}\int_{0}^{\infty} \dover{ds}{s}e^{-s/B} \, \Lambda_{\sigma} (s) \quad ,
 \label{eq:calM_i_form}
\end{equation}
where
\begin{eqnarray}
   \Lambda_1(s) \hspace{-6pt}& = & \hspace{-6pt} 
   \left(-\frac{3}{4s}+\frac{s}{6}\right)\frac{\cosh s}{\sinh s}+\frac{3+2s^2}{12\sinh^2s}+\frac{s\cosh s}{2\sinh^3s} \quad , \nonumber\\[-5.5pt]
 \label{eq:Lambda_s_def}\\[-5.5pt]
   \Lambda_2(s) \hspace{-6pt} & = & \hspace{-6pt} 
   \frac{3}{4s}\frac{\cosh s}{\sinh s}+\frac{3-4s^2}{4\sinh^2s}-\frac{3s^2}{2\sinh^4s} \quad ,\nonumber
\end{eqnarray}
forms that were first identified by \cite{Adler71}.  In these integrals, 
and throughout this paper, the magnetic field \teq{B}
is expressed in dimensionless form, being scaled by the 
quantum critical field \teq{B_{\rm cr}=m_e^2c^3/(e\hbar) \approx 4.41\times 10^{13}} Gauss,
at which the electron cyclotron and rest mass energies are equal.
In the limit of \teq{B\ll1}, \teq{{\cal{M}}_1\approx 26/315} and 
\teq{{\cal{M}}_2\approx 16/105} are both independent of the 
field strength, while for highly-supercritical fields, \teq{B\gg1}, 
one can quickly determine that \teq{{\cal{M}}_1\approx 1/(6B^3)} 
and \teq{{\cal{M}}_2\approx 1/(3B^4)}.  Since the splitting rates are 
proportional to the \teq{({\cal M}_{\sigma})^2}, it is apparent that the 
\teq{{\cal M}_{\sigma}} constitute strong-field modification factors.

The approximate splitting rates can be scaled by \teq{1/c} and thereby expressed 
as inverse attenuation lengths, i.e., {\it attenuation coefficients}
\teq{{\cal R}^{\rm sp}}, using these two functions.  For the 3 CP-permitted modes, 
photons of dimensionless energy \teq{\erg} propagating at 
angle \teq{\thetakB} to the local field direction have coefficients
\begin{eqnarray}
   {\cal R}^{\rm sp}_{\perp\to\parallel\parallel} &=& 
   \dover{\fsc^3}{60\pi^2\lambar_c}\, \erg^5\, B^6\, {\cal M}_1^2 \, \sin^6\thetakB
   \; =\; \dover{1}{2}\, {\cal R}^{\rm sp}_{\parallel\to\perp\parallel} \quad ,
  \nonumber\\[-5.5pt]
  &&  \label{eq:splitt_atten} \\[-5.5pt]
   {\cal R}^{\rm sp}_{\perp\to\perp\perp} &=& 
   \dover{\fsc^3}{60\pi^2\lambar_c }\, \erg^5\, B^6\, {\cal M}_2^2 \, \sin^6\thetakB \quad .\nonumber
\end{eqnarray}
These forms are integrated over the energies of the produced photons, 
which sum to \teq{\erg m_ec^2}.
Here \teq{\fsc =e^2/\hbar c} is the fine-structure constant, and \teq{\lambar_c=\hbar/m_ec} 
is the reduced Compton wavelength of the electron. 
The \teq{\thetakB=\pi /2} specialization of Eq.~(\ref{eq:splitt_atten}) is listed in \cite{BH01}, and the 
\teq{\sin^6\thetakB} angle dependence captures Lorentz transformation 
properties for boosts along \teq{\Bvec} in this special 
\teq{\erg\sin\thetakB \ll 2} case.  Observe that the \teq{1/2} factor multiplying 
\teq{{\cal R}^{\rm sp}_{\parallel\to\perp\parallel}} in Eq.~(\ref{eq:splitt_atten}) accounts 
for the two possibilities \teq{\parallel \to \perp\parallel} and \teq{\parallel \to \parallel\perp}, 
which have identical rates \citep[see Eq.~(23) of ][]{Adler71}; thus this first relation 
encapsulates crossing symmetries for splitting.  The strong dependence on \teq{\erg} 
guarantees that photon splitting can be very effective at attenuating 
soft gamma-rays provided that \teq{\thetakB} is not small.  For 
unpolarized photons, Eq.~(\ref{eq:splitt_atten}) can be averaged to yield 
\begin{equation}
   {\cal R}^{\rm sp}_{\rm ave} \; =\;  \dover{\fsc^3}{120\pi^2\lambar_c }\, \erg^5\, B^6\,
    \Bigl( 3{\cal M}^2_1+{\cal M}^2_2\Bigr) \, \sin^6\thetakB \quad .
 \label{eq:split_pol_ave}
\end{equation}
Such a form is useful if all three CP-permitted modes of splitting are assumed to operate, 
in which repeated splittings are possible, and a splitting cascade can develop, 
producing numerous photons of lower energy with potentially discernible
spectral bumps near the escape energy: see \cite{Baring95,HB96,HBG97}.
Note also that for \teq{\erg\sin\thetakB \lesssim 2} near the pair threshold,
\cite{BH97} provided a compact, empirical approximation to the attenuation 
coefficient for the \teq{\perp\to\parallel\parallel} mode.

One-photon pair production \teq{\pairs} is a first-order QED process that is 
forbidden in field-free regions because four-momentum cannot be conserved. 
An external magnetic field can absorb momentum perpendicular to
\teq{\Bvec}, and so the pair conversion of a single photon can proceed.  Early
determinations of the rate by \cite{Toll52} and \cite{Klep54} revealed 
that it rises exponentially with
increasing photon energy and magnetic field strength, becoming
significant for gamma-rays above the absolute threshold, \teq{\erg =
2/\sin\thetakB}, and for fields approaching \teq{B_{\rm cr}}. This threshold 
applies to the parallel (\teq{\parallel}) polarization, with that for the 
\teq{\perp} polarization being a factor of \teq{1+\sqrt{1+2B}} higher: 
see Eq.~(\ref{eq:calFpp_perp}).
In general, the produced pairs occupy excited Landau 
levels in a magnetic field, and for each Landau level configuration of the pairs, 
the attenuation coefficient \teq{{\cal R}^{\rm pp}} for the reaction exhibits a divergent resonance at
each pair state threshold, producing a characteristic sawtooth structure
\citep[e.g.,][]{DH83,BK07}.  These mathematically-complicated forms involve 
sums over associated Laguerre functions and are often unwieldy in numerical 
analyses such as opacity computations.  As
the divergences are integrable in the photon energy dimension,
mathematical approximations to the rate can 
be developed using proper-time techniques originally due to \cite{Schwin51} 
that effectively average over ranges of \teq{\erg} somewhat broader than the 
separation of neighboring resonances.  The most 
widely-used expressions of this genre are those derived in \cite{Klep54},
\cite{Erber66} and  \cite{TE74}; see also the book of \cite{ST68}.  When
expressed as attenuation coefficients, they assume the form 
\begin{equation}
   {\cal R}^{\rm pp}_{\parallel,\perp} \; =\; \dover{\fsc}{\lambar_c} 
   \, B\sin\thetakB\,  {\cal F}_{\parallel,\perp} \left(\erg_\perp,\, B\right)
   \; ,\quad
   \erg_{\perp}\; =\; \erg \sin\thetakB \quad .
  \label{eq:pp_general}
\end{equation}
As in Eq.~(\ref{eq:splitt_atten}), covariant structure under Lorentz boosts 
parallel to \teq{\Bvec} is apparent, since \teq{\erg_{\perp} = \erg \sin\thetakB} 
is a conserved quantity under such transformations.  Thus, the appearance 
of the extra \teq{\sin\thetakB} factor captures time dilation or length contraction information
in such boosts.  The specific forms first obtained by \cite{Erber66} were 
\begin{eqnarray}
   {\cal F}_{\perp} & = & \dover{1}{2}\, {\cal F}_{\parallel} 
   \; =\; \dover{2}{3}\, \calFErber\quad ,\nonumber\\[-5.5pt]
 \label{eq:Erber_asymp}\\[-5.5pt]
   && \calFErber \left(\erg_\perp,\, B\right)\; =\; 
              \dover{3\sqrt{3}}{16\sqrt{2}} \exp \left(-\dover{8}{3 \erg_{\perp} B}\right)\quad .\nonumber
\end{eqnarray}
These are applicable when the produced pairs are 
in high Landau states, namely when \teq{\erg_\perp^2/2B \gg 1} and \teq{\erg_\perp \gg 1}.
Integrated over the momenta of the produced pairs, these are
valid only in the domain \teq{\erg_{\perp} B\lesssim 0.1}, for which the 
pairs are ultra-relativistic, and the exponential factor introduces a profound 
sensitivity to the values of \teq{\erg}, \teq{B} and \teq{\thetakB}.  
The approximations in Eq.~(\ref{eq:Erber_asymp}) become inaccurate 
near the pair threshold of \teq{\erg_{\perp} =2}, a domain that is important 
in fields \teq{B\gtrsim 0.5}, as pointed out by \cite{BH01}.  Nevertheless, 
they are often extremely useful, and have been employed widely 
in the pulsar literature, including in the seminal papers of 
\cite{Sturr71,RS75} and \cite{DH82}.

For the purposes of opacity determinations, it is necessary to employ
improvements of the asymptotic formulae in Eq.~(\ref{eq:Erber_asymp})
that accommodate reduced rates for pair production near threshold.  Here we adopt 
the protocol of \cite{SB14}, who used the analysis of threshold corrections
provided by \cite{BK07}, specifically their Eq.~(3.4), yielding the form
\begin{equation}
   \calFTH \left(\erg_\perp,\, B\right) \;=\;
                \dover{3\erg_\perp^2-4}{2\erg_\perp^2 \,
               \sqrt{ {\cal L}(\erg_{\perp})\, \phi (\erg_{\perp}) }}
               \, \exp \left\{ -\dover{\phi(\erg_\perp)}{4B}  \right\}\quad ,
 \label{eq:B88_asymp}
\end{equation}
valid for \teq{\erg_{\perp} > 2}, where 
\teq{\phi (\erg_{\perp}) \; =\; 4\erg_{\perp} -{\cal L}(\erg_{\perp})} and
\begin{equation}
    {\cal L}(\erg_{\perp}) \; =\;  \left( \erg_{\perp}^2 - 4 \right) \log_e\left( \dover{\erg_{\perp}+2}{\erg_{\perp}-2}\right)\quad .
 \label{eq:calL_def}
\end{equation}
This analytic result improves upon the Erber form in Eq.~(\ref{eq:Erber_asymp}) by several orders
of magnitude near the pair threshold \teq{\erg_{\perp}\sim 2}, and reduces to Erber's result in the limit
\teq{\erg_{\perp}\gg 1}.  It is a more accurate approximation than the very similar form obtained 
in \cite{B88}.  Eq.~(\ref{eq:B88_asymp}) incorporates domains where the created pairs are 
mildly-relativistic, and can be reliably applied in fields up to \teq{B\sim 0.5} when 
\teq{\erg_{\perp} B\gg 1}. 
\citet{BK07} also presented in their Eq.~(B.5) polarization-dependent forms to account
for near-threshold modifications to the polarized rate.  Their results satisfied
\teq{{\cal F}_{\perp}\approx (\erg_{\perp}^2-4)/(2\erg_{\perp}^2)\, {\cal F}_{\parallel}}, 
from which one obtains
\begin{equation}
   {\cal F}_{\perp} \left(\erg_\perp,\, B\right) \;=\;
                \dover{\erg_\perp^2-4}{\erg_\perp^2 \,
               \sqrt{ {\cal L}(\erg_{\perp})\, \phi (\erg_{\perp}) }}
               \, \exp \left\{ -\dover{\phi(\erg_\perp)}{4B}  \right\}\quad ,
 \label{eq:BK07_perp_asymp}
\end{equation}
and a partner form for \teq{{\cal F}_{\parallel}} that can be applied even fairly close to its 
\teq{\erg_{\perp}=2} threshold.  These results are used in our computations, 
though we observe that since Eq.~(\ref{eq:BK07_perp_asymp}) does not contain 
terms dependent on \teq{1+\sqrt{1+2B}}, it may not be precise near the 
pair threshold for \teq{\perp} photons.

To provide an accurate treatment of the polarization dependence of pair thresholds,
\cite{SB14} followed \cite{HBG97} in adding a ``patch''  using the exact 
attenuation coefficients from Eq.~(6) of \cite{DH83} at sufficiently low energies, namely
 \teq{\erg_{\perp} < 1 + \sqrt{1+4 B}}.  We adopt an identical 
procedure here, for which photons with parallel and perpendicular
polarization produce pairs only in the ground (0,0) and first excited
(0,1) and (1,0) states respectively.  Here \teq{(j,k)} denotes the Landau
level quantum numbers of the produced pairs.   Thus, including
only the (0,0) pair state for \teq{\parallel} polarization gives
\begin{equation}
   {\cal F}^{\rm pp}_{\parallel} \; =\; \dover{2}{\erg_{\perp}^2  \vert p_{\hbox{\sevenrm 00}}\vert}
   \,\exp\left(-\dover{\erg_{\perp}^2}{2B}\right)
   \quad , \quad 
   \erg_{\perp} \geq 2\quad ,
  \label{eq:calFpp_par}
\end{equation}
and only the sum of the (0,1) and (1,0) states for $\perp$ polarization yields
\begin{equation}
   {\cal F}^{\rm pp}_{\perp} \; =\; \dover{2E_0(E_0+E_1)}{\erg_{\perp}^2  \vert p_{\hbox{\sevenrm 01}}\vert}
   \,\exp\left(-\dover{\erg_{\perp}^2}{2B}\right)
   \; , \; 
   \erg_{\perp} \geq 1 + \sqrt{1+ 2B} \; .
  \label{eq:calFpp_perp}
\end{equation}
In these expressions, two energies of pairs appear:
\begin{displaymath}
   E_0  = (1 + p_{01}^2)^{1/2}\quad , \quad  
   E_1  = (1 + p_{01}^2 + 2B)^{1/2}\quad ,
\end{displaymath}
where
\begin{displaymath}
   \vert p_{jk}\vert  = \left[ \dover{\erg_{\perp}^2}{4} - 1 - (j+k)B + 
   \left( \dover{(j-k)B}{\erg_{\perp}} \right)^2\right]^{1/2}  \quad ,
\end{displaymath}
describes the magnitude of the momentum parallel to \teq{\Bvec}.
The pair production rate in this regime behaves like a wall at the 
thresholds given in Eqs.~(\ref{eq:calFpp_par}) and~(\ref{eq:calFpp_perp}), 
and photons will create pairs as soon as they satisfy these kinematic
restrictions.  Note that we performed numerical tests that incorporated the 
next pair state threshold (\teq{p_{01}}) for the \teq{\parallel} polarization to 
add to the form Eq.~(\ref{eq:calFpp_par}), and found that the 
attenuation opacities were insensitive to this inclusion. 
Also, in the detailed analysis of pair opacities 
in pulsars and magnetars, \cite{SB14} observed that escape energies were 
not sensitive to the polarization character of this patch for pulsar fields, 
but that the polarization dependence of the threshold was influential 
for the supercritical field domain of magnetars, the main focus here.
The importance of near-threshold treatment of pair opacity when 
\teq{B >1} will become apparent in due course.

\section{Photon Splitting Attenuation Lengths}
 \label{sec:split_atten_lgth}
In this section we compute the attenuation length for the photon splitting only.  While escape energies are
the prime focus of the paper, a brief presentation of attenuation lengths is motivated in validating the
numerical determination of opacities. The reason why discussion of pair creation is omitted in this Section
is that it is extensively documented in the works by \cite{HBG97}, \cite{BH01} and \cite{SB14}.  In
particular, the analytic results herein for splitting attenuation of photons emitted above the stellar
surface have not been developed elsewhere. In considering opacities for pair creation, \cite{SB14} developed
an analytical approach to the integral of optical depth in Eq.~(\ref{eq:opt_depth}) for photons initially
emitted parallel to the field line near the magnetic axis, and for quasi-polar emission locales. Here we
adopt a similar procedure and produce analogous analytic approximations for photon splitting. 
Flat Minkowski spacetime is presumed throughout this Section.

Using the propagation angle \teq{\eta} as the integral variable instead of the pathlength \teq{s}, the optical 
depth for photon splitting can be stated as
\begin{equation}
   \tau_{\sigma}(l) \;=\; \dover{\fsc^3\,\erg^5}{60\pi^2\lambar_c}\,\int_{0}^{\eta(l)}\,B^6\,{\cal{M}}^2_{\sigma}\,\sin^6{\thetakB}\frac{ds}{d\eta}\,d\eta\quad.
 \label{eq:tau_sigma}
\end{equation}
Here the subscript \teq{\sigma} denotes different splitting modes and the \teq{{\cal{M}}_{\sigma}} is 
the corresponding amplitude coefficient, as defined in Eq.~(\ref{eq:calM_i_form}).  For situations 
where the emitted photon population is unpolarized, the optical depth can be expressed as an average
of the \teq{\perp} and \teq{\parallel} polarizations for the incident (attenuated) photon
\begin{equation}
   \tau_{\rm ave} \;=\; \dover{1}{2} \, \left(\tau_{\perp\to\parallel\parallel}+\tau_{\perp\to\perp\perp}+\tau_{\parallel\to\parallel\perp}\right) \quad ,
 \label{eq:tau_ave}
\end{equation}
where the \teq{\perp} state photons have two splitting channels.

The upper limit \teq{\eta(l)} is the inversion of Eq.~(\ref{eq:eta_def}) for path length \teq{s=l}; in general, 
it is algebraically non-trivial.  Note that since both the \teq{\parallel\to\perp\,\parallel} and 
\teq{\parallel\to\parallel\,\perp} orderings of the produced photons are incorporated in \teq{{\cal R}^{\rm sp}_{\parallel\to\perp\parallel}} 
in Eq.~(\ref{eq:splitt_atten}), they are also included in \teq{\tau_{\parallel\to\parallel\perp}} 
in Eq.~(\ref{eq:tau_ave}), leading to \teq{2 {\cal M}_1^2} contributions identifiable in 
Eq.~(\ref{eq:split_pol_ave}), and Eq.~(\ref{eq:atten_length_ave}) below.

\begin{figure*}
 \begin{minipage}{17.5cm}
\vspace*{-10pt}
\centerline{\hskip 10pt \includegraphics[height=7.4cm]{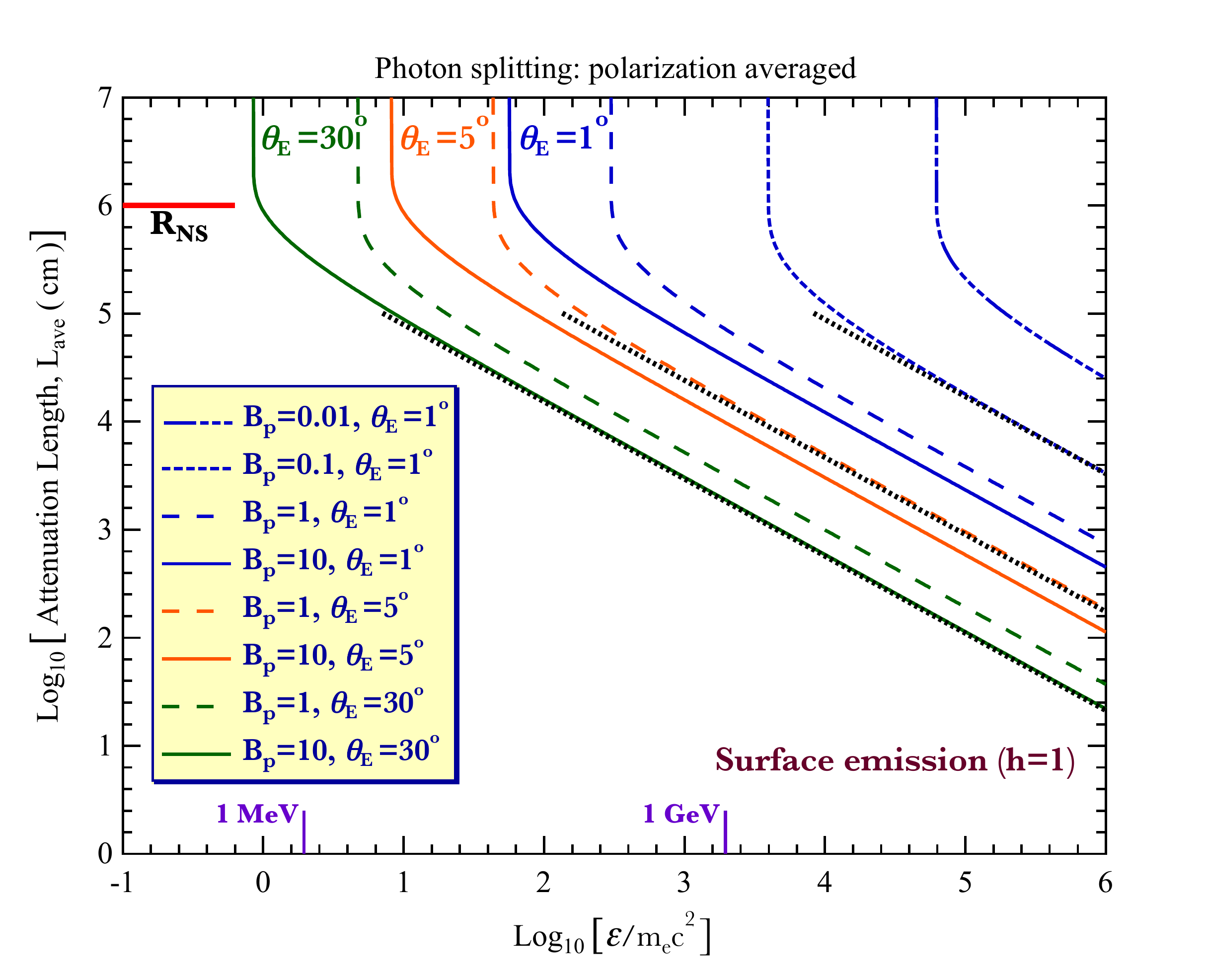}
   \hskip -15pt\includegraphics[height=7.4cm]{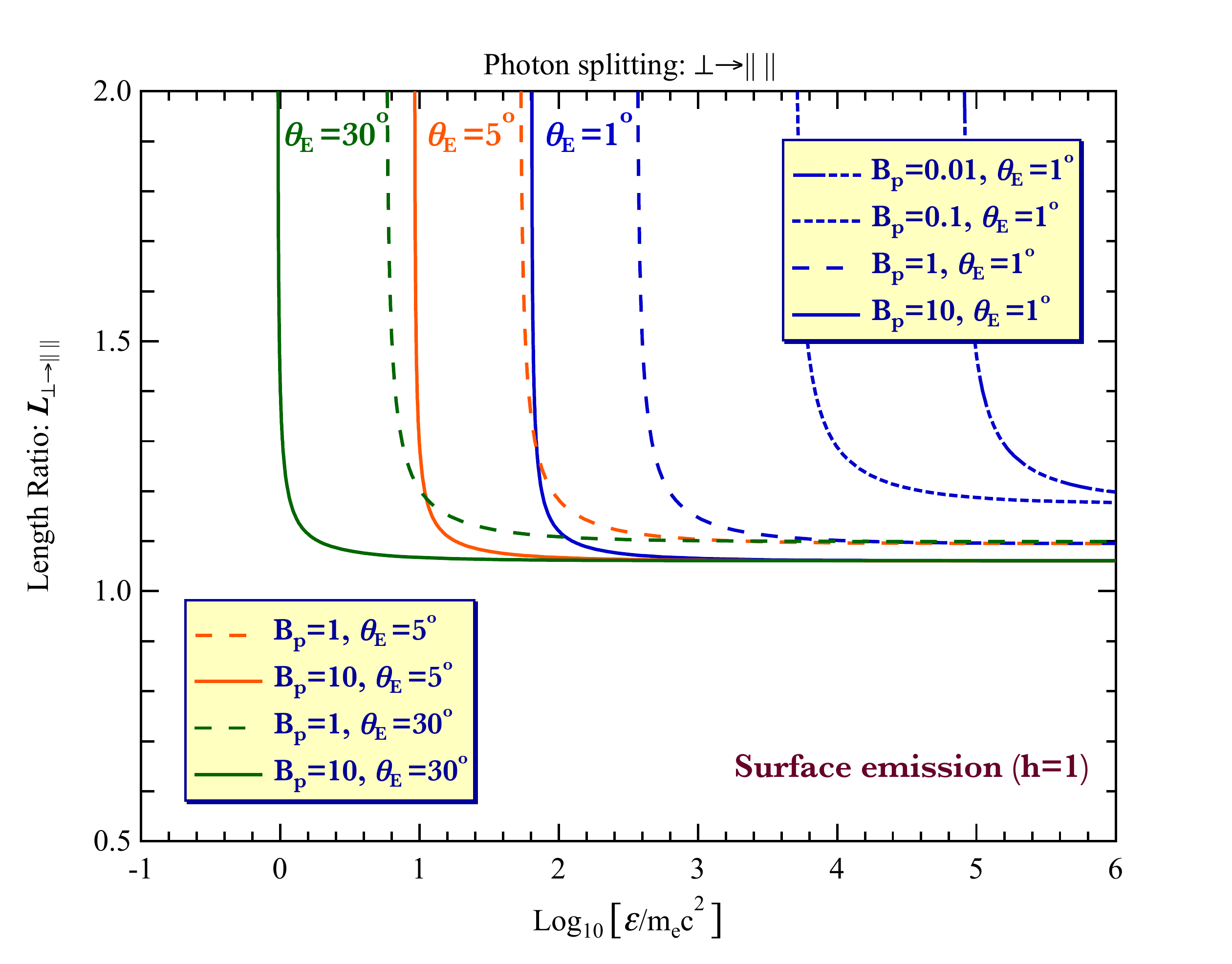}}
\vspace*{-5pt}
\caption{{\it Left Panel}: Attenuation lengths \teq{L_{\rm ave}}, averaged over all 
polarization modes for photon splitting, for light emitted from the neutron star surface \teq{(h = 1)} 
and propagating in flat spacetime, initially parallel to the local field, plotted as a function of 
photon energy (in units of \teq{m_ec^2}).  Curves constitute numerical results obtained from Eq.~(\ref{eq:tau_ave}), 
and are color-coded for emission colatitudes \teq{\thetaE =1^{\circ}} (blue), 
\teq{\thetaE =5^{\circ}} (orange) and \teq{\thetaE =30^{\circ}} (green), as indicated.
Different polar field strengths \teq{B_p} are represented, labelled in units of \teq{B_{\rm cr}}. 
The dotted straight lines represent the analytic approximation from Eq.~(\ref{eq:atten_length_ave}), for three cases:
\teq{B_p=10} and \teq{\thetaE =30^{\circ}}, \teq{B_p=1} and \teq{\thetaE =5^{\circ}}, and
\teq{B_p=0.1} and \teq{\thetaE =1^{\circ}}.  The solid red line on the upper left marks the 
neutron star radius \teq{\rns}, the pertinent scale for field line curvature near the surface.
{\it Right Panel}: The ratio \teq{\mathbfit{L}_{\perp\to\parallel\parallel}} of the attenuation 
length for the \teq{\perp\to\parallel\parallel} splitting mode from Eq.~(\ref{eq:tau_sigma}) to the average 
value \teq{L_{\rm ave}} derived from computation of Eq.~(\ref{eq:tau_ave}).  The sequence 
of \teq{\thetaE } and \teq{B_p} values and associated color coding matches that on the left.
See the text for a discussion of the two asymptotic domains at large \teq{\varepsilon} and large 
\teq{\mathbfit{L}_{\perp\to\parallel\parallel}}.
 \label{fig:attenlen_split_surface}}
\end{minipage}
\end{figure*}

Considering only the case of a photon emitted at small colatitude, \teq{\thetaE \ll 1}, and the  direction of 
the photon is exactly parallel to the local field line (i.e., \teq{\Thetae =0}), \teq{\eta} is well approximated 
by \teq{l/(l+\rE)}. In this case, we also have
\begin{equation}
   \deltaE\;\approx\; \frac{1}{2}\thetaE\quad,
 \label{eq:deltaE_approx}
\end{equation}
as the emission angle relative to the local radial direction. Eq.~(\ref{eq:B_dipole}) and 
Eq.~(\ref{eq:thetakB_plane}) can be approximated as
\begin{equation}
   B \;\approx\; B_p\,\dover{(\deltaE-\eta)^3}{h^3\deltaE^3} 
   \quad \hbox{and} \quad
   \sin{\thetakB} \;\approx\; \frac{3}{2}\eta \quad ,
 \label{eq:B_thetakB_polar}
\end{equation}
where \teq{h=\rE/\rns} is the emission altitude scaled  by the radius of the star.
In addition, the Jacobian
\begin{equation}
   \dover{ds}{d\eta} \;\approx\; \dover{\rE\deltaE}{(\deltaE-\eta)^2}
 \label{eq:ds_deta}
\end{equation}
can be deduced with the help of the sine law adapted from Eq.~(\ref{eq:eta_def}).
These results can be inserted into Eq.~(\ref{eq:tau_sigma}), yielding an approximation for the 
optical depth. By changing the integration variable into \teq{x=\eta/\delta_E}, 
and writing \teq{\xmax = l/(l+\rE )}, one arrives at
\begin{equation}
   \tau_{\sigma} \;\approx\; \dover{243\fsc^3}{81920\pi^2}  \dover{B_{p}^6 \erg^5\thetaE^6}{h^{17}} 
   \dover{\rns}{\lambar_c}\int_{0}^{x_{\hbox{\fiverm max}}} {\cal{M}}^2_{\sigma}(1-x)^{16}x^6\,dx\,.
 \label{eq:tau_polar_thetaE}
\end{equation}
Note that the magnetic field strength \teq{B\approx B_p{(1-x)^3}/h^3} is the argument of 
\teq{{\cal{M}}_{\sigma}} and so appears inside the integral. 

If one considers emission at different 
colatitudes but along a same field line, one has \teq{\sin^2{\thetaE}=h \sin^2{\theta_f}}. 
Here \teq{\theta_f} is the footpoint colatitude, which is the colatitude of the field line when it 
intersects the surface of the star.  In the small colatitude limit corresponding to polar locales, 
\teq{\thetaE\approx\theta_f\sqrt{h}}.  Then the optical depth integral assumes the form
\begin{equation}
   \tau_{\sigma}\; \approx\;\dover{243\fsc^3}{81920\pi^2}  \dover{B_{p}^6 \erg^5\theta_f^6}{h^{14}} 
   \dover{\rns}{\lambar_c}\int_{0}^{x_{\hbox{\fiverm max}}} {\cal{M}}^2_{\sigma}(1-x)^{16}x^6\ dx \, .
 \label{eq:tau_polar_thetaf}
\end{equation}
For small mean free paths for photon splitting, \teq{l\ll h\rns}, the upper limit \teq{\xmax \approx l/\rE} of the integral 
is much less than unity.  Then the magnetic field does not change much in both magnitude and direction 
along the path of the photon.  Consequently, \teq{{\cal M}_{\sigma}} can be approximated as the value at the 
emission point, i.e. \teq{{\cal M}_{\sigma}\approx {\cal M }_{\sigma} (B_{\hbox{\sixrm E}})}, and be placed outside 
the integration.  The magnitude of the field at the emission point is \teq{B_{\hbox{\sixrm E}}=B_p \sqrt{4-3 h\sin^2\theta_f}/(2h^3)} 
and can be approximated by \teq{B_{\hbox{\sixrm E}} \approx B_p/h^3} for quasi-polar locales. 
The leading order evaluation of the integral is then trivial, yielding 
\begin{equation}
   \tau_{\sigma} \;\approx\; \dover{243\fsc^3}{573440\pi^2}\,
   \dover{\rns}{\lambar_c}\,{\cal{M}}^2_{\sigma}B_{p}^6 \erg^5\theta^6_f\left( \dover{l}{h^3\rns}\right)^7\quad.
 \label{eq:tau_polar_small_L}
\end{equation}
By setting \teq{\tau_{\sigma}=1}, the attenuation length \teq{l\to L_{\sigma}} is obtained as
\begin{equation}
   L_{\sigma} \; \approx\; h^3 \rns \left\lbrack \dover{573440\pi^2}{243\,\fsc^3} 
   \dover{\lambar_c}{{\cal M}_{\sigma}^2 B_p^6 \theta_f^6 \rns} \right\rbrack^{1/7} \erg^{-5/7}\quad .
 \label{eq:atten_length_polar}
\end{equation}
This is the analytic approximation for polar emission locales that serves as the photon splitting 
analog of the pair creation result in Eq.~(30) in \cite{SB14}.  All modes possess attenuation 
lengths with energy dependence \teq{\erg^{-5/7}}, which applies at energies well below pair 
threshold.  It is also a simple matter 
to sum the modes according to the prescription in Eq.~(\ref{eq:tau_ave}) to generate 
the polarization-averaged attenuation length \teq{L_{\rm ave}}, which can be simply related to 
the conversion length for the \teq{\perp\to\parallel\parallel} mode in Eq.~(\ref{eq:atten_length_polar}):
\begin{equation}
   L_{\rm ave} \; =\; \left[ \dover{ 2 {\cal M}_1^2}{ 3 {\cal M}_1^2 +{\cal M }_2^2} \right]^{1/7} \, L_{\perp\to\parallel\parallel} \quad .
 \label{eq:atten_length_ave}
\end{equation}
Note again that in all of Eqs.~(\ref{eq:tau_polar_small_L}), (\ref{eq:atten_length_polar})
and~(\ref{eq:atten_length_ave}), the emission point values \teq{{\cal M}_{\sigma}\approx {\cal M }_{\sigma} (B_{\hbox{\sixrm E}})}
are assumed by the splitting amplitude coefficients.

The left panel of Fig.~\ref{fig:attenlen_split_surface} illustrates how the attenuation lengths vary with
photon energy for different polar magnetic fields and footpoint colatitudes, averaged over polarizations,
i.e., using Eq.~(\ref{eq:tau_ave}). The Figure is restricted to surface emission locales, so that \teq{h=1}
and \teq{\thetaE = \theta_f}.  The neutron star radius \teq{\rns} (marked) serves as the scale for 
curvature of field lines proximate to the surface, and separates the two main regions for the morphology of the 
attenuation length curves. For \teq{L\ll\rns},
the curves have a power-law behavior as \teq{L\propto \varepsilon^{-5/7}}, which is inferred qualitatively
in \cite{HBG97}, and analytically here. 
Since the dipole magnetosphere is scale invariant in terms of its angular geometry, emission at
higher altitudes reduces the local field according to \teq{B_{\hbox{\sixrm E}} \approx B_p/h^3},
and the emission colatitude scales as \teq{\thetaE\approx\theta_f\sqrt{h}}.  Thus, the 
simple approximate result \teq{L_{\sigma} \propto h^3} {\it along a given field line} emerges in Eq.~(\ref{eq:atten_length_polar}),
with some additional \teq{h} dependence implicit in the \teq{{\cal M}_{\sigma}} factors 
that is generally weak in subcritical fields that are encountered when \teq{h \gtrsim 3-4}.

The attenuation length curves are declining functions of the photon energy. 
In the \teq{L\gg\rns} domains, the attenuation lengths diverge and
the vertical \teq{L\rightarrow\infty} asymptotes define the escape energies \teq{\eesc}, the subject of
Section~\ref{sec:escape}.  Below these energies, the magnetosphere is transparent to photon splitting. The
analytic approximation in Eq.~(\ref{eq:atten_length_ave}) is also depicted as discrete dots, providing a
precise match to the numerical results when \teq{L\ll\rns}. This analytic protocol also verifies the
numerics of each of the splitting modes individually, but not illustrated graphically.  The attenuation
lengths decline with increasing \teq{B_p} since the attenuation coefficients are increasing functions of the
field strength.  The attenuation lengths and escape energies decrease with \teq{\thetaE}, because at
larger emission colatitudes, the radii of field line curvature are smaller, and photons require a shorter
propagation distance in order to realize significant \teq{\thetakB} values.

The right panel of Fig.~\ref{fig:attenlen_split_surface} shows the ratio \teq{\mathbfit{L}_{\perp\to\parallel\parallel}
= L_{\perp\rightarrow\parallel\parallel} / L_{\rm ave}} as functions of the photon energy
\teq{\varepsilon} and so serves as a proxy for \teq{L_{\perp\rightarrow\parallel\parallel}}, whose curves
would resemble those on the left panel.  The vertical upturns in the curves at low energies illustrate that
the escape energies for \teq{\perp\rightarrow\parallel\parallel} mode is generally greater than that of the
polarization-averaged result, an obvious consequence of adding modes of opacity. The ratio curves become
flat in the high energy domains since the opacities for all three modes scale as \teq{\erg^5}. Using
Eq.~(\ref{eq:atten_length_ave}), the asymptotic limits of the curves can be expressed  as
\teq{L_{\perp\rightarrow\parallel\parallel}/L_{\rm ave} =
\left[{{(3\cal{M}}_1^2+{\cal{M}}_2^2)}/{{2\cal{M}}_1^2}\right]^{1/7}}, and the numerical results reproduce
this analytic result well when \teq{\erg} is sufficiently high, i.e. a factor of ten or so above the escape energy.  
Monotonic declines of the length ratios and escape energies with emission colatitude \teq{\thetaE} 
and polar field strength \teq{B_p} are obvious and expected.

\begin{figure}
\vspace*{-10pt}
\centerline{\hskip 10pt\includegraphics[width=9.5cm]{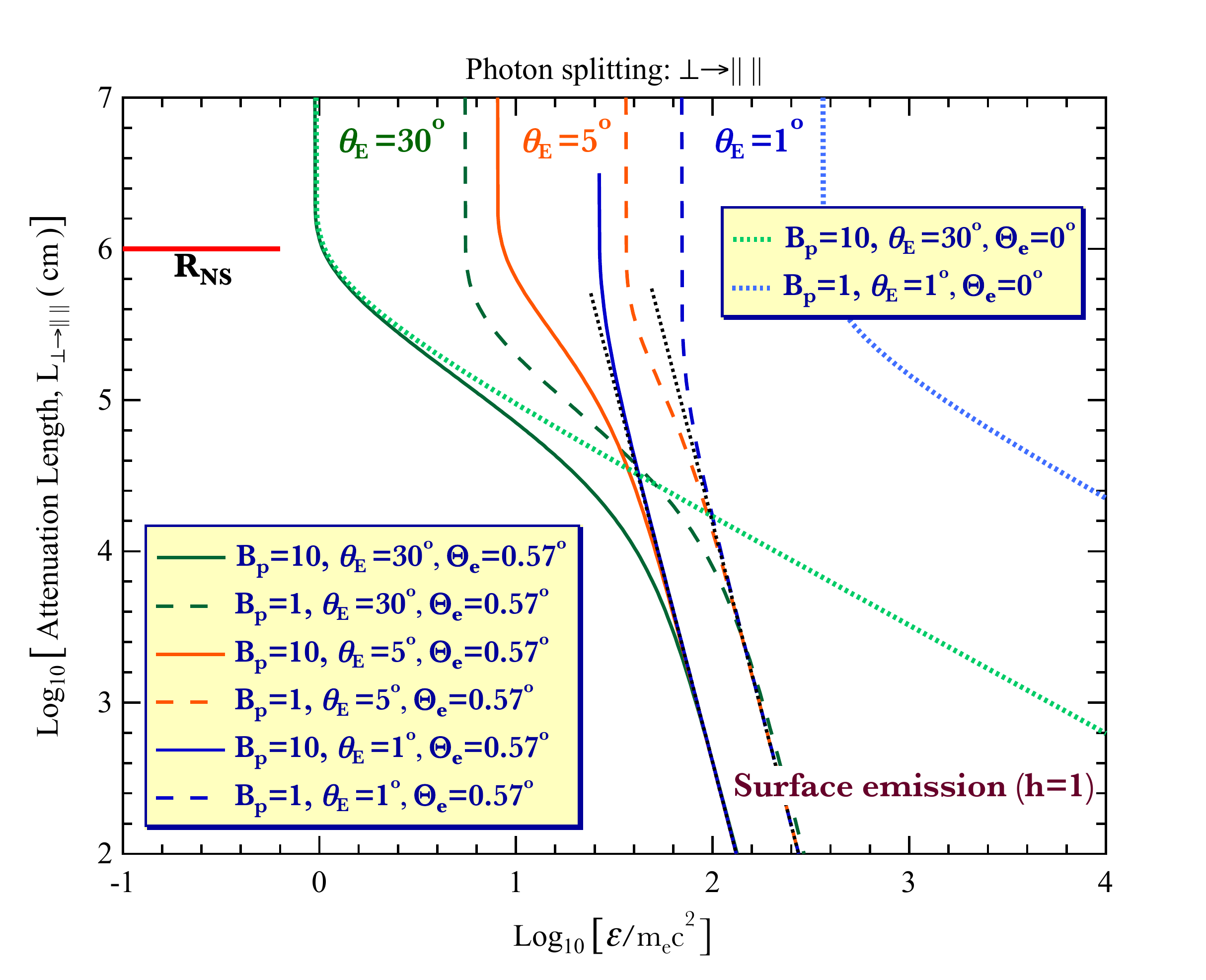}}
\vspace*{-5pt}
\caption{Flat spacetime attenuation lengths \teq{L_{\perp\to\parallel\parallel}} for 
the splitting mode \teq{\perp\to\parallel\parallel}, as functions of 
photon energy (in units of \teq{m_ec^2}).  As in Fig.~\ref{fig:attenlen_split_surface}, 
the light is emitted from the neutron star surface.   However, here 
most examples are for the light wavevector {\bf k} initially at a non-zero angle 
\teq{\Thetae\equiv \thetakB =1}radian to the local magnetic field vector {\bf B}.
Different polar field strengths \teq{B_p} are represented, labelled in units of \teq{B_{\rm cr}}. 
The dotted straight black lines represent the analytic approximation from Eq.~(\ref{att_length_unparl}), for 
two field choices \teq{B_p=10} and \teq{B_p=1}, a result that is independent of
the value of \teq{\thetaE }.  The light green and blue dotted curves depict 
cases where \teq{\theta_0=0}, i.e. light initially moving parallel to {\bf B}.
The neutron star radius scale \teq{\rns} is marked at the upper left.
 \label{fig:attenlen_split_surface_obq}}
\end{figure}

It is often the case that the photons are not emitted exactly parallel to the local field. Relativistic electrons 
with finite Lorentz factors have a Doppler cone of emission due to aberration with a beaming angle 
\teq{\Thetae\sim 1/\gammae}.  This applies to a wide variety of mechanisms, including 
curvature radiation, synchroton and synchrotron self-Compton emission in 
gamma-ray pulsar models \citep[e.g.,][]{DH96,Romani96,HK15}
and resonant inverse Compton scattering in scenarios for generating hard X-ray tail signals 
in magnetars \citep{BH07,FT07,Beloborodov13,Wadiasingh18}.
The development of an analytic approximation becomes rather simple if non-zero  \teq{\Thetae} is assumed. 
In this case, the opacity is achieved generally on scales much smaller than the radius of field line curvature, 
i.e. so that \teq{L\ll h\rns}. Accordingly, a uniform field can be assumed along the trajectory 
until \teq{\tau=1} is realized, and the optical depth formula in Eq.~(\ref{eq:tau_sigma}) yields,
to leading order in \teq{L/(h\rns )}, 
\begin{equation}
   \tau_{\sigma} \;\approx\; \dover{\fsc^3}{60\pi^2}
   \,{\cal{M}}^2_{\sigma}\,B^6_{\hbox{\sixrm E}} \,\erg^5\, \sin^6{\Thetae}\, \dover{l}{\lambar_c} \quad .
 \label{eq:tau_thetae_neq0}
\end{equation}
Inverting this, the attenuation length is expressed as
\begin{equation}
   L_{\sigma} \;\approx\; \dover{60\,\pi^2\lambar_c}{\fsc^3}\,{\cal{M}}^{-2}_{\sigma} \,B^{-6}_{\hbox{\sixrm E}}
   \,\erg^{-5}\,\sin^{-6}{\Thetae} \quad ,
 \label{att_length_unparl}
\end{equation}
where again, \teq{{\cal M}_{\sigma}} denotes \teq{{\cal M }_{\sigma} (B_{\hbox{\sixrm E}})}.
Numerical evaluations of the attenuation lengths for the \teq{\perp\rightarrow\parallel\,\parallel} 
mode attenuation lengths with a non-zero \teq{\Thetae} are illustrated in Fig.~\ref{fig:attenlen_split_surface_obq}. 
The \teq{\Thetae=0} cases and the analytic approximation from Eq.~(\ref{att_length_unparl}) 
are also plotted for comparison. The non-zero \teq{\Thetae} cases exhibit a steep decrease in
attenuation length in the high energy domain, with a \teq{L\propto \varepsilon^{-5}} 
dependence when \teq{L\ll \rns}, evinced in Eq.~(\ref{att_length_unparl}).  
For large footpoint colatitudes, \teq{\thetaE \gtrsim 20^{\circ}}, a non-zero \teq{\Thetae}  
does not change the escape energy much from \teq{\Thetae=0} values, a consequence of 
the comparatively short radius of curvature of field lines local to the emission point.  However, when the 
footpoint colatitude is small, \teq{\thetaE \lesssim 10^{\circ}}, the escape energy is strongly reduced, 
making it more difficult for high energy photons to escape.

\section{Escape Energies for Photon Splitting and Pair Creation}
 \label{sec:escape}
In this section the escape energies for both processes are explored.  
These energies form the prime focus of this paper, with a particular 
emphasis on their determination along field loops in the inner magnetosphere.  
This case more directly connects to magnetar emission models.
The exposition starts with analytics in the small colatitude domain 
and then moves to illustrations of numerical results. The analytic derivations provide 
quantitative behavior of the escape energies that enhances the understanding of
the numerics, and serves as checks on them.  Flat Minkowski spacetime is assumed throughout.

\subsection{Analytic Approximations: Small Colatitudes}
 \label{sec:eesc_small_theta}

For pair creation, an analytic approximation for escape energies in 
the small colatitude domain \teq{\theta_f\ll 1} was derived in \cite{SB14}, 
using Erber's asymptotic attenuation coefficient in Eq.~(\ref{eq:Erber_asymp}). 
Since this is an exponential form that is sensitive to the photon 
propagation angle \teq{\thetakB} along the path, they used the method of 
steepest descents to evaluate the opacity integral, yielding
\begin{equation}
   \eesc \;=\; \dover{2^{13}h^{5/2}}{3^5B_p\theta_f}
   \bigg\{\log \Bigl(\dover{3^6}{2^{19}}\dover{\fsc\rns}{\lambar_c} \Bigr) 
   + \dover{1}{2} \log \Bigl( \dover{3 \pi \eesc B_p^3 \theta_f^3}{2h^{11/2}} \Bigr) \bigg\}.
 \label{eq:eesc_pairs_small_thetaf}
\end{equation}
This result assumes that the photons are emitted parallel to the local field, 
and that light propagates in straight lines.  Since it is 
based on Erber's approximation, it is applicable to 
subcritical fields and energies well above pair threshold.
The second logarithmic term on the right hand side is only weakly dependent on its argument,  
so one can infer that \teq{\eesc\propto1/B_p} and \teq{\eesc\propto1/\theta_f},
approximately. These couplings are direct consequences of the 
appearance of the product of \teq{\erg B \sin\thetakB } in the argument of 
Erber's exponential form, which generates the correlation 
\teq{\eesc B_p \theta_f\approx \rm{constant}} at quasi-polar colatitudes. 

For photon splitting we modify the analytic approach employed by \cite{SB14} 
again treating the domain of small colatitudes.  Eq.~(\ref{eq:tau_polar_thetaf}) 
serves as a natural starting point for photons emitted parallel to {\bf B}. 
For escape to an observer at infinity, we set \teq{r\rightarrow \infty}, 
so that \teq{\xmax\rightarrow 1}, yielding the form 
\begin{equation}
   \tau_{\sigma} \;\approx\;\frac{3^5\, \fsc^3}{2^{14}\cdot5\, \pi^2}\frac{\rns}{\lambar_c} 
   \frac{B_{p}^6 \erg^5\theta^6_f}{h^{14}} \,
   {\cal{I}_{\sigma}} \quad ,
 \label{eq:tau_polar_thetaf_infy}
\end{equation}
for 
\begin{equation}
   {\cal{I}_{\sigma}} =  \int_{0}^{1} {\cal{M}}^2_{\sigma}(1-x)^{16}x^6\, dx \quad .
 \label{eq:define_cal_i}
\end{equation}
Setting \teq{\tau_{\sigma} = 1}, the escape energy is then expressed as
\begin{equation}
   \eesc\;\approx\;\left[\dover{2^{14}\cdot5 \,\pi^2}{3^5\,\fsc^3}\dover{\lambar_c}{\rns}
   \dover{h^{14}}{ B_{p}^6\,{\cal{I}_{\sigma}} }\right]^{1/5}\theta^{-6/5}_f \quad .
 \label{eq:eesc_split_polar}
\end{equation}
This is the analytic approximation for the photon splitting escape energy in the 
\teq{\theta_f\ll 1} domain that is analogous to Eq.~(\ref{eq:eesc_pairs_small_thetaf}), 
and is derived here for the first time.
The energy dependence \teq{\eesc\propto \theta^{-6/5}_f} emerges regardless of the surface
field strength and it applies to all splitting modes.  Note the argument of \teq{{\cal{M}}_{\sigma}} 
is \teq{B\approx B_p{(1-x)^3}/h^3}, which comes from Eq.~(\ref{eq:B_thetakB_polar}). 
Generally the integral in the denominator can only be solved numerically.

Observe that when \teq{B_p/h^3<1}, \teq{{\cal{M}}_{\sigma}(x)} grows slowly 
with increasing \teq{x}, as the field strength declines along an outward 
photon trajectory.  Then \teq{{\cal{M}}_{\sigma}} can 
be approximated by its value at the peak of the \teq{(1-x)^{16}x^6} profile, i.e at \teq{x=3/11}, 
and can be placed outside the integration. Then Eq.~(\ref{eq:tau_polar_thetaf_infy}) simplifies to
\begin{equation}
   \tau_{\sigma} \;\approx\; \dover{\mu_p\fsc^3}{\pi^2} \dover{\rns}{\lambar_c} 
   \dover{B_{p}^6 \erg^5\theta^6_f}{h^{14}}{\cal{M}}^2_{\sigma,p}\quad,
 \label{eq:tau_polar_thetaf_infy_low_b}
\end{equation}
where 
\begin{equation}
   \mu_p \; =\; \dover{3^4}{2^{14}\cdot 2860165} 
   \quad \Rightarrow\quad
   \dover{\mu_p\fsc^3}{\pi^2} \dover{\rns}{\lambar_c} \;\approx\; 1.762
\end{equation}
for \teq{\rns = 10^6}cm.  Also,
\begin{equation}
   {\cal{M}}_{\sigma,p}\;=\; {\cal{M}}_{\sigma}\left[B\left(x=\frac{3}{11}\right)\right]\; \approx\;  
   {\cal{M}}_{\sigma}\left(\frac{512\ B_p}{1331\ h^3}\right)\quad
 \label{eq:m_peak}
\end{equation}
is the value of \teq{{\cal{M}}_{\sigma}} at \teq{x=3/11}. Setting Eq.~(\ref{eq:tau_polar_thetaf_infy_low_b}) 
equal to unity, the escape energy can be expressed as
\begin{equation}
   \eesc\;\approx\;\left[\dover{\pi^2}{\mu_p\, \fsc^3}\dover{\lambar_c}{\rns}
   \dover{ h^{14}}{ B_{p}^6{\cal{M}}^2_{\sigma,p}}\right]^{1/5}
    \theta^{-6/5}_f  \quad,
 \label{eesc_split_small_thetaf}
\end{equation}
or equivalently expressed in terms of the emission colatitude \teq{\thetaE\approx\theta_f\sqrt{h}} :
\begin{equation}
   \eesc\;\approx\;\left[\dover{\pi^2}{\mu_p\, \fsc^3}\dover{\lambar_c}{\rns}
   \dover{ h^{17}}{ B_{p}^6{\cal{M}}^2_{\sigma,p}}\right]^{1/5}
    \thetaE^{-6/5}  \quad.
 \label{eesc_split_small_thetaE}
\end{equation}
The asymptotic approximations in Eq.~(\ref{eesc_split_small_thetaf}) and 
Eq.~(\ref{eesc_split_small_thetaE}) have not been derived before in the literature, and 
obviously coincide for surface emission (i.e., \teq{h=1}).  Both
results apply for \teq{\Thetae =0} at the point of emission.  We remark that they 
are most useful for locales where the emission point samples near-critical or 
subcritical fields.  For highly supercritical values of \teq{B_p/h^3}, these two forms 
are less practical because the \teq{(1-x)^3} dependence of the \teq{{\cal M}_{\sigma}^2}
factors shifts the peak of the \teq{{\cal I}_{\sigma}} integral away from \teq{x=3/11}.
In such circumstances, direct numerical evaluation of Eq.~(\ref{eq:eesc_split_polar}) 
is desirable, and is indeed the protocol adopted in the illustrations of 
Section~\ref{sec:eesc_numerics}.

The form in Eq.~(\ref{eesc_split_small_thetaE}) needs to be adapted when \teq{h} 
is not a constant parameter.  Such is the situation where photons are emitted at 
various points along a specific field loop. The corresponding evaluation is then 
germane to the study of \cite{Wadiasingh18} where the spectra of resonant 
inverse Compton scattering resulting from monoenergetic electrons moving along 
field loops is calculated. That work pertains to the persistent hard X-ray tails of magnetars, 
which will form part of the discussion in Section~\ref{sec:resComp}. The polar forms for 
these dipole field loops can be expressed as
\begin{equation}
   \dover{r}{\rns}\;=\;{\rmax}\,\sin^2{\theta}\quad,
 \label{eq:field_loop}
\end{equation}
where \teq{\rmax} is the maximum (equatorial) radius of the loop rescaled by \teq{\rns}. 
These dipolar loops can also be parameterized by their footpoint colatitudes via \teq{\sin{\theta_f}=1/\sqrt{\rmax}}, 
and the colatitude for a specific loop ranges from \teq{\theta_f} to \teq{\pi-\theta_f}.
The analytic approximation can be obtained by setting \teq{h\approx r_{\hbox{\sevenrm max}}\,\sin^2\thetaE
\approx r_{\hbox{\sevenrm max}}\,\thetaE} in Eq.~(\ref{eesc_split_small_thetaE}):
\begin{equation}
   \eesc \;\approx\; \left[\dover{\pi^2}{\mu_p\, \fsc^3}\dover{\lambar_c}{\rns}
   \dover{ {r_{\hbox{\sevenrm max}}}^{17}}{ B_{p}^6{\cal{M}}^2_{\sigma,p}
   }\right]^{1/5}
    \thetaE^{28/5}  \quad. \label{eq:esc_energy_field_loop_low_b}
\end{equation}
Here 
\begin{equation}
   {\cal{M}}_{\sigma,p}\; \approx\;  
   {\cal{M}}_{\sigma}\left(\dover{512}{1331}\dover{B_p}{{\thetaE^6r_{\hbox{\sevenrm max}}}^3}\right)\quad,
 \label{eq:m_peak_loop}
\end{equation}
and it needs to be evaluated at each colatitude \teq{\thetaE} along a loop.
Eq.~(\ref{eq:esc_energy_field_loop_low_b}) is applicable to quasi-polar locales with small colatitudes \teq{\thetaE\ll 1}, 
corresponding to a large value for \teq{r_{\hbox{\sevenrm max}}} for a particular field loop.
The strong dependence of the 
argument of the \teq{{\cal M}_{\sigma, p}} function on \teq{\thetaE} yields significant modifications to 
the \teq{\thetaE^{28/5}} dependence in supercritical fields near the stellar surface. 

Analytic approximations for escape energies with a non-zero emission angle 
\teq{\Thetae} are routinely derived in the meridional 
plane corresponding to \teq{\zetae=0} : see the right panel of Fig.~\ref{fig:geometry}.
The simplest path to the approximation is by replacing \teq{\deltaE} with \teq{\deltaE+\Thetae}.
In the small colatitude domain where \teq{\thetaE\ll\Thetae}, 
Eq.~(\ref{eq:B_dipole}) and Eq.~(\ref{eq:thetakB_plane}) can be approximated as
\begin{equation}
   B \;\approx\; B_p\,\dover{(\Thetae-\eta)^3}{h^3\Thetae^3} 
   \quad \hbox{and} \quad
   \sin{\thetakB} \;\approx\; \frac{3}{2}\eta-\Thetae \quad ,
    \label{eq:B_thetakB_polar_unparl}
\end{equation}
with the Jacobian
\begin{equation}
   \frac{ds}{d\eta}\;\approx\;  \frac{\rE\Thetae}{(\Thetae-\eta)^2}\quad.
\end{equation}
Again, these results can be inserted into Eq.~(\ref{eq:tau_sigma}), yielding an approximation 
for the optical depth
\begin{equation}
   \tau \;\approx\;\frac{3^5\fsc^3}{2^8\cdot5\pi^2}\dover{B^6_p\erg^5\Thetae^6}{h^{17}}
   \dover{\rns}{\lambar_c}
   {\cal{J}_\sigma}\quad
\end{equation}
with
\begin{equation}
   {\cal{J}_\sigma} \; =\; \int_{0}^{1} {\cal M}^2_{\sigma}(1-x)^{16} \left( x- \dover{2}{3} \right)^6\, dx \quad .
 \label{eq:calJ_def}
\end{equation}
Here the integration variable is \teq{x=\eta/\Thetae}, and the upper limit is \teq{\xmax \approx1}. 
Setting \teq{\tau(\infty) = 1}, one arrives at
\begin{equation}
   \eesc\approx\left[\dover{2^8\cdot5\, \pi^2}{3^5\, \fsc^3}\dover{\lambar_c}{\rns}\dover{h^{17}}{ B_{p}^6
   {\cal{J}_\sigma}
   }\right]^{1/5}
   \Thetae^{-6/5}.
 \label{eq:esc_energy_small_colati_unparl}
\end{equation}
One can numerically solve the integral and find the analytic approximation for the 
escape energy in the small \teq{\Thetae} domain.  When \teq{B_p/h^3<1}, 
the same logic as for the evaluation of Eq.~(\ref{eq:eesc_split_polar}) 
applies and one can extract the \teq{{\cal M}_\sigma} and evaluate the integral.  This results in a form
\begin{equation}
   \eesc \;\approx\; \left[ \dover{2196606720 \, \pi^2}{13^2 \cdot 8273\, \fsc^3}
   \dover{\lambar_c}{\rns}\dover{h^{17}}{ B_{p}^6{\cal{M}}^2_{\sigma,0}} \right]^{1/5}   \Thetae^{-6/5} \quad .
 \label{eq:esc_energy_small_colati_unparl_lowb}
\end{equation}
Here the \teq{{\cal M}_{\sigma,0}} is evaluated at \teq{x=0}, which is the peak 
for the \teq{(1-x)^{16}\left(x-2/3\right)^6} profile.  Accordingly, the argument 
of \teq{{\cal M}_{\sigma,0}} scales with \teq{B_p/h^3} and depends slightly on the 
colatitude \teq{\thetaE} of emission.

\subsection{Numerical Results}
 \label{sec:eesc_numerics}

\begin{figure*}
 \begin{minipage}{17.5cm}
\vspace*{-10pt}
\centerline{\hskip 10pt \includegraphics[height=7.4cm]{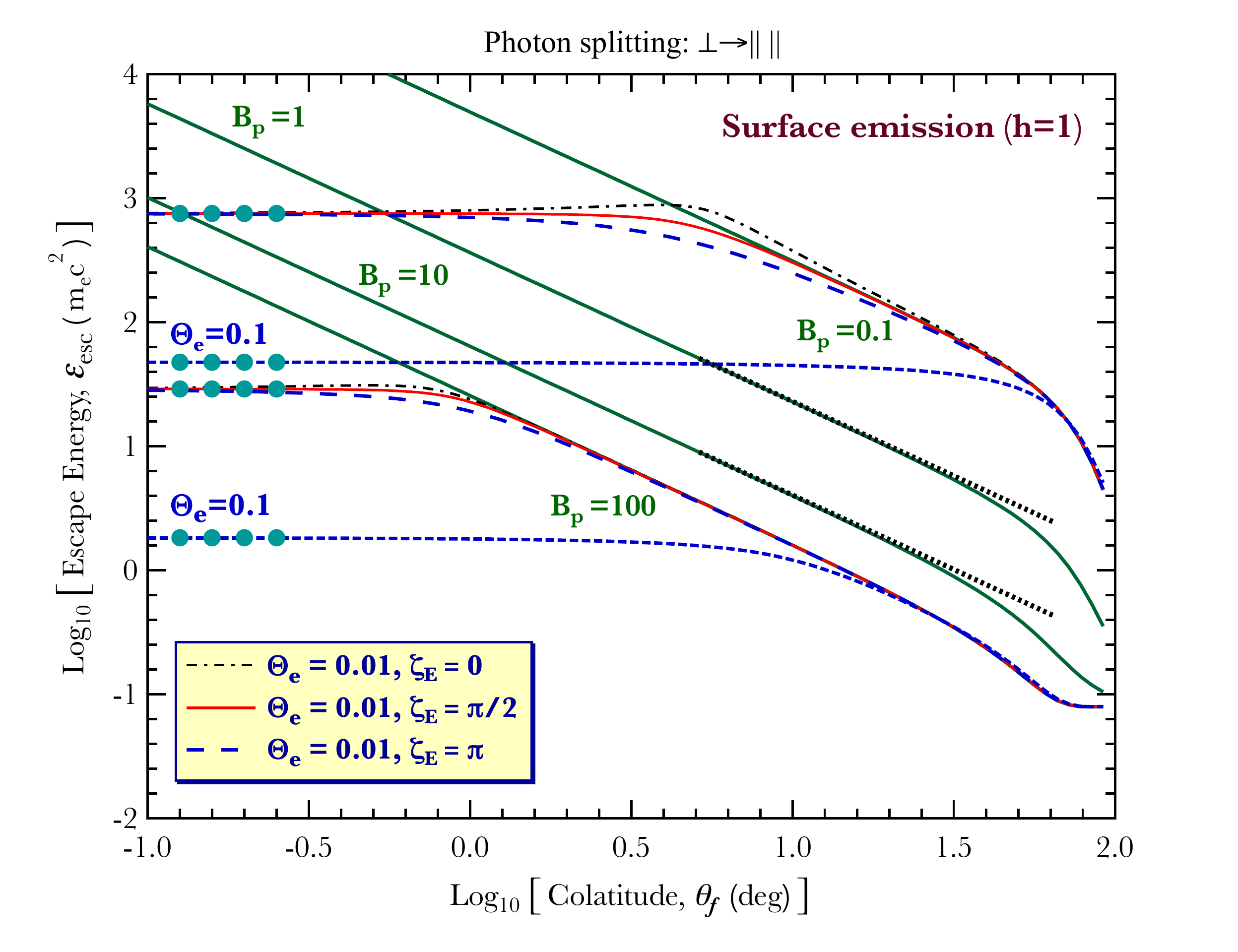}
   \hskip -30pt\includegraphics[height=7.4cm]{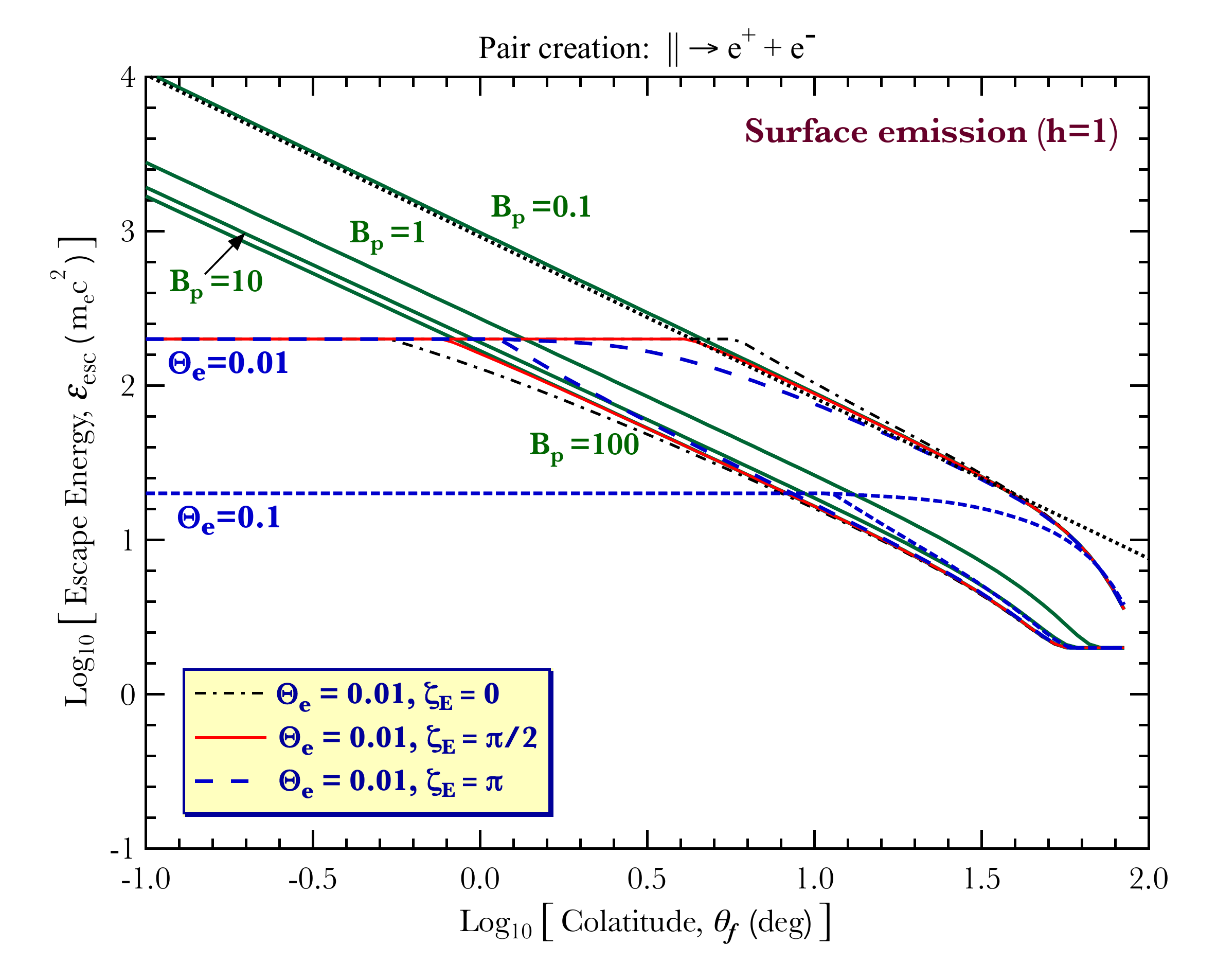}}
\vspace*{-5pt}
\caption{The escape energies \teq{\eesc} for photon splitting \teq{\perp\to\parallel\parallel} 
(left panel) and single photon pair production \teq{\parallel \to e^{\pm}} (right panel) 
for light emitted from the neutron star surface \teq{(h = 1)} and propagating in flat spacetime.  
They are numerical results obtained by setting  \teq{L\to\infty} in Eq.~(\ref{eq:atten_length_def}), 
and are plotted as functions of magnetic colatitude \teq{\thetaE \equiv\theta_f} 
for photon emission both along {\bf B} (solid green curves; \teq{\Thetae=0}) 
and angles \teq{\Thetae =0.1\,\hbox{rad} \approx 5.7^{\circ}} and \teq{0.01\,\hbox{rad} \approx 0.57^{\circ}}
to the field (dashed blue, solid red and dash-dot black curves for \teq{B_p=0.1, 100} cases only: see the legend).  Consult 
Fig.~\ref{fig:geometry} and Section~\ref{sec:geometry} for the definition of the azimuthal angle \teq{\zetae}. 
The curves are grouped and labelled by their polar field strengths \teq{B_p=0.1, 1, 10, 100}.  
The escape energies for each process are monotonically
decreasing functions of \teq{B_p} for the range of parameters shown.  The 
\teq{\Thetae=0} curves have slopes of -6/5 (splitting) and -1
(pair creation) at small \teq{\thetaE}, as identified by Harding, Baring
\& Gonthier (1997), and diverge near the polar axis, where the field
line radius of curvature becomes infinite.  The dotted black lines are the analytic approximations 
in Eq.~(\ref{eq:eesc_pairs_small_thetaf}) for pair creation and Eq.~(\ref{eq:eesc_split_polar}) for photon splitting.
The filled acqua circles in the left panel denote evaluations using the 
analytic approximation in Eq.~(\ref{eq:esc_energy_small_colati_unparl}).
 \label{fig:eesc_split_pairs_surface}}
\end{minipage}
\end{figure*}

The escape energies as functions of footpoint colatitude for the \teq{\perp\to\parallel\parallel} 
mode splitting are illustrated in the left panel of Fig.~\ref{fig:eesc_split_pairs_surface}.
The chosen polar field strengths \teq{B_p} encompass much of the range of interest 
for both magnetars and highly-magnetized pulsars.
Note that the field strength at the emission point is \teq{B_f = B_p [\sqrt{1+3 \cos^2\theta_f}\, ]/2}.
These results are for surface emission of outward-propagating photons only, and they are obtained from 
the \teq{l\to\infty} limit of Eq.~(\ref{eq:tau_sigma}), using a bisection 
decision algorithm to isolate the divergence of the attenuation length.
The solid green curves represent the numerical results for the parallel emission cases,
i.e. \teq{\Thetae=0}, for \teq{B_p=0.1,1,10,100}. 
The escape energies are declining functions of \teq{\theta_f}, realizing a power-law behavior 
\teq{\eesc\propto\theta_f^{-6/5}} at small colatitudes that was first identified by \cite{HBG97}. 
This power-law character is also obtained in the analytic approximation in
Eq.~(\ref{eq:eesc_split_polar}), which is displayed as the dotted straight diagonal lines in the Figure.
The excellent precision of the approximation is evident, and was obtained 
specifically for numerical determination of the \teq{I_1} integral.  Note that only colatitudes
up to the equator \teq{\theta_f=90^{\circ}} are plotted, with cases of inward propagation 
into the optically (Thomson) thick surface layers in the other hemisphere being omitted.

\newpage

In the left panel of Fig.~\ref{fig:eesc_split_pairs_surface}, non-zero \teq{\Thetae} cases are also displayed for 
\teq{B_p=0.1} and \teq{B_p=100} as curves for different azimuth angle \teq{\zetae}. 
The most striking feature of these loci is that non-zero \teq{\Thetae} reduces 
\teq{\eesc} substantially at small colatitudes, a property identified by 
\cite{BH01} that is also evident in the attenuation length plot in 
Fig.~\ref{fig:attenlen_split_surface_obq}. The origin of this \teq{\eesc} reduction is that magnetic field lines 
curve only slightly near the magnetic pole, so a non-zero value for \teq{\Thetae} 
dramatically increases the \teq{\perp\to\parallel\parallel}
attenuation coefficient due to its strong dependence on \teq{\sin{\thetakB}}.
The choices of \teq{\Thetae=0.1} and \teq{0.01} correspond roughly to Lorentz cone 
angles for electrons of energies 5 and 50 MeV, respectively; these are appropriate 
for Compton upscattering models of magnetar hard X-ray emission 
\citep[e.g. see][and references therein]{Wadiasingh18}.  These curves coalesce 
with the \teq{\Thetae=0} examples when the emission colatitude exceeds \teq{\Thetae} 
and field line curvature quickly establishes sufficiently large values for \teq{\thetakB} 
to govern the opacity phase space.  Observe that the escape energy is generally insensitive 
to the value of \teq{\zetae} at both small and large emission colatitudes; this is no longer 
true when the emission angle \teq{\Thetae} relative to {\bf B} exceeds about 10 degrees.  When \teq{\theta_f\ll \Thetae}, 
the escape energies ``plateau'' at values independent of the choice of \teq{\zetae}. 
These values match those given by the analytic approximation in Eq.~(\ref{eq:esc_energy_small_colati_unparl})
that are depicted as filled aqua circles in the plot.
Note that the morphology of escape energy curves for the other 
CP-permitted splitting modes is similar, albeit with slightly different values for \teq{\eesc}.

\begin{figure*}
 \begin{minipage}{17.5cm}
\vspace*{-10pt}
\centerline{\hskip 10pt \includegraphics[height=7.4cm]{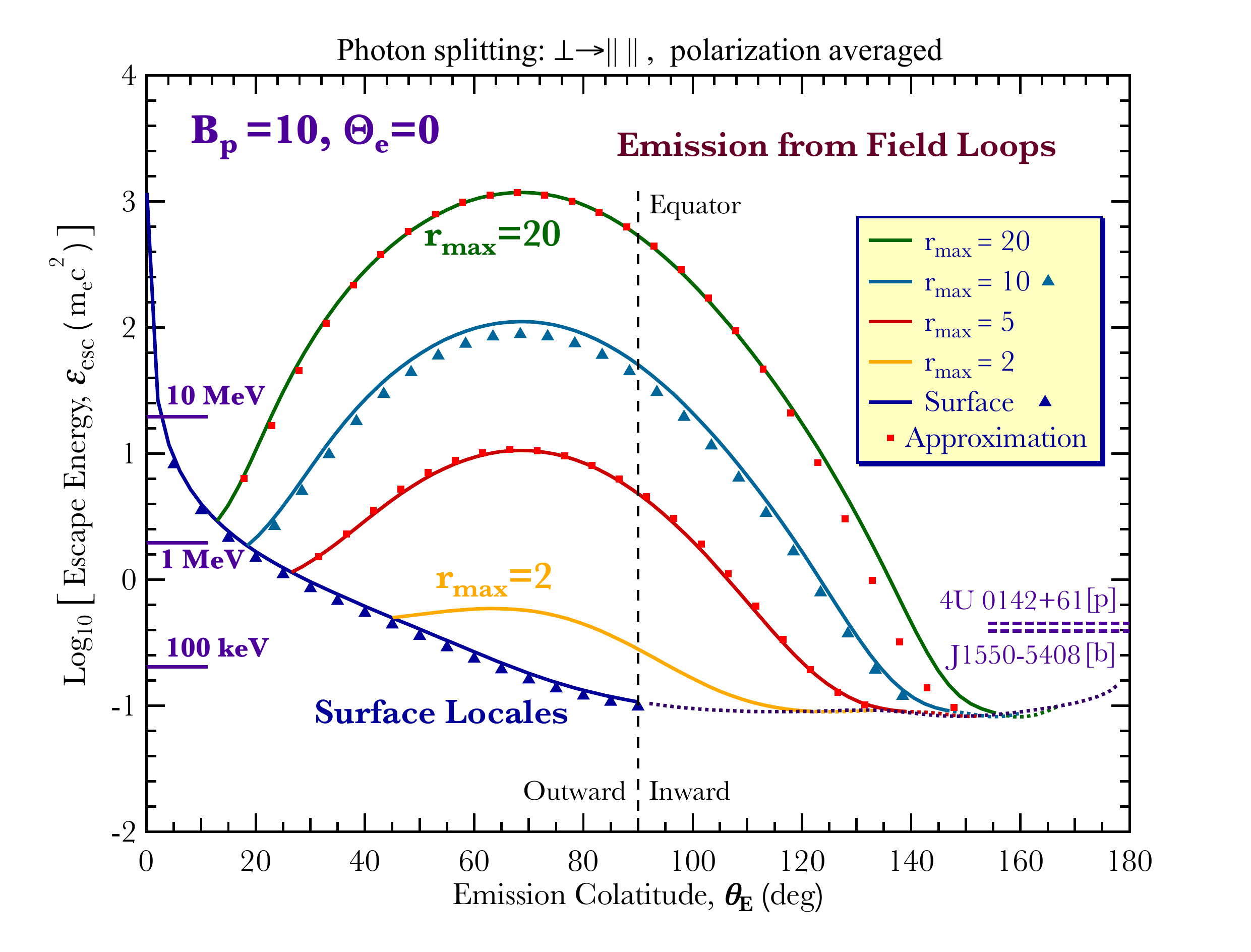}
   \hskip -25pt\includegraphics[height=7.4cm]{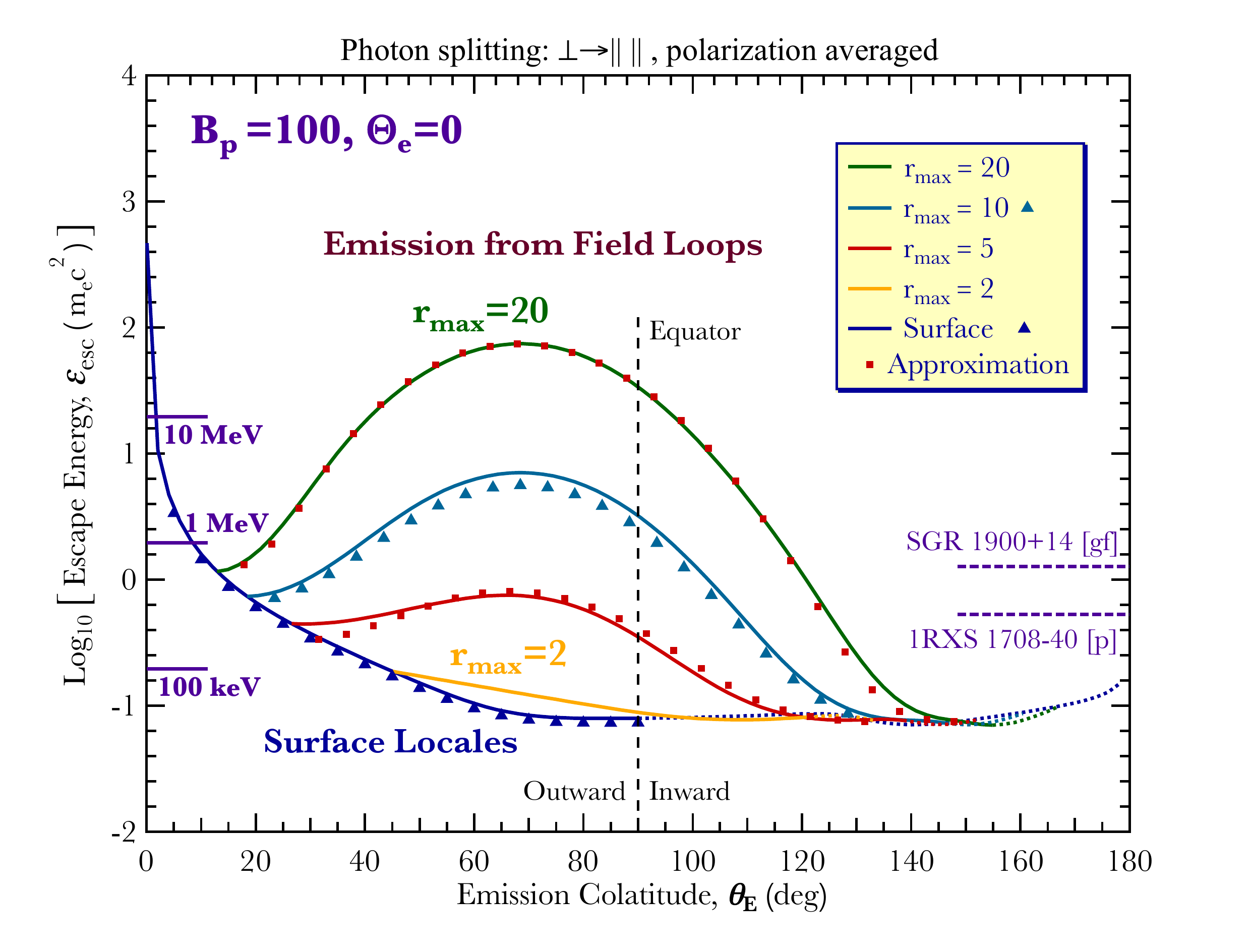}}
\vspace*{-5pt}
\caption{Photon splitting escape energies in flat spacetime for the mode \teq{\perp\to \parallel\parallel} 
(curves) and also averaged over polarization modes (triangles),
for emission initially parallel to the local magnetic field line (\teq{\Thetae =0}).
The left and right panels are for surface polar fields \teq{B_p=10} and \teq{B_p=100},
respectively.  The abscissa is the emission colatitude \teq{\thetaE}, spanning outward 
propagation cases to the left of the equatorial marker line, to inward propagation
to the right of this vertical dashed line.  Four of the \teq{\perp\to \parallel\parallel} 
curves are for magnetospheric emission at points along 
dipolar magnetic field loops, labelled by \teq{r_{\rm max}=2,5,10,20}, the maximum loop altitude 
in units of \teq{\rns}.  In contrast, the dark blue curves for both panels are not for loop emission, 
but are surface emission cases that are displayed in the left hand panel of 
Fig.~\ref{fig:eesc_split_pairs_surface}.  All curves include dotted portions 
in the inward trajectory hemisphere that demarcate cases where photons 
would impact the stellar surface if not attenuated beforehand; these are generally 
near \teq{\thetaE\sim 180^{\circ}} for magnetospheric loop examples. 
In addition, escape energies for polarization-averaged opacities are exhibited 
as triangles for the surface emission and \teq{r_{\rm max}=10} cases only.
At the lower right of each panel are marker energies (purple dashed lines) 
signifying the approximate maximum energy \underline{observed} in 
several magnetars with polar fields somewhat close to the illustrated values, 
SGR J1550-5408 (bursts, [b]), AXPs 4U 0142+61 and 1RXS 1708-40 (persistent emission, [p]), 
and SGR 1900+14 (giant flare, [gf]); see text for details. 
 \label{fig:eesc_split_loops}}
\end{minipage}
\end{figure*}

The right panel of Fig.~\ref{fig:eesc_split_pairs_surface} displays the escape energies as functions of 
surface colatitude for the \teq{\parallel\to e^++e^-} mode of pair creation, 
the polarization mode with the lowest pair threshold.  Here the solid numerical 
curves are obtained by the pair creation physics elements summarized in Section~\ref{sec:split_pair_phys}, 
specifically inserting Eq.~(\ref{eq:BK07_perp_asymp}) and Eq.~(\ref{eq:calFpp_par}) 
into Eq.~(\ref{eq:pp_general}). The \teq{\Thetae=0} curves display a power-law 
behavior as \teq{\eesc\propto\theta_f^{-1}}, which is slightly flatter than that for photon splitting,
character first highlighted by \cite{HBG97}.  The analytic approximation in 
Eq.~(\ref{eq:eesc_pairs_small_thetaf}) is represented as the dotted straight diagonal line 
for the \teq{B_p=0.1}, noting that it becomes inaccurate at lower energies near threshold conditions.
For low field cases like \teq{B_p=0.1} or \teq{B_p=1},
the escape energies of pair creation are lower than those for photon splitting,
thereby indicating the greater efficiency of the pair conversion process.  In contrast,
for \teq{B_p=10} or \teq{B_p=100}, photon splitting escape energies drop significantly 
due to the strong dependence of the splitting rate on the field strength, and 
hence dominate pair creation opacity.  This division of the competitiveness of the 
two processes based on field strength was highlighted in the study by \cite{BH98}. 
Non-zero \teq{\Thetae} cases are also depicted for pair creation. 
It is noticeable that curves with the same \teq{\Thetae} but different \teq{B_p} 
merge in the small colatitude plateau domain, \teq{\theta_f\ll \Thetae}, because
pair conversion \teq{\gamma\to e^+e^-} is very effective and photons will produce pairs as soon 
as they satisfy the kinematic threshold restriction in Eq.~(\ref{eq:calFpp_par}) 
during their propagation through the magnetosphere.
Accordingly, the asymptotic values in the \teq{\theta_f\ll \Thetae} regime only depend on \teq{\Thetae}
via the threshold criterion \teq{\eesc=2/\sin{\Thetae}}.
A horizontal tail appears in the equatorial regime on the right hand side of the Figure. 
This tail marks the absolute threshold \teq{\eesc =2} for pair creation, 
realized because \teq{\thetakB} approaches \teq{\pi/2} rapidly during
photon passage through the magnetosphere.  

It is noteworthy that the general properties of the escape energy curves in 
both panels of Fig.~\ref{fig:eesc_split_pairs_surface} reproduce those exhibited 
in Fig.~1 of \cite{BH01} for \teq{0.1 \leq B_p \leq 3.16}, though the values of \teq{\eesc} were lower by 
about factors of 1.3--1.8 therein because of their treatment of 
general relativistic effects for propagation from the surface.   The focus on flat spacetime results here is 
adopted to facilitate comparison with compact analytic approximations.

\vspace{-10pt}
\subsubsection{Emission Along Magnetic Field Loops}

A key aspect of the magnetar opacity problem that has not been discussed 
before in the literature concerns emission along field lines at low 
altitudes, an element that connects directly to models for their 
magnetospheric emission.  Emission persists for many dynamical (light crossing) times, 
even in transient flare activity, so plasma trapping by closed field 
lines is naturally presumed.   Therefore, an investigation of opacity in 
closed field regions above the stellar surface is motivated.
One anticipates that general relativistic
influences are likely to have limited impact on the escape energies 
at altitudes \teq{r \gtrsim 2\rns}, an element that is addressed in Sec.~\ref{sec:GR}.   

Fig.~\ref{fig:eesc_split_loops} illustrates the 
photon splitting escape energies \teq{\eesc} as functions of emission colatitude \teq{\thetaE} 
for photons emitted along specific magnetic field loops, i.e. \teq{\Thetae=0}.  
The surface polar field strengths chosen, \teq{B_p=10} or \teq{B_p=100},
are clearly appropriate for magnetars, and suggest a monotonic decline of escape energies 
with increasing \teq{B_p}.  Extrapolating to lower field values would lead to corresponding increases
in the values of \teq{\eesc}.
The bell-shaped curves represent the escape energies for \teq{\perp\to\parallel\parallel} mode splitting 
corresponds to field loops labelled by\teq{\rmax=2,5,10,20}, color-coded as in the legend. 
The solid blue curves display the \teq{\eesc} for surface emission, 
and are identical to the corresponding curves in the left panel of Fig.~\ref{fig:eesc_split_pairs_surface}, 
though extended here to \teq{\theta_f > 90^{\circ}} colatitudes via the dotted portions.
As with Fig.~\ref{fig:eesc_split_pairs_surface}, the magnetosphere is opaque 
to photons with energies exceeding \teq{\eesc} and transparent to those with energies below this value.

\begin{figure*}
 \begin{minipage}{17.5cm}
\vspace*{-10pt}
\centerline{\hskip 10pt \includegraphics[height=7.4cm]{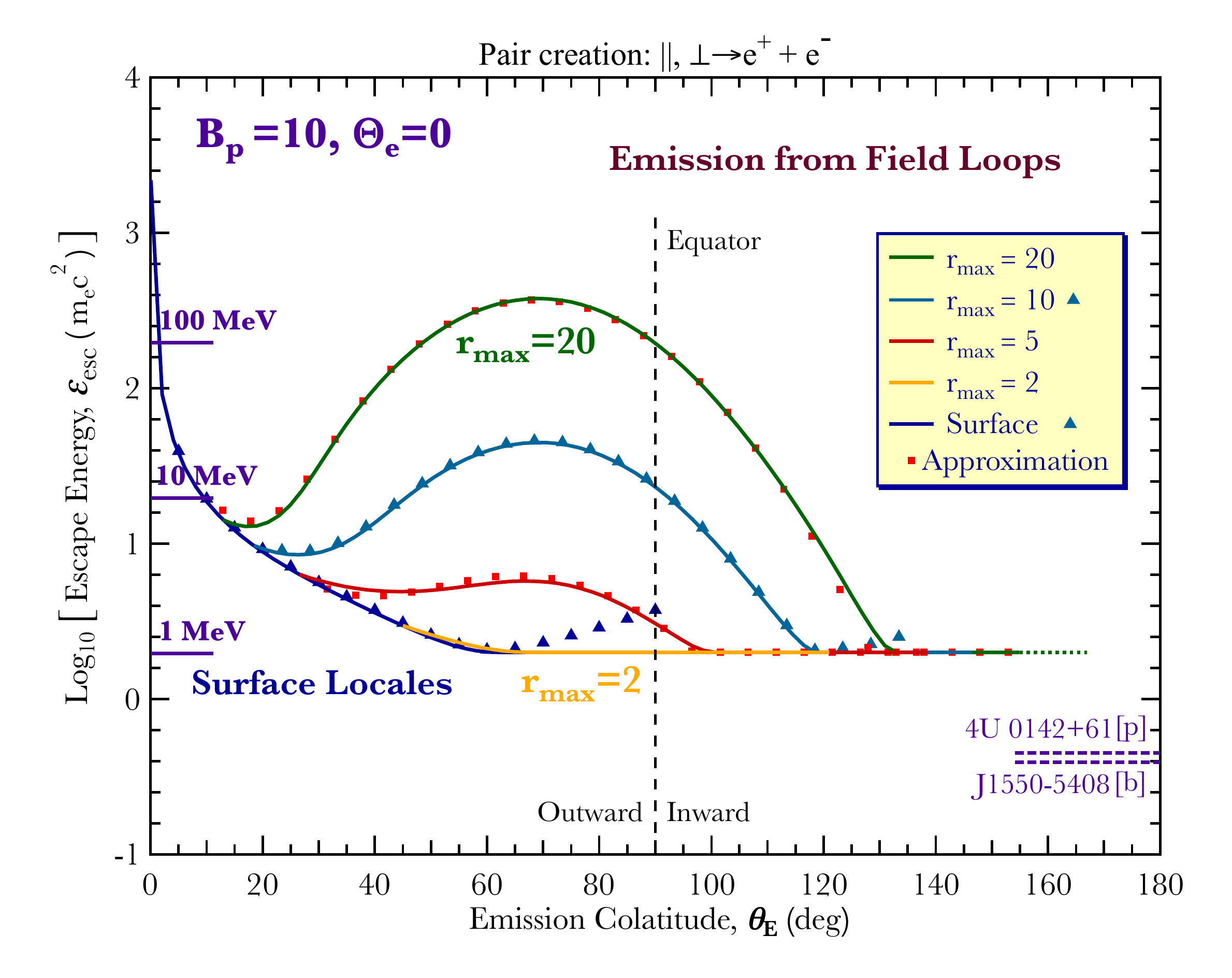}
   \hskip -15pt\includegraphics[height=7.4cm]{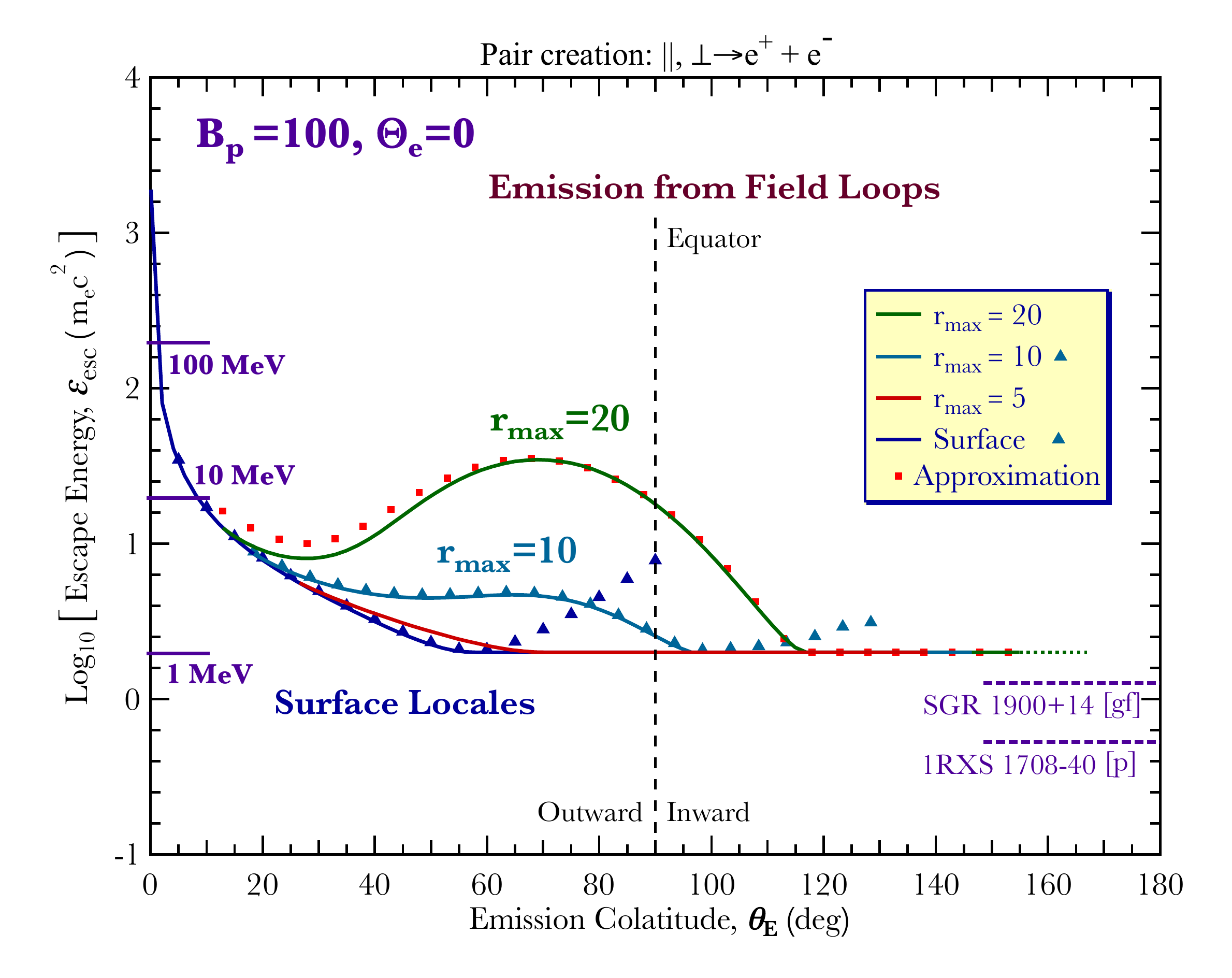}}
\vspace*{-5pt}
\caption{Pair creation escape energies in flat spacetime as functions of emission colatitude \teq{\thetaE}, 
for the polarization modes \teq{\parallel\to e^+e^-} (curves) and \teq{\perp\to e^+e^-} (triangles),
for the meridional case of light propagation initially parallel to the local magnetic field line (\teq{\Thetae =0}).
The left and right panels are for surface polar fields \teq{B_p=10} and \teq{B_p=100},
respectively.  The \teq{r_{\rm max}} choices and other details are as for Fig.~\ref{fig:eesc_split_loops}.
The dark blue curves for both panels are again for
surface emission cases that are displayed in the right panel of Fig.~\ref{fig:eesc_split_pairs_surface}.
The escape energies for \teq{\perp\to e^+e^-} opacities, exhibited 
as triangles for the surface emission and \teq{r_{\rm max}=10} cases only, 
are higher than those for \teq{\parallel\to e^+e^-} because of the higher threshold 
and lower conversion rate.  Note that the purple dashed lines defining marker 
maximum \underline{observed} energies 
of various magnetars (see Fig.~\ref{fig:eesc_split_loops}) are lower than the absolute threshold of pair
creation, \teq{2m_ec^2}. 
 \label{fig:eesc_pairs_loops}}
\end{minipage}
\end{figure*}

Loop emission curves begin and end on the surface emission curve at the two footpoint colatitudes.
Each of these loop emission curves exhibits a peak at around \teq{\thetaE \sim 70\degree}, 
either side of which \teq{\eesc} decreases sharply.  To aid identification of the magnetic hemispheres,
the vertical dashed lines represent the colatitude for equator, dividing the regions where photons are emitted outward and inward. 
The escape energy curves are asymmetric about the equator mostly because of inherent 
differences in field strengths and \teq{\thetakB} values for outward and inward trajectories.
The dotted parts of the inward portions of all curves (at the largest colatitudes) 
denote the shadowing cases in which emitted photons would impact the stellar surface;
in such circumstances, whether photon splitting happens or not is largely irrelevant.
Observe that near the lower polar regions \teq{\thetaE\sim 180^{\circ}}, there is some slightly non-monotonic 
behaviour for the curves, originating in the complex interplay between field magnitudes and 
field line curvature in determining splitting attenuation lengths.
Escape energies for polarization-averaged emission are also depicted as triangles for surface emission and loop emission with \teq{\rmax=10}. 
These ``loci'' are at slightly lower \teq{\eesc} than the \teq{\perp\to\parallel\parallel} case, 
since \teq{{\cal R}^{\rm sp}_{\rm ave}} is always greater than \teq{{\cal R}^{\rm sp}_{\perp\to\parallel\parallel}}. 
Finally the red squares in the Figure represent an empirical approximation for loop emission, 
which is generally applicable for moderate altitudes.  This is detailed in the Appendix, 
with the red dots representing computations using Eq.~(\ref{eq:emp_split}), and is useful for 
implementation in magnetar emission models: see Sec.~\ref{sec:resComp}.

Fig.~\ref{fig:eesc_split_loops} also includes markers signifying the approximate maximum energies 
detected for four different magnetars, sampling both quiescent and flaring emission.  
These serve as lower bounds to implied turnover energies, and
provide a benchmark for interpreting how photon splitting can impact the spectrum emergent 
from the magnetosphere.  The magnetars were selected so that their polar field values, as
inferred from pulsation timing, were relatively proximate to the 
\teq{B_p=10,100} choices in the respective panels.  The timing ephemerides enable estimates of
\teq{B_p\sin\alpha \sim 6.4\times 10^{19} \sqrt{P \dot{P}}} in the vacuum rotator case \citep{ST83},
serving as lower bounds to \teq{B_p}.  Here \teq{\alpha} is the inclination angle 
between the stellar rotation and magnetic moment axes.  The
inferred field strengths are also impacted by plasma loading of the 
magnetosphere, where currents contribute as well as Poynting flux in transporting 
angular momentum to infinity \citep[e.g., see][]{HCK99}.  For various magnetars,
equatorial field values \teq{B_p/2} for the case of orthogonal rotators (\teq{\alpha =90^{\circ}}) are 
listed in the McGill magnetar catalog \citep{OK14}.  The maximum energies for the persistent 
emission data in pulsed hard X-ray tails are 230 keV for AXP 4U 0142+61
\citep{Hartog08a}, which has \teq{B_p = 2.7 \times 10^{14}}G, and 
270 keV for AXP 1RXS 1708-40 \citep{Hartog08b} with \teq{B_p = 9.3 \times 10^{14}}G, both sources 
being observed by INTEGRAL. The representative maximum energies for flaring activity 
are 200 keV for SGR J1550-5408 (\teq{B_p = 4.3 \times 10^{14}}G) during its storm of 
bursts dating from January 2009 \citep[see][for {\it Fermi}-GBM observations]{vdH12}, 
and 650 keV for SGR 1900+14 (\teq{B_p = 1.4 \times 10^{15}}G) during its sole detected giant flare on 27 August 1998 
\citep[see][for {\it BeppoSAX} data]{Feroci99}.

\newpage

Comparison between the escape energy contours and these observational results 
constrains the probable magnetospheric location of hard X-ray emission 
in the selected magnetars --- similar inferences can be made for other such ultra-magnetized neutron stars.
For \teq{B_p=10} cases, 
the observed maximum energy for persistent emission from 4U 0142+61 is comparable to the peak value of the \teq{\rmax=2} curve.
Thus the \teq{\perp} polarization state photons cannot be emitted from the region with \teq{\rmax<2}, 
otherwise the maximum observed energy would be attenuated by photon splitting.
For the \teq{\rmax=5} loop, the escape energy curve exceeds the maximum observed energy for \teq{\thetaE<115\degree}.
Therefore, for loop emission with \teq{\rmax=5} colatitudes greater than 115\teq{\degree} are 
forbidden by the 4U 0142+61 observations as described in \cite{Hartog08a}; 
these correspond to inward propagation at emission points.  Similarly, for \teq{B_p=100} cases, 
magnetospheric regions with \teq{\rmax \lesssim 4} are completely forbidden 
by the maximum observed energy for the persistent emission of 1RXS 1708-40, 
while the \teq{\rmax=5} loop is permitted at low colatitudes (\teq{\thetaE<85\degree}).
The reason for this \teq{\thetaE} asymmetry is that the inward emission 
at high colatitudes precipitates propagation into regions with strong fields and shorter radii of field curvature, 
so that consequently the escape energies decline.
Polarization-averaged conclusions about permitted emission locales are similar, 
i.e. one would make similar inferences for the locales of origin of \teq{\parallel} photons if they are allowed to split.
Polar colatitudes, corresponding to very large \teq{\rmax}, are generally transparent to photon splitting, 
because the field line curvature is small, and photons generally propagate almost parallel to \teq{\bf B} for long distances.
This property can be deduced from Eq.~(\ref{eq:esc_energy_field_loop_low_b}).

Fig.~\ref{fig:eesc_pairs_loops} is the pair creation analog of the photon splitting results in Fig.~\ref{fig:eesc_split_loops}.
Solid curves here represent the escape energies for \teq{\parallel\to e^+e^-} mode pair creation for both loop emission and surface emission.
The \teq{\rmax=2} curve is omitted in the \teq{B_p=100} panel because it overlaps the surface emission curve, being essentially visually indistinguishable.
In contrast to the photon splitting plots, the pair creation escape energy curves here decrease slightly at first with increasing emission colatitude, 
then increase for the cases with large \teq{\rmax}.
For the \teq{\rmax=20} case in the left panel in Fig.~\ref{fig:eesc_pairs_loops}, 
the escape energy curve declines from the footpoint colatitude to around 17\teq{\degree}.
This is because the attenuation coefficient rate only weakly depends on the field strength 
in the super-critical field region (see Fig.~\ref{fig:eesc_split_pairs_surface}).
The escape energy curves are then determined primarily by the curvature of the dipolar field, 
and approach the surface emission curves for small \teq{\rmax} cases. 
Such character was identified by \cite{SB14}.  As \teq{\thetaE} continues to increase, then the 
drop in \teq{\BEmag} with altitude impacts the opacity as the field becomes sub-critical, and 
\teq{\eesc} then rises and eventually peaks before beginning to fall as the equator is approached.
Eventually  at large emission colatitudes \teq{\thetaE}, all the escape energy curves 
realize the horizontal plateau corresponding to the absolute pair threshold \teq{2m_ec^2} 
This is because photons quickly satisfy the pair threshold criterion \teq{\eesc \sin\thetakB > 2} for inward propagation into stronger, converging fields.

Escape energies for \teq{\perp\to e^+e^-} mode creation are displayed as triangles for surface emission and loop emission with \teq{\rmax=10}.
Their behavior imitates the corresponding \teq{\parallel\to e^+e^-} curves at first, 
but then they rise near the colatitudes where the \teq{\parallel\to e^+e^-} curves realize the threshold tails,
since the threshold for \teq{\perp\to e^+e^-} is not constant, but an increasing function of field
strength --- see Eq.~(\ref{eq:calFpp_perp}).  Analogous to Fig.~\ref{fig:eesc_split_loops}, 
the red squares in the figure represent an empirical approximation for pair creation, 
detailed in the Appendix --- see Eq.~(\ref{eq:emp_pairs}).
This approximation is fine for most loop colatitudes, but is not very accurate for low altitude cases 
where the escape energy curves decrease with increasing emission colatitude.
Notwithstanding, both this empirical form, and that for photon splitting, 
provide tools useful for assessing the importance of attenuation in emission models for magnetars, 
an element that will be addressed in Sec.~\ref{sec:resComp}.

\begin{figure*}
 \begin{minipage}{17.5cm}
\vspace*{-10pt}
\centerline{\hskip 10pt \includegraphics[height=7.4cm]{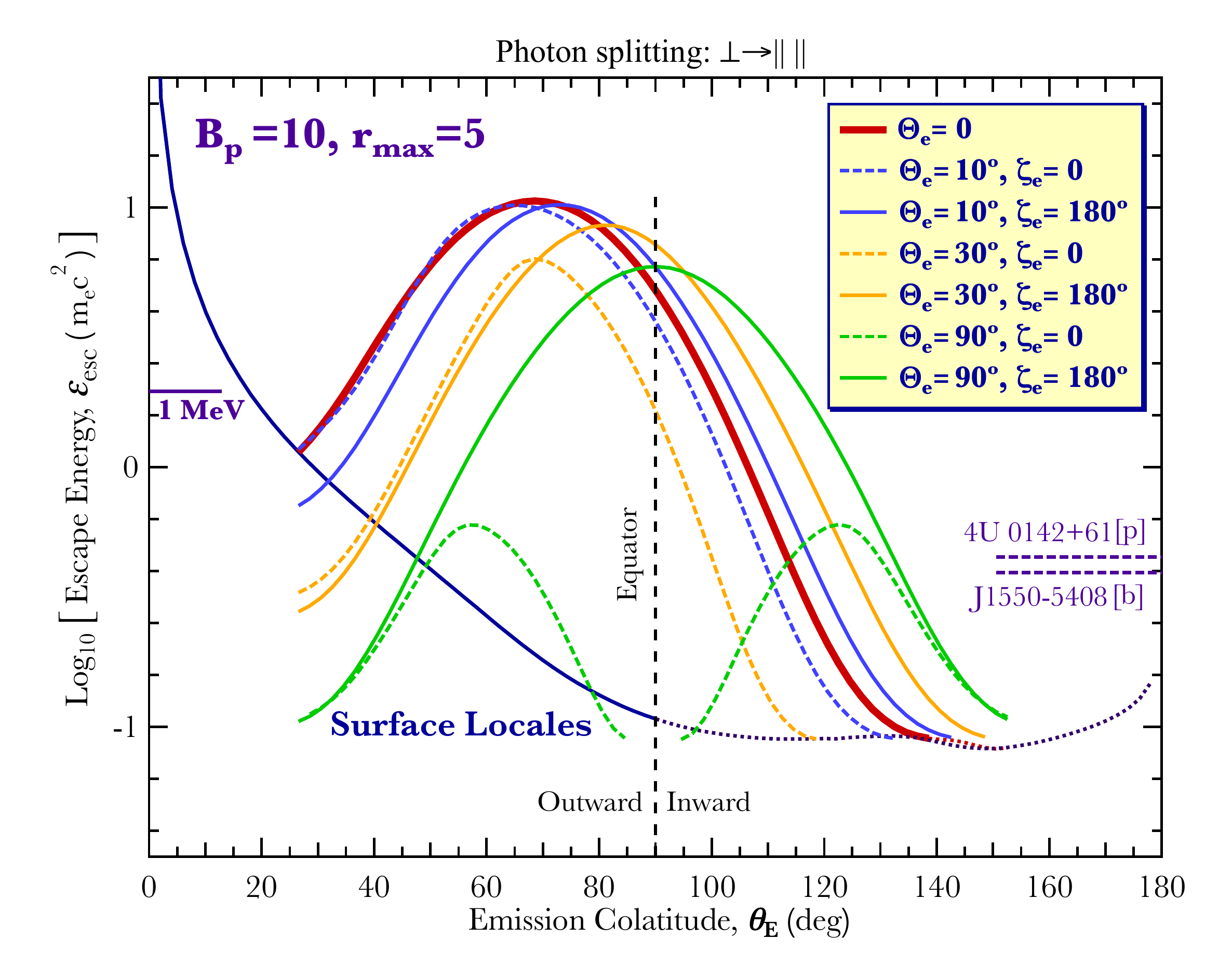}
   \hskip -15pt\includegraphics[height=7.4cm]{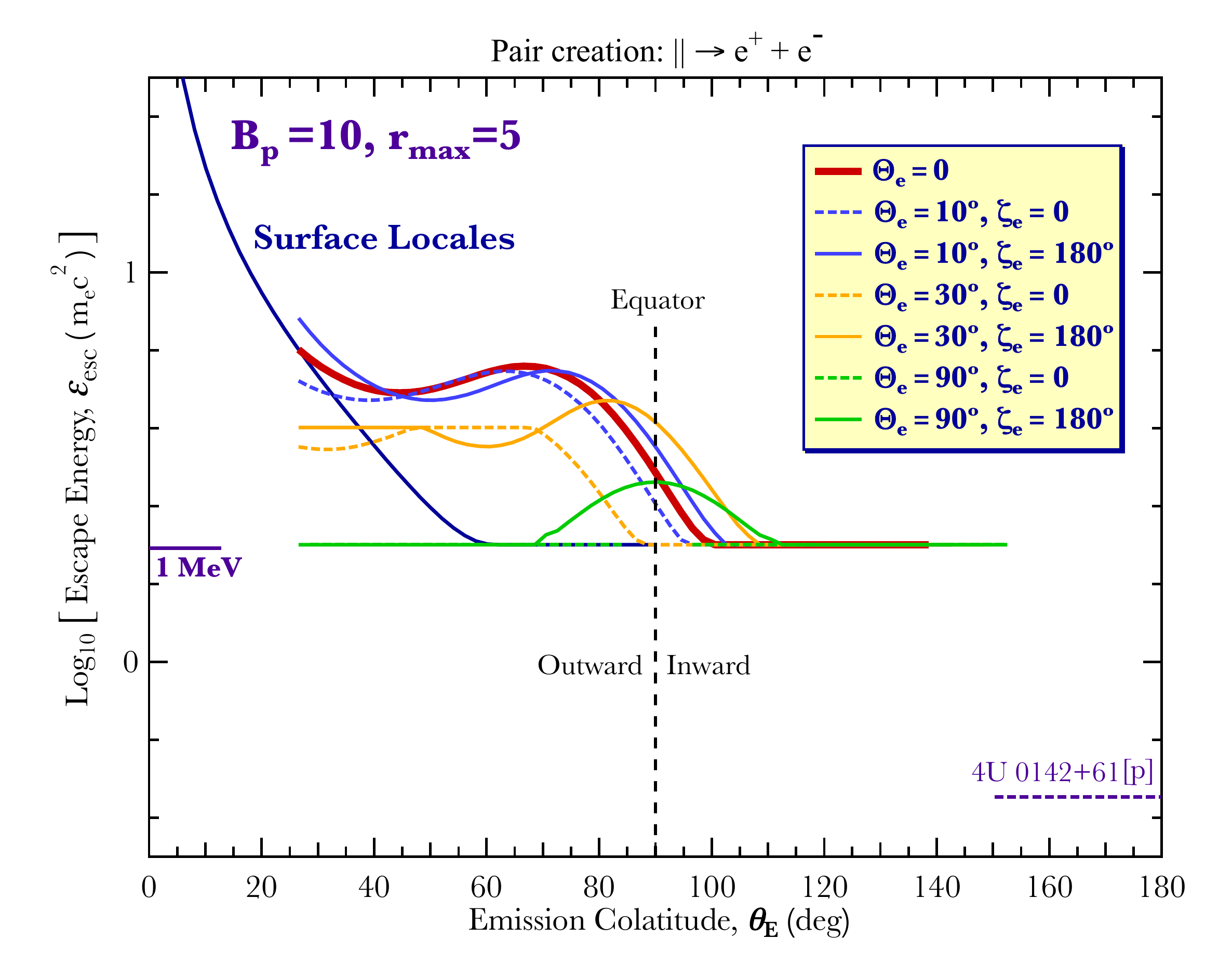}}
\vspace*{-5pt}
\caption{The escape energies \teq{\eesc} for photon splitting \teq{\perp\to\parallel\parallel} 
(left panel) and single photon pair production \teq{\parallel \to e^+e^-} (right panel) 
as functions of emission colatitude \teq{\thetaE}, 
for magnetospheric emission at points along the dipolar magnetic field loop with \teq{\rmax=5} and \teq{B_p=10}, propagating in flat spacetime.  
Curves are color-coded for initial emission angle \teq{\Thetae=0} (dark red, heavyweight), 
\teq{\Thetae=10\degree} (blue), \teq{\Thetae=30\degree} (orange) and \teq{\Thetae=90\degree} (green). 
The \teq{\Thetae=0} cases are those displayed in the left panel of 
Fig.~\ref{fig:eesc_split_loops} and Fig.~\ref{fig:eesc_pairs_loops}, respectively.
Different azimuth angles \teq{\zetae} (see section \ref{sec:geometry}) are represented by 
different line styles: dashed curves for \teq{\zetae=0^{\circ}} and solid curves for \teq{\zetae=180^{\circ}}.
The dark blue curves in each panel are for surface emission cases that are displayed in 
Figs.~\ref{fig:eesc_split_pairs_surface}, \ref{fig:eesc_split_loops} and \ref{fig:eesc_pairs_loops}.
 \label{fig:eesc_split_pairs_loops_obq}}
\end{minipage}
\end{figure*}

Identical to Fig.~\ref{fig:eesc_split_loops}, the maximum observed energies \teq{\emax}
from typical magnetars are also marked in the bottom right of each panel,
again serving to define lower bounds to \teq{\eesc}.
As these energies are lower than the absolute threshold of pair creation, 
it is obvious that pair creation opacity cannot be responsible for attenuating photons at or below the maximum energies 
observed in either persistent or flaring magnetar emission.
Yet, being generally below around 10--20 MeV when \teq{\rmax \leq 10}, 
the \teq{\emax} do provide important constraints that may explain why {\it Fermi}-LAT 
has not detected persistent emission from magnetars above 100 MeV \citep{Fermi_mag_bounds,Lietal2017}.
In comparing these pair creation results with escape energies for photon splitting, 
it is evident that the peak \teq{\eesc} value of each bell-shaped curve here is 
often lower than that of their photon splitting counterparts in Fig.~\ref{fig:eesc_split_loops}. 
This is because the field strength drops rapidly with the increase of the colatitude for those 
emission loops possessing \teq{\rmax > 5}.  So if we combine both photon splitting and pair creation, 
the escape energy curves for loop emission should be dominated by pair creation for quasi-equatorial regions near the peak escape energy 
and the ``wings" of the curves near the field line footpoint colatitudes are dominated by photon splitting.
This generic division applies to both \teq{B_p=10} and 100 cases;
the combined opacity is presented in Fig.~\ref{fig:eesc_split_pairs_loops_GR} in Sec.~\ref{sec:GR}.

Non-zero emission angles \teq{\Thetae} are required by most magnetar flare models, 
wherein the hard X-ray signals are produced by hot magnetospheric pair plasma.
Fig.~\ref{fig:eesc_split_pairs_loops_obq} illustrates the escape energies \teq{\eesc} as functions of 
emission colatitude for the \teq{\rmax=5} loop with mostly non-zero emission angle \teq{\Thetae},
for \teq{\perp\to\parallel\parallel} mode photon splitting and \teq{\parallel\to e^+e^-} mode pair creation. 
The  \teq{\Thetae=0} and surface emission cases are also depicted for comparison.
For each \teq{\Thetae} value, two azimuth angle (\teq{\zetae=0,180\degree}) cases are plotted. 
Curves with other azimuth angles lie between these two curves and are not depicted here to enhance clarity of the Figure.
For the \teq{\perp\to\parallel\parallel} mode photon splitting (left panel), 
a non-zero \teq{\Thetae} will generally lower the escape energy curves and 
tends to shift the peaks to higher colatitudes (\teq{\zetae=180\degree} cases) 
or lower colatitudes (\teq{\zetae=0} cases). 
For some \teq{\zetae=0} cases, a second peak appears at a high colatitude.
For photons emitted with \teq{\zetae=0} propagating inward, 
the increase of the field strength and the field line curvature strongly reduces the escape energy 
at quasi-equatorial colatitudes, thus creating the valley and the second peak for the curve.
This feature is evident in the \teq{\Thetae=90\degree}, \teq{\zetae=0} curve, which realizes a symmetric
double peak shape with respect to the equator because of the symmetry of the dipole field; 
a similar symmetry is possessed by the \teq{\Thetae=90\degree}, \teq{\zetae=180^{\circ}} curve.
The quasi-equatorial points for the \teq{\Thetae=90\degree}, \teq{\zetae=0} curve are omitted, 
generating gaps in the curve corresponding to photon trajectories that impact the star surface.
For the inward region of the Figure (\teq{\thetaE>\pi/2}), 
a non-zero \teq{\Thetae}, \teq{\zetae>90\degree} emission direction actually increases the escape energy above the \teq{\Thetae=0} values, 
since the photon trajectory is then directed away from the polar region where the field strength is very strong.
Clearly \teq{\zeta_e=0} cases have increased colatitude domains that are excluded as 
emission zones for the highest energy photons observed from 
4U 0142+61 (persistent signal) and SGR J1550-5408 (bursts).

The \teq{\parallel \to e^+e^-} mode escape energies are illustrated in the right panel of Fig.~\ref{fig:eesc_split_pairs_loops_obq}.
The general behavior of the escape energy curves resembles that of their \teq{\perp\to\parallel\parallel} counterparts. 
Curves with a large \teq{\Thetae = 30^{\circ}, 90^{\circ} } realize plateaux when \teq{\thetaE \lesssim 50^{\circ}}.
This is because these small colatitudes correspond to relatively low altitude and strong field strength, 
so that pair conversion is very efficient and a photon will create pairs as soon as it satisfies the threshold criteria.
Then a non-zero \teq{\Thetae} establishes an effective threshold of \teq{\eesc=2/\sin{\Thetae}}, 
and photons with higher energies will be attenuated once emitted.
This upper bound is clear for the \teq{\Thetae = 30, 90} cases, 
but not for the \teq{\Thetae}=10 cases since their entire escape energy curves are below that bound.
This upper bound is broken at large colatitude regions, around \teq{\thetaE \sim 68^\circ} for \teq{\rmax=5}, 
because of the reduction of the field strength. Then the bell shape appears.
For the \teq{\Thetae=90\degree} cases, the upper bound coincides with the \teq{2m_ec^2} absolute threshold, 
thus at most colatitudes the escape energy curves are horizontal lines.
In addition, for the \teq{\Thetae=90\degree}, \teq{\zetae=0} case, 
the escape energy curve is slightly distorted near the absolute threshold. 
This is because the pair creation coefficients \teq{{\cal R}^{\rm pp}_{\parallel,\perp}} 
are not monotonically-increasing functions
of photon energy when occupation of only the first or ground Landau levels is possible:
see Eqs.~(\ref{eq:calFpp_par}) and~(\ref{eq:calFpp_perp}) for details.

\subsection{Photon Splitting Opacity and Resonant Compton Upscattering in Magnetars}
 \label{sec:resComp}

To forge a more direct connection between the opacity determinations and magnetar observations, it is
insightful to consider a specific emission model.  An appropriate choice is the resonant inverse Compton
scattering scenario, which is regarded to be the dominant mechanism \citep{BH07,FT07,bwg11,Nobili08,Beloborodov13,Wadiasingh18}
for the generation of persistent hard X-ray emission.  As noted in the
Introduction, such resonant Compton scattering of soft thermal X-rays from magnetar surface layers or atmospheres by
relativistic electrons/positrons in activated regions of their magnetospheres is extremely efficient due to
the predominance of scatterings in the fundamental cyclotron resonance, i.e. at \teq{\erg = B} in the
electron rest frame. In that model, since  \teq{(\boldsymbol{\Omega} \times \boldsymbol{r} ) \times
\boldsymbol{B} } drift velocities are small in the inner magnetospheres of slowly-rotating magnetars, the
relativistic \teq{e^{\pm}} of dimensionless momenta \teq{\boldsymbol{p}_e = \gamma_e \boldsymbol{\beta}_e}
move along magnetic field lines, i.e. \teq{\boldsymbol{p}_e \boldsymbol{\cdot} \BEtotvec /(|\boldsymbol{p}_e |
| \BEtotvec |) = \pm 1 }. This is because transverse energies of these pairs are radiated away extremely
efficiently, with cyclotron cooling timescales much shorter than \teq{10^{-16}} seconds.  The interacting
leptons thus can be safely assumed to occupy their ground state Landau level in a quantum description.  Note
that throughout this discussion, \teq{\BEtotvec} denotes the magnetic field vector at the point of scattering.

The impact of photon splitting and pair creation in attenuating such emission is obviously greatest for the
highest energies. The approximate maximum photon energy in this upscattering mechanism differs from
traditional invocations of inverse Compton scattering due to the kinematics of the process. The kinematics
for resonant Compton scattering \citep[e.g.,][]{DH86} does not match that for
field-free Compton scattering, since momentum transverse to an external field does not have to be conserved
in QED.  Yet, energy conservation still holds in the presence of the magnetic field, and the upscattered
photon energy is ultimately bounded by \teq{\gamma_e m_e c^2}. For incident soft photon energies \teq{m_e
c^2 \erg_s \sim kT \sim 0.5} keV emanating from magnetar surface layers, the scattering generally leaves the
electron in its ground Landau state \citep{gonthier00,Gonthier-2014-PRD}. Moreover, owing to
relativistic aberration when \teq{\gamma_e \gg 1}, the incident photon is essentially parallel to
\teq{\Bvec} in the electron/positron rest frame. Under these assumptions, and restricting the discussion to
Minkowski spacetimes, the maximum {\it{resonant}} scattered photon energy in the observer frame is
\citep{BH07,bwg11,Wadiasingh18}
\begin{equation}
   \varepsilon_{f}^{\rm max} \; =\; \gamma_e ( 1+ \beta_e) \left[\frac{\BEmag }{1 + 2 \BEmag } \right] 
   \quad ,\quad
   \BEmag \; =\; \vert \BEtotvec\vert \quad .
 \label{eq:efmax}
\end{equation}
The maximum outgoing photon momentum \teq{\boldsymbol{k}_f} (\teq{\equiv \boldsymbol{k}}) is directed
parallel to the lepton momentum at the interaction point \teq{\boldsymbol{p}_e \boldsymbol{\cdot}
\boldsymbol{k}_f /(|\boldsymbol{p}_e | |\boldsymbol{k}_f |) = 1}, i.e., \teq{\Thetae \approx 0} in the notation 
of Section~\ref{sec:eesc_small_theta}. Eq.~(\ref{eq:efmax}) dictates the effective cut-off for the inverse 
Compton upscattered spectrum, for any instantaneous viewing angle with respect to the magnetic axis \teq{\muvec}. 
An exception to this occasionally arises when the scattering samples soft photons deep in the Wien tail of
the quasi-Planck spectrum radiating from the stellar surface, and the emission is exponentially suppressed.
There is also some low-level, non-resonant emission appearing a little above \teq{\efmax}.

Specializing to the dipole field adopted throughout this paper, one may specify \teq{\BEmag} in terms of
magnetic colatitude \teq{\thetaE } via \teq{\BEmag = B_p \rns^3\sqrt{1 + 3\cos^2 \thetaE }/(2 r^3)}, where
\teq{r/\rns = \rmax \sin^2 \thetaE  } using a loop's altitude parameter \teq{\rmax}. If
\teq{\nvec} is the unit vector signifying the direction to an observer 
(i.e., \teq{\nvec = \boldsymbol{k}_f /\vert \boldsymbol{k}_f\vert }), 
in uniformly-activated magnetospheres, only specific colatitudes at meridional
and antimeridional field loops locales satisfy the alignment condition
\begin{equation}
   \frac{\boldsymbol{p}_e \cdot \nvec}{ |\boldsymbol{p}_e|} \; =\; 1 \quad ,
 \label{eq:tangents}
\end{equation}
which is required for Eq.~(\ref{eq:efmax}) to be operable. Accordingly, the conditions for strong 
Doppler-boosted and collimated upscattering emission are readily identifiable geometrically. The 
meridional/antimeridional plane is that defined by the two vectors \teq{\muvec} and \teq{\nvec},
and therefore rotates with the star as it spins and \teq{\muvec} traces out a conical surface.
For leptons moving along field lines from the southern to the northern magnetic footpoints, 
the colatitude  \teq{\thetaE} of interaction points
satisfying Eq.~(\ref{eq:tangents}) is found in Eq.~(34) of \cite{Wadiasingh18}, 
wherein \teq{\vartheta_0} corresponds to \teq{\thetaE} here:
\begin{equation}
   \cos \thetaE  \;\approx\;  \pm \sqrt{\left[ 2 + \cos^2 \theta_v - \cos \theta_v \sqrt{8 + \cos^2 \theta_v} \right]/6} \quad .
 \label{eq:scattpoint}
\end{equation}
The dot product \teq{\muvec \cdot \nvec = \cos \theta_v} captures information on the instantaneous viewing angle
\teq{\theta_v} to the magnetic axis.   The two branches of this relation correspond to antimeridional
(\teq{+}) and meridional (\teq{-}) loops, respectively, when {\it the leptons traverse from south to north}.  
These two branches demarcate two hemispheres: antimeridional loops sample \teq{0 < \thetaE < \pi/2}, while
meridional ones correspond to \teq{\pi /2 < \thetaE  <\pi }. Thus, for an observer staring down towards the
north pole, the emission points for the meridional loop are in the southern hemisphere, and in the northern
hemisphere for the antimeridianal loop. The range of allowed colatitudes for a field line above the stellar
surface is bounded by its footpoint colatitudes: \teq{ \thetamin < \thetaE  < \thetamax} where
\teq{\thetamin = \arcsin \sqrt{1/\rmax} } and \teq{\thetamax = \pi - \thetamin}. When the leptons move in
the opposite direction along the loops, from north to south, the hemispheres comprising the meridional and
antimeridional loops are interchanged, as are the applicable signs for the approximate identity in
Eq.~(\ref{eq:scattpoint}).

If pulsating magnetars are inclined rotators, with an angle \teq{\alpha} between the spin (\teq{\Omegavec})
and magnetic (\teq{\muvec}) axes that satisfies \teq{\cos\alpha = \Omegavec \cdot \boldsymbol{\hat{\mu} } },
then as the star rotates through spin phases \teq{\Omega t}, an observer samples a range of 
instantaneous viewing angles \teq{\theta_v} during its period:
\begin{equation}
   \cos \theta_v \; =\; \sin \alpha \cos (\Omega t) \sin \zeta + \cos \alpha \cos \zeta 
   \; , \quad 
   \zeta \; =\; \alpha + \theta_{v0} \; .
 \label{eq:modulation}
\end{equation}
In this relation, \teq{\zeta} is the observer viewing angle
satisfying \teq{\cos\zeta = \Omegavec \cdot \nvec}. Thus, \teq{\theta_{v0}} is the viewing angle relative to
\teq{\muvec} at rotational phases \teq{\Omega t = 2n\pi}, with \teq{n=0, \pm 1, \pm 2\dots\;}. It is now
apparent that when this result is inserted into Eq.~(\ref{eq:scattpoint}), and the result used to specify
\teq{\BEmag = \vert\BEtotvec\vert}, that the phase dependence of the expression for the effective maximum energy
due to resonant Compton scattering in Eq.~(\ref{eq:efmax}) can be mapped throughout the pulse period \teq{0
\leq t \leq 2\pi /\Omega}.  Clearly, this maximum upscattering energy is identical for both meridional and
antimeridional loops.  Yet, the model spectra and pulse profiles of hard X-rays in magnetars may not possess
such a symmetry if the activation of the magnetosphere is not uniform in magnetic latitude --- such is the
case if the densities and/or energies of electrons emanating from footpoints in the two hemispheres differ. 
For the purposes of the discussion here, it is assumed that the magnetosphere is uniformly activated on a
toroidal surface defined by field loops with the same value of \teq{\rmax}, with the intersection of the toroid 
and the stellar surface forming two circles of footpoints at colatitudes \teq{\thetamin} and \teq{\pi - \thetamin}.

\begin{figure*}
 \begin{minipage}{17.5cm}
\vspace*{-3pt}
\centerline{\includegraphics[width=0.99\textwidth]{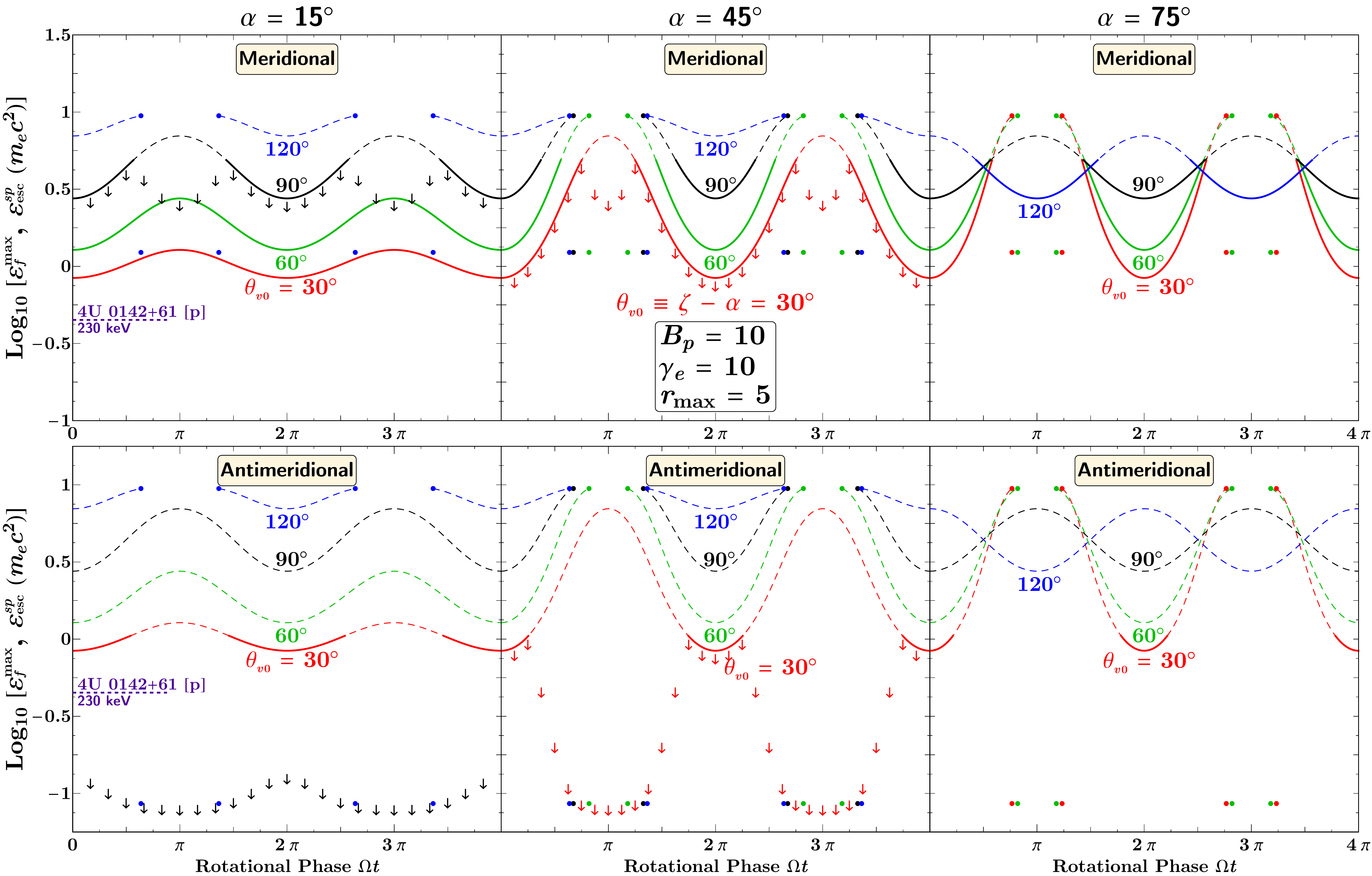}}
\vspace*{-5pt}
\caption{Photon splitting escape energies (dotted curves; \teq{\eescsp}) for
\teq{\perp\to\parallel\parallel} and resonant Compton maximum cutoff energy
\teq{\efmax} (solid and long-dashed) as functions of spin phase \teq{\Omega t},
for oblique rotators with \teq{\alpha \equiv \arccos ( \Omegavec \cdot \boldsymbol{\hat{\mu} }) = \{15^\circ, 45^\circ, 75^\circ \}}.  
The first and second rows of panels correspond to meridional and antimeridional loops,
respectively. In each panel, curves are depicted for several choices of the
particular observer viewing angle \teq{\theta_{v0} = \zeta - \alpha} at phase zero
(\teq{\cos \Omega t = 1}), as labeled and color-coded. The \teq{\efmax} emission
energies are depicted as solid curves if \teq{\efmax < \eescsp } and long-dashed
loci when the inequality is violated and \teq{\perp} photons are attenuated in a
range of energies below \teq{\efmax}. Abrupt termination of curves at the dots
marks emission at field line footpoints, demarcating gaps that constitute
subsurface locales. Also, the downward arrows mark four sample profiles for
effective maximum energies observed in the \teq{\perp} state, signifying
\teq{\min \{ \efmax ,\, \eescsp \}}.  See text for details.}
\label{fig:modulation}
\end{minipage}
\vspace{-10pt}
\end{figure*}

To illustrate the potential role of photon splitting in attenuating X-rays and $\gamma$-rays produced
by the resonant Compton upscattering mechanism, Fig.~\ref{fig:modulation} depicts the maximum resonant
Compton energy \teq{\efmax} according to Eq.~(\ref{eq:efmax}),  applicable to both polarization states,
\teq{\perp} and \teq{\parallel}, together with the escape energies \teq{\eescsp} for the
\teq{\perp\to\parallel\parallel} mode of photon splitting. The splitting escape energies are computed
specifically at the resonant Compton interaction colatitudes in Eq.~(\ref{eq:scattpoint}) where the Doppler
boosting leading to emission up to around \teq{\efmax} is obtained; these \teq{\eescsp} values were
determined using the empirical formula given in Eq.~(\ref{eq:emp_split}) in the Appendix, which applies to curves
in Fig.~\ref{fig:eesc_split_loops}. We focus here on photon splitting, since it can attenuate at energies below the
absolute pair threshold \teq{2m_ec^2} that are pertinent to the turnovers in magnetar persistent tail
emission: for comparison, the maximum energy of \teq{\sim 230}keV detected from 4U 0142+61 is marked on the left axis.
The choice of \teq{\gamma_e} and \teq{B_p} is identical to that used for the resonant Compton \teq{\efmax}
modulation profiles exhibited in Fig.~5 of \cite{Wadiasingh18}.   However, a lower value of
\teq{r_{\rm max} = 5} is chosen here to connect to Fig.~\ref{fig:eesc_split_loops} and to accentuate the potential
influence of splitting attenuation on phase-resolved emission signatures.   Results are exhibited for three
different rotator magnetic inclination angles \teq{\alpha =\{15^\circ,45^\circ, 75^\circ\}}.

Solid curves represent the \teq{\efmax} traces when unattenuated, and the dashed portions of these
constitute rotational phases where \teq{\eescsp < \efmax} and attenuation of \teq{\perp} polarizations is
guaranteed.  The dotted curves define \teq{\eescsp} for splitting.  The rotational modulations of both
\teq{\efmax} and \teq{\eescsp} are illustrated for five choices of the observer viewing angle \teq{\zeta},
as labelled using \teq{\theta_{v0}=\zeta - \alpha}.  For select values of \teq{\theta_{v0}}, both sets of
curves possess gaps, constituting phases where the Doppler-boosted emission on meridional/antimeridional
field lines samples zones inside the neutron star, i.e. when \teq{\thetaE  > \thetamax} or \teq{\thetaE  <
\thetamin \approx 0.1476\pi \approx 26.57^{\circ}} (for \teq{\rmax=5}).  The dots at the extremities of
these gaps (the ends of the curves) thus signify scattering locales that are coincident with field line
footpoints on the stellar surface. Note that there will still be emission at the phases corresponding to the
gaps in the curves, but that it will be primarily at energies substantially below the values of \teq{\efmax}
at the footpoints, since it will correspond to signals not Doppler-boosted along the field lines. Observe
also that the particular case of \teq{\alpha = 15^\circ}, the \teq{\theta_{v0} = 150^\circ} result permits
no solutions outside the neutron star, for either hemisphere.  This is because the rotator is almost
aligned, and the observer's viewing angle is almost polar throughout the period so that scattering locales with
the required Doppler beaming along \teq{\nvec} are restricted to very small radii with
supercritical fields.

The various loci possess a wealth of information worth highlighting. Two distinctive features of the plots are
immediately apparent.  The first is that between the meridional and antimeridional cases, for each of the
\teq{\alpha} values there is a large disparity in photon splitting escape energies when averaged over spin
phase.  This asymmetry between northern and southern hemispheres is expected from the skewness of the curves
exhibited in Fig.~5 that address emission tangent to field loops: antimeridional scattering points with
\teq{\thetaE > \pi /2} will generally possess lower escape energies since they sample downward propagation
into regions with stronger fields and shorter radii of field curvature. A consequence of this property is
that even if the two hemispheres are equally activated with electron distribution profiles that are
symmetric in magnetic latitude and uniform in magnetic longitude, an observer will discern a phase
dependence in the spectral shape near the maximum energies that signals the action of attenuation due to
photon splitting.  If Adler's selection rules for splitting apply, then there will also be a strong
polarization dependence to this effect --- the phase-dependent attenuation will emerge in the \teq{\perp}
state only, and this happens to be the dominant polarization state for resonant upscattering near
\teq{\efmax} \citep{BH07,Wadiasingh18}.  In contrast, if all polarization modes of
splitting can proceed, the splitting \teq{\eescsp} curves will move down slightly (see Fig.~\ref{fig:eesc_split_loops}
for a general indication).  Then the spectral attenuation at select pulse phases will be more pronounced,
yet the polarization dependence of this shaping of a spectral turnover will be weaker. These attenuation
nuances for the resonant Compton model are being explored in Wadiasingh et al. (in prep.).  Note that since
turnovers are not cleanly observed in X-ray tail data for various magnetars, the statistical quality of
phase-resolved observations at the maximum energies is presently insufficient to explore these types of
signatures.  Accordingly, probes of this information define potential science agendas for planned missions
such as AMEGO\footnote{see {\tt https://asd.gsfc.nasa.gov/amego/index.html}.} and
e-ASTROGAM \citep[see][]{DeAngelis17}; see the case argued in detail in \cite{Wadiasingh2019}.

The second noticeable characteristic of Fig.~\ref{fig:modulation} is the strong anti-correlation between
\teq{\efmax} and \teq{\eescsp}: phases where the maximum upscattering resonant energy is greatest coincide
with phases where the splitting escape energy is the lowest, i.e. opacity is the highest.  This feature is
easily understood as a consequence of large \teq{\efmax} in Eq.~(\ref{eq:efmax}) corresponding to scattering
locales in high fields, i.e. comparatively near the footpoints of field lines, precisely the locations where
the escape energies for splitting are the lowest in Fig.~\ref{fig:eesc_split_loops}.  This anti-correlation enhances the
ability of photon splitting to attenuate hard X-rays generated by resonant Compton upscattering, especially
for the antimeridional hemisphere; the same coupling would arise for pair creation escape energies, albeit
at energies above \teq{2m_ec^2}.  Inspection of the combined splitting/pair creation
escape energy depictions in Fig.~\ref{fig:eesc_split_pairs_loops_GR} indicates that pair creation lowers
the overall \teq{\eesc} values at high altitudes along field loops near the equator.  In the phase-modulated 
illustration in Fig.~\ref{fig:modulation}, this would manifest itself as a modest lowering of the dotted zones 
around the peaks of the \teq{\eesc} curves.  Thus, the main impact of adding pair creation opacity 
is to slightly broaden the phase intervals where opacity curtails the maximum possible energy  in \teq{\eesc > 2} domains, 
and then essentially exclusively in the meridional examples.

A third, related property is that as the rotator obliquity increases from \teq{\alpha = 15^{\circ}} to 
\teq{\alpha = 45^{\circ}}, the amplitude of the \teq{\efmax} and \teq{\eescsp} modulations increases, though
it appears to saturate at even larger obliquities.  This is a signature of how a viewer selectively samples
the magnetosphere at different rotational phases subject to the Doppler beaming restriction 
in Eq.~(\ref{eq:tangents}).  As \teq{\alpha} increases from small values, larger portions of the
\teq{\rmax = 5} loops provide access to the condition of tangency for the \teq{\nvec} vector, thereby
generating more extensive ranges of field strengths sampled.  This manifests itself as an increase in the
modulations, but once a full sampling of the toroidal loop surface is enabled, at \teq{\alpha \sim
45^{\circ}}, no further rise in the scale of the modulations is observed when the rotator obliquity
continues to increase. While not illustrated here, modulation traces for larger values of \teq{\rmax} do
exhibit (i) swings with greater ranges of \teq{\efmax} \citep[see Fig.~5 of][]{Wadiasingh18} and
\teq{\eescsp} due to the increase in field strength ranges spanned by a loop, and (ii) both lower mean
values of \teq{\efmax} and higher means for \teq{\eescsp} because the larger loops possess lower fields on
average.  As a result, the extent of attenuation by splitting is reduced substantially for meridional cases
when \teq{\rmax} is somewhat larger, but only marginally so for the antimeridional hemisphere.

It is also insightful to focus on the character of the modulation curves for individual choices of the
viewer ``impact angle'' \teq{\theta_{\rm v0}}.  Generally, at a particular spin phase, three cases
pertaining to opacity configurations are identifiable: (O: opaque) both antimeridional and meridional escape
energies are lower than \teq{\efmax}, and thereby the \teq{\perp}-mode is attenuated near the cut-off
\teq{\efmax}, (S: semi-opaque) only the antimeridional emission is split while the meridional one escapes,
and (T: transparent) neither meridional/antimeridional escape energies are low enough to be constraining,
and the emission from the toroidal surface is transparent to photon splitting. As noted above, if only the
\teq{\perp} polarization splits, cases O and S should evince relatively high polarization degree, nearly \teq{\sim
100\%}, at the spectral cut-off.  Unsurprisingly, choices of higher \teq{B_p} or lower \teq{r_{\rm max}}
than depicted in Fig.~\ref{fig:modulation} generally result in cases O and S presenting throughout most
rotational phases.  As an intermediate example, case O transpires for the \teq{\theta_{v0} = 90^\circ}
choice for a range of phases near \teq{\cos \Omega t = -1} for all the  \teq{\alpha} columns of
Fig.~\ref{fig:modulation}, but not for other rotational phases. At these phases, the instantaneous
\teq{\theta_v} is generally large and the emission colatitude in Eq.~(\ref{eq:scattpoint}) is proximate to
the magnetic footpoints.  Likewise, the \teq{\theta_{v0} = 30^\circ} example exhibits phases of case S and
case T, accompanied by phase-resolved polarization degrees that vary strongly as the signal transitions from
transparency to opacity for the photon splitting mode \teq{\perp\to\parallel\parallel}.  Interestingly, a
pulse profile may present all three configurations O, S and T depending on rotational phase, an obvious
example being provided by \teq{\theta_{v0}=30^{\circ}} for the \teq{\alpha = 45^{\circ}} case.

As a cautionary note accompanying the results presented in this subsection, the highlighted signatures
should not be over-interpreted.  In reality, resonant Compton upscattering emission involves a convolution
of various \teq{r_{\rm max}} toroidal surfaces \citep[e.g., see][]{Wadiasingh18} and Lorentz factors for
electrons populating the magnetosphere, both of which will blur the features identified here. Moreover, for
some rotational phases the resonant spectrum is effectively suppressed because the local resonance condition
samples soft photon energies \teq{\erg_s} above the Planck mean of \teq{3 kT}, and this can move the
luminous part of the upscattering signal to well below the escape energy \teq{\eescsp}. In addition,
excursions from dipolar field morphology such as those predicted in twisted field models \citep[e.g.][]{cb17} 
will complicate the phase dependence of \teq{\efmax} and \teq{\eescsp} substantially,
introducing higher-order asymmetries. Notwithstanding, if the highlighted phase-resolved polarimetric characteristics
are detected by future hard X-ray spectrometers and polarimeters, they would help confirm the action of
photon splitting in Nature.  In particular, signatures such as ones highlighted here offer a path to
constraining both the magnetar geometry parameters \teq{\{ \alpha, \zeta \}}, and the particle Lorentz
factor \teq{\gamma_e} distribution, and also which \teq{r_{\rm max}} bundles contribute to the total
emission. 

\vspace{-5pt}
\section{General Relativistic Influences}
 \label{sec:GR}

To complete this exposition on photon splitting and pair creation opacity in magnetar magnetospheres,
it remains to explore how curved spacetime at low altitudes impacts the escape energies.  
Since such general relativistic (GR) effects have already been identified at length in the papers of
\cite{HBG97,BH01,SB14}, the presentation here will be limited.  While light bending and its change in frequency 
imposed by the curvature of spacetime are familiar concepts, the non-Euclidean metric distorts the 
mathematical form of a dipole magnetic field \citep{Petterson74,WS83,GH94}, altering the field line shape 
and increasing the strength of the field.  All these modifications combine to alter the rates for 
any QED process involving photons in strongly-magnetized neutron star environs.  
The interplay between the GR magnetic geometry and the GR photon ray-tracing is important 
to identify, since this captures complete information on the vector values of the field 
\teq{\BGR} and the photon momentum \teq{\kGR} at each locale that are needed for 
precise determination of the opacity.

\vspace{-5pt}
\subsection{General Relativistic Opacity Construction}

The general relativistic construction in this paper follows that laid out in detail in \cite{SB14}.  
The optical depth is obtained as an integration over the geodesic path length variable \teq{s}, with the attenuation 
coefficient specified at each point in the local inertial frame (LIF) in terms of the variables 
\teq{B}, \teq{\omega} and \teq{\thetakB}.  The expressions for these are posited without derivation.
The simplest quantity to specify is obviously the LIF photon energy \teq{\omega},
\begin{equation}
   \omega\; =\; \dover{\erg}{\sqrt{1-\Psi}}\quad ,\quad
   \Psi\; =\; \dover{r_s}{r}\; \equiv\; \dover{2GM}{c^2r}
 \label{eq:blueshift}
\end{equation}
at radius \teq{r}, where \teq{r_s=2GM/c^2} is the Schwarzschild radius
of a neutron star of mass \teq{M}.  This expresses the blueshift of the photon energy
\teq{\erg} observed at infinity.  Throughout this section, we will set \teq{M = 1.44M_{\odot}}.
The proxy radius parameter \teq{\Psi} facilitates the specification of all pertinent general 
relativistic quantities, and will be used as the variable for the path integration.

The general relativistic form of a dipole magnetic field in a Schwarzschild metric was 
expounded in \citet{WS83,Muslimov86,GH94}.  Its evaluation in the LIF can be expressed in a compact form 
\begin{equation}
   \BGR\; =\; 3\dover{B_p\Psi^3}{r_s^3} \left\{ \xi_r (\Psi)\,\cos\theta\, \hat{r}
      + \xi_{\theta} (\Psi)\,\sin\theta\, \hat{\theta}\right\}
 \label{eq:dipole_GR}
\end{equation}
by defining field component functions 
\begin{eqnarray}
   \xi_r (x) & = & - \frac{1}{x^3} \left[ \log_e(1-x) + x + \frac{x^2}{2} \right] \nonumber\\[-5.5pt]
 \label{eq:xi_r_theta_def}\\[-5.5pt]
   \xi_{\theta} (x) & = & \frac{1}{x^3\sqrt{1-x}} \left[ (1-x)\, \log_e(1-x) + x - \frac{x^2}{2} \right] \;\; .\nonumber
\end{eqnarray}
%
%
For Euclidean geometry, corresponding to \teq{\Psi\ll 1}, the leading terms of Taylor series expansions
yield \teq{\xi_r(\Psi )\approx 1/3} and \teq{\xi_{\theta} (\Psi )\approx 1/6}, 
so that then Eq.~(\ref{eq:dipole_GR}) reproduces the familar result in 
Eq.~(\ref{eq:B_dipole}).  When \teq{\Psi} is not small, in general the 
\teq{\xi_r, \xi_\theta} exceed these values and the magnitude of the magnetic field 
in the LIF is increased by the curvature of spacetime.

The curved photon geodesic can be identified once the initial momentum at the emission point
is specified.  To illustrate the key influences of general relativity on 
splitting and pair creation opacity determinations, we restrict considerations 
to light propagation in a plane constituted by a single magnetic longitude, a protocol adopted by \cite{SB14}.  
The momentum therefore possesses no \teq{\phi} component and can be 
specified at any point by \teq{\hat{r}} and \teq{\hat{\theta}} coordinate vectors, 
just like the magnetic field in this plane.  This provides a convenient simplification 
of the GR \teq{(\boldsymbol{k}, \boldsymbol{B})} geometry algebra.  
For this configuration, and indeed for all photon orbits 
in a Schwarzschild geometry, a photon's path at infinity moves parallel to
a straight line drawn from the center of the star, with the path displaced from it by a distance \teq{b}.  
This impact parameter \teq{b} is proportional to the ratio of the orbital angular momentum
and the energy, both being conserved along the photon trajectory.  
Typical geometry is illustrated in \citet{PCF83}, or in textbooks such as \citet{ST83}.

\begin{figure}
\vspace*{-5pt}
\centerline{\hskip -10pt\includegraphics[width=9.25cm]{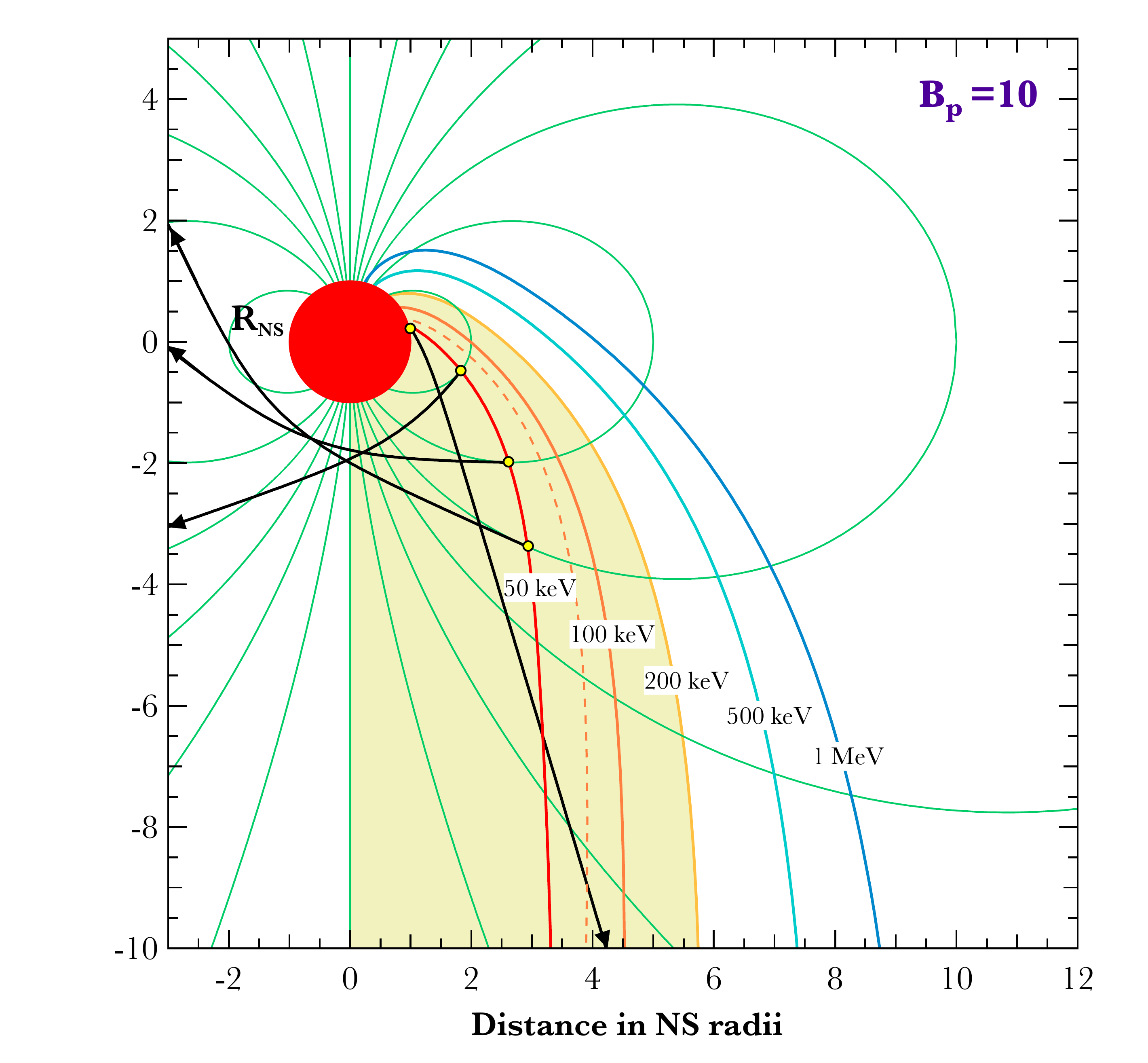}}
\vspace*{-7pt}
\caption{A ``splittosphere'' diagram of opacity for a magnetar with \teq{M=1.44M_\odot}, \teq{R=10^6}cm and \teq{B_p=10},
for photon propagation in the Schwarzschild metric that is coplanar with a nested set of dipolar field lines (depicted in light green). 
The \teq{(x,z)} coordinates (as in Fig.~\ref{fig:geometry}) are scaled by neutron star radii and the central red solid unit circle represents the neutron star. 
Colored contours represent the lowest possible 
emission altitude for transparency at a given colatitude, for a photon of a fixed energy (as labeled) that is emitted parallel to the local 
\teq{\Bvec} and away from the upper pole.  Inside these contours, the magnetosphere
is opaque to the \teq{\perp\to\parallel\parallel} mode of photon splitting; the yellow shaded region indicates the zone of opacity for photons 
with an energy of 200 keV.  The dashed orange contour is a flat spacetime version of the 100 keV case.
Photon trajectories are plotted for selected emission points on the 50 keV contour 
as black curves with arrows. The curvature of these geodesics in general relativity is determined 
using Eq.~(\ref{eq:curved_traj}) and its adaptation beyond periastron passage.
\label{fig:splittosphere}}
\vspace{-20pt}
\end{figure}

Scaling \teq{b} by the Schwarzschild radius, as we have with \teq{r}, 
introduces a new trajectory parameter \teq{\Psi_b= r_s/b} that can
be related to emission point parameters \teq{\PsiE} and \teq{\thetaE} 
for the specific case of the dipole field via
\begin{equation}
   \Psi_b \; =\;  \PsiE\, \sqrt{ (1-\PsiE)\, \left\{ 1+\left[\xi (\PsiE)\right]^2 \cot^2\thetaE \right\}  }\quad ,
 \label{eq:Psi_b_def}
\end{equation}
where
\begin{equation}
    \xi (\Psi )\; =\; \dover{\xi_r(\Psi )}{\xi_{\theta}(\Psi )}\quad .
 \label{eq:xi_def}
\end{equation}
This functional relationship \teq{\Psi_b (\PsiE,\, \thetaE )} applies only for the 
special case of photon emission parallel to the local magnetic field, the major focus 
of Section~\ref{sec:eesc_numerics} that is motivated therein.
The textbook photon orbit equation is an integral equation that expresses the 
change in polar angle in terms of the path's radial parameter \teq{\Psi}, namely
\begin{equation}
   \theta(\Psi) \; \equiv\; \thetaE + \Delta\theta \; =\; 
   \thetaE \pm \int_{\Psi}^{\PsiE} \frac{d\Psi_r}{\sqrt{\Psi_b^2-\Psi_r^2(1-\Psi_r)}}\quad .
 \label{eq:curved_traj}   
\end{equation}
This expresses the dependence \teq{\theta (r)} as viewed by an observer 
at infinity.  There are two branches to this solution of the geodesic equation, and 
these are bifurcated by the periastron point \teq{\Psi = \Psi_p} where the radius 
is a minimum.  This location corresponds to \teq{d\theta /dr = - (\Psi /r) d\theta /d\Psi \to \infty}, 
so that \teq{\Psi_b^2 = \Psi_p^2 ( 1 - \Psi_p)}.  For an outward-propagating photon, 
the branch with the positive sign in Eq.~(\ref{eq:curved_traj}) is applicable.
For inward propagation, the negative sign is chosen at first.  In that case, if the
photon passes through periastron, the subsequent portion of outward propagation 
employs the positive sign, starting from \teq{\Psi_p} instead of \teq{\PsiE}, and 
adds two \teq{\Delta \theta} contributions in a piecewise fashion.  Examples of three 
periastron transit trajectories where this construction is applied are illustrated in Fig.~\ref{fig:splittosphere}.

The exact calculation of the photon trajectory requires a time-consuming numerical integration,
motivating the use of suitable analytic approximations.  While the series 
evaluation technique of \cite{SB14} works well for quasi-polar emission locales, something 
more widely applicable is needed to treat periastron passage.  The ``cosine relation'' provided by 
\cite{Beloborodov02} gives an accurate approximation, relating the photon's radius, 
its direction and the polar angle, from which we can obtain an analytic form of \teq{\theta(\Psi)}. 
Beloborodov's formula works with \teq{\sim 1}\% accuracy for \teq{\Psi <0.5}, which suffices
for our calculations here.

\newpage

From the foregoing elements, one can assemble the last important ingredient 
for opacity calculations, namely the angle \teq{\thetakB} between the photon 
momentum \teq{\kGR} and the magnetic field \teq{\BGR} in the LIF.  The photon 
momentum is posited in Eq.~(A1) of \cite{HBG97}, and is directly derivable 
from the path equation via \teq{k_{\theta}/k_r = d\theta /dr}, the approach of
\cite{SB14}.  Thus, for photons moving away from the star (\teq{0 < \thetaE < \pi /2}),
\begin{equation}
   \kGR\; =\; \dover{\erg}{\Psi_b\sqrt{1-\Psi }}
    \left\{ \sqrt{\Psi_b^2 - \Psi^2 (1-\Psi )}\, {\hat r} + \Psi \sqrt{1-\Psi} \, {\hat \theta} \right\}
 \label{eq:momentum_LIF}
\end{equation}
in the local inertial frame, with a magnitude that captures the blueshift of the photon energy 
relative to that perceived at infinity.   Forming the cross product of this with 
Eq.~(\ref{eq:dipole_GR}), one soon arrives at 
\begin{equation}
   \sin\thetakB \; =\; \dover{\sqrt{\Psi_b^2 - \Psi^2 (1-\Psi )}
        - \Psi \sqrt{1-\Psi}\, \xi (\Psi ) \, \cot\theta }{\Psi_b\, \sqrt{  1+ \left[\xi (\Psi )\right]^2 \cot^2\theta } }
 \label{eq:sinthetakB_GR}
\end{equation}
an expression that is also obtained by rearranging Eq.~(37) of \cite{GH94}.  
Employing the form for \teq{\Psi_b} in Eq.~(\ref{eq:Psi_b_def})
quickly reveals that when \teq{\Psi=\PsiE}, this expression yields
\teq{\sin\thetakB=0}.  For light emitted in the lower hemisphere, 
\teq{\pi /2 < \thetaE < \pi}, a change of sign appears in the \teq{\hat{r}} term in Eq.~(\ref{eq:momentum_LIF}),
and thus also in the first term of the numerator of Eq.~(\ref{eq:sinthetakB_GR}).  Since 
then \teq{\cot\theta} is negative, one again observes that \teq{\sin\thetakB=0} at the point 
of emission.

The opacity integral can be formed using the pathlength element \teq{ds = \sqrt{1-\Psi}\, c\, dt} 
in the LIF, in which the coordinate time element \teq{dt} can be expressed in terms of observer 
frame coordinates using formulae for Shapiro delay.  A change of variables to \teq{\Psi} 
is expedient, and the details are given in \cite{SB14}.  The optical depth in this general relativistic 
formulation takes the form
\begin{equation}
   \tau (\Psi )\; =\; r_s \int_{\Psi}^{\PsiE}
   \dover{{\cal R}(\omega,\, \sin\thetakB, \vert\BGR\vert )
         \, \Psi_b\, d\Psi_r}{\Psi_r^2 \sqrt{(1-\Psi_r ) \left\{\Psi_b^2-\Psi_r^2(1-\Psi_r )\right\} }}\quad ,
 \label{eq:tau_Psi_GR}
\end{equation}
where the arguments of the quantum pair creation attenuation coefficient
\teq{{\cal R}} are given by Eqs.~(\ref{eq:blueshift}),~(\ref{eq:dipole_GR})
and~(\ref{eq:sinthetakB_GR}).  One then formally defines the attenuation 
length \teq{L} as in \citet{HBG97} and \citet{BH01} via
\begin{equation}
   \tau \left(\PsiL\right) \; =\; 1
   \quad ; \quad 
   s\left(\PsiL\right) \; =\; L \quad .
 \label{eq:atten_length_GR}
\end{equation}
\teq{L} is approximately the cumulative LIF distance that a photon
will travel from its emission point before being attenuated.  As before, the escape energy 
\teq{\eesc} is defined by the condition \teq{L\to \infty}, i.e. \teq{\PsiL \to 0}.

Using this formalism, a variety of opacity results were generated, most of which were 
designed as checks on the construction and numerics and are not presented here, 
since our focus is on highlighting the main influences of general relativity.  First among 
these checks were comparisons with the results of \cite{SB14}: the new C codes 
developed here for computing the opacity in Eq.~(\ref{eq:tau_Psi_GR}) reproduced 
the pair creation attenuation lengths presented in Fig.~5 and the escape energies 
in Figs.~6 and~7 of that paper with excellent precision.  To appraise the photon splitting GR 
opacity numerics, we compared escape energies with those illustrated for surface 
emission in Fig.~1 of \cite{BH01}, which were systematically around 20\% lower than our 
determinations at colatitudes greater than around 5--10 degrees when photons were emitted
initially parallel to \teq{\Bvec} (\teq{\Thetae = 0}).  Approximately the same offset applied 
to pair creation opacity also, and our numerical results again agreed with those 
generated by \cite{SB14}.   In contrast, for \teq{\Thetae =0.01}, when \teq{\thetaE\lesssim 1^{\circ}}, 
there was good agreement between our computations and the values of \teq{\eesc} in Fig.~1 of \cite{BH01}
for both attenuation processes.  

The numerical evaluation of \teq{\sin\thetakB} 
was suggested by \cite{SB14} as the possible origin of the discrepancy at  \teq{\thetaE\gtrsim 1^{\circ}}, since their 
determinations of \teq{\sin\thetakB} agreed with those in Fig.~5 of \cite{GH94}, yet are smaller by around 20\% 
relative to those in \cite{HBG97}.  Our codes here nicely reproduce the \teq{\sin\thetakB} results of both \cite{GH94}
and \cite{SB14}.  A very recent inspection of the old propagation codes used in \cite{HBG97,BH01} 
has revealed a factor of 2 bug in the photon ray-tracing (in Eq.~(A2) of Harding et al. (1997), 
the first \teq{m} should be \teq{2m}).  This has been corrected, thereby bringing
the numerical evaluations from this old independent code into excellent agreement with the 
escape energies presented here and in \cite{SB14}.

\subsection{Opacity Volumes and Escape Energies}

The general relativistic construction was used to develop
an illustration of the geometry of regions of opacity for photon splitting in the 
magnetosphere of a \teq{B_p=10} magnetar, which is presented in Fig.~\ref{fig:splittosphere}.
This represents a planar (meridional) section of azimuthally-symmetric volumes
of opacity/transparency, for the case of photons emitted parallel to field lines 
at positions \teq{x>0} and directed away from the upper (northern) pole.  
For a selection of photon escape energies \teq{\eesc = 50, 100, 200, 500, 1000} keV
as observed at infinity, the depicted contours separate the zones of opacity in the interior
from the outer regions of transparency where the field is lower on average.  The opacity 
volumes, termed the ``{\bf splittosphere},'' extend down to the magnetic axis of symmetry at large 
colatitudes, or the stellar surface at small colatitudes. A cross section of the splittosphere for the 200 keV case 
is identified by the shaded yellow region inside the corresponding surface contour.
Photons emitted inside these regions with energies 
larger than the labeled energies are subject to splittings \teq{\perp\to\parallel\parallel}. 
Similar volumes would arise for the other modes of splitting if they are permitted.  
These volumes, which somewhat resemble cometary tails, decrease in flat spacetime 
(see the dashed curve for 100 keV) and obviously decrease with the increase of photon energy 
in accord with the rates in Eq.~(\ref{eq:splitt_atten}).

\begin{figure*}
 \begin{minipage}{17.5cm}
\vspace*{-10pt}
\centerline{\hskip 10pt \includegraphics[height=7.4cm]{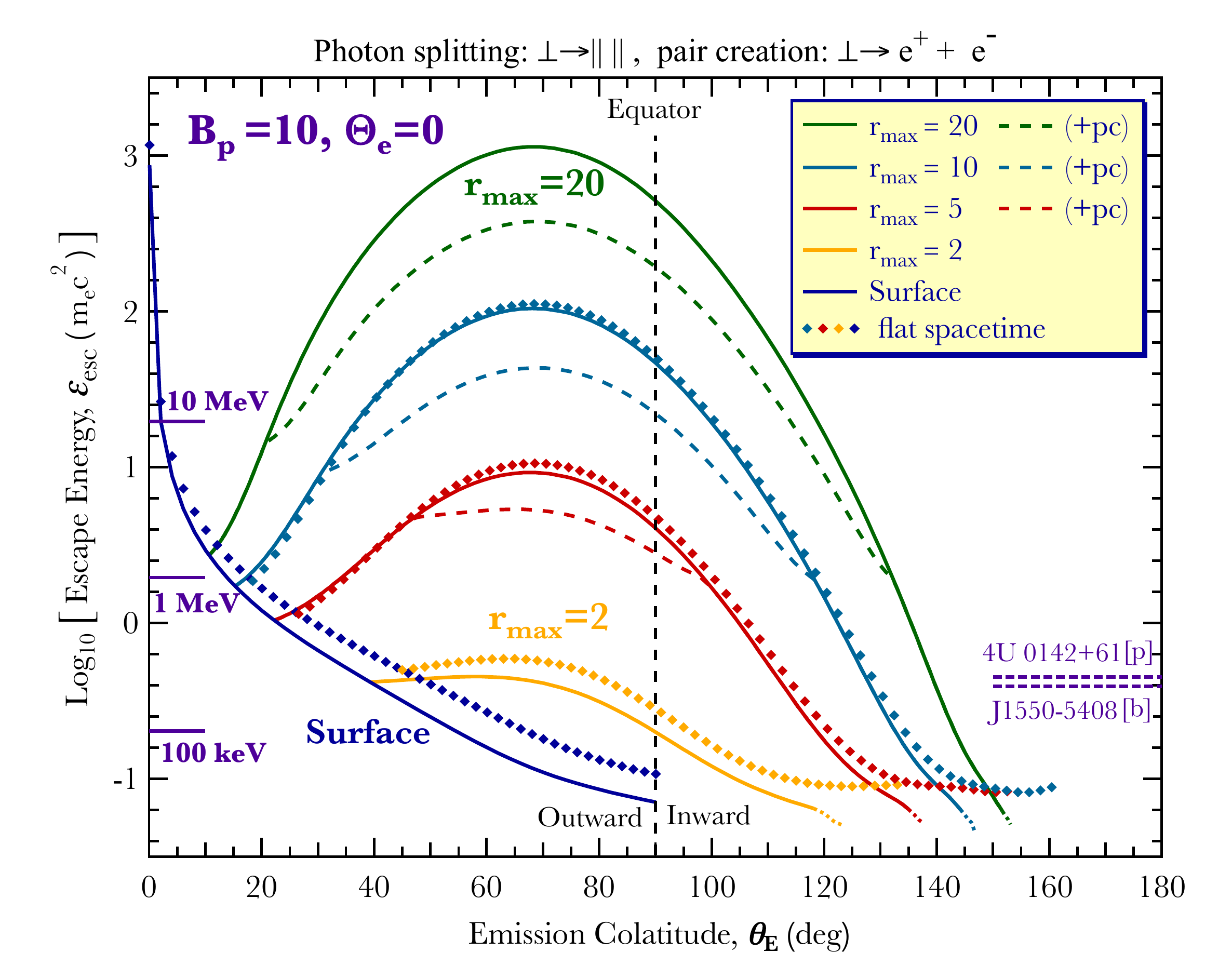}
   \hskip -15pt\includegraphics[height=7.4cm]{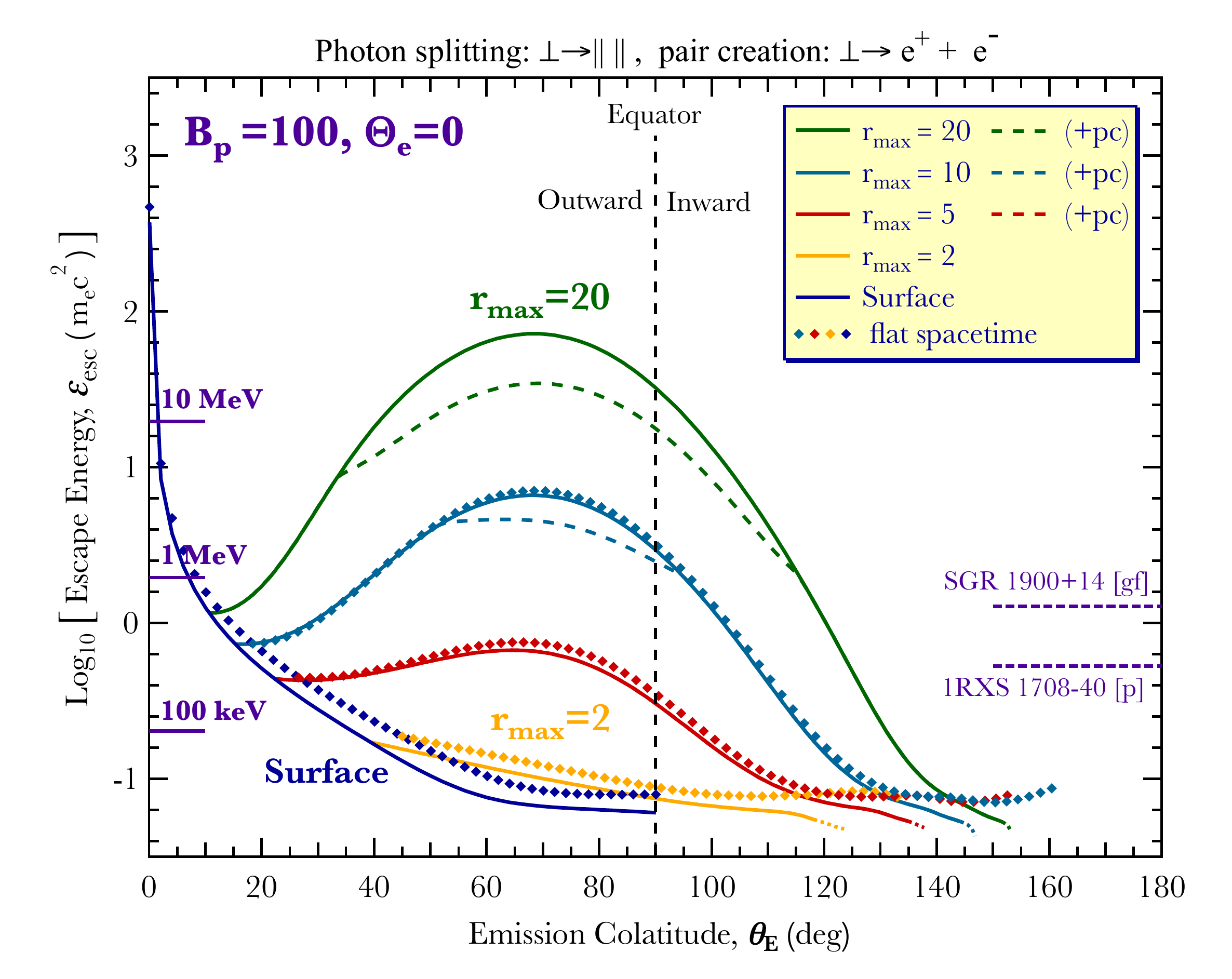}}
\vspace*{-5pt}
\caption{Escape energies for photon splitting (mode \teq{\perp\to \parallel\parallel},
solid curves) and also when adding pair creation of the \teq{\perp} state to the 
total opacity ($+$pc, dashed curves),
for maximum loop altitudes  \teq{r_{\rm max}=2,5,10,20} in units of \teq{\rns}.
All determinations are for general relativistic propagation, except for those denoted as being for flat spacetime,
namely the traces of diamonds for the surface and \teq{\rmax =2, 5, 10} examples.
A meridional specialization with photon emission initially parallel to the local magnetic field line 
(\teq{\Thetae =0}) is adopted, as in Figs.~\ref{fig:eesc_split_loops} and \ref{fig:eesc_pairs_loops},
with the left and right panels for surface polar fields \teq{B_p=10} and \teq{B_p=100},
respectively.  The surface emission curves for both curved and flat spacetimes are 
truncated at the equator (outward emission only).   Other aspects of the display such as 
shadowing portions and the magnetar marker energies are as in 
Figs.~\ref{fig:eesc_split_loops} and \ref{fig:eesc_pairs_loops}.  As \teq{\thetaE} increases 
in transiting from footpoint to the opposite footpoint, the escape energy traces first a splitting evaluation  (solid), 
then to a lower pair creation determination (dashed, at equatorial colatitudes), and back to a 
splitting evaluation (inward propagation as depicted in Fig.~\ref{fig:splittosphere}); see text for details. 
 \label{fig:eesc_split_pairs_loops_GR}}
\end{minipage}
\end{figure*}

The splittosphere contours are asymmetric about the \teq{x} axis, because it is 
assumed that the leptons flow from north (upper) pole to south (lower) pole. 
If leptons transit field loops in both directions, a mirror image opacity zone would also arise
with a reflection of the one in Fig.~\ref{fig:splittosphere} about the \teq{x} axis.  
It is noticeable that each contour here realizes an 
asymptote that is approximately parallel to the \teq{z} axis and extends downwards to infinity
at colatitudes \teq{\pi > \thetaE > \pi /2} where inward emission prevails.
While the field strength at the emission point drops well below the critical value in these locales, the 
photon trajectory passes close to the stellar surface where the field is intense and 
\teq{\thetakB} is also large.  In such cases, the dominant contribution to the opacity arises near
and at periastron passage and can be substantial.  This is the reason why the escape energy 
contours do not merge with the \teq{z}-axis at finite values of the \teq{x} coordinate. 

While Fig.~\ref{fig:splittosphere} applies to photon emission along field lines as 
typically arises for Doppler-boosted mechanisms, one can quickly discern the general character of
modifications to the splittosphere morphology for non-zero emission angles \teq{\Thetae}.  
Inspection of the left panel of Fig.~\ref{fig:eesc_split_pairs_loops_obq} reveals that at low colatitudes 
somewhat near the north pole, the escape energies generally drop as \teq{\Thetae} 
increases above zero, while in the lower magnetic hemisphere \teq{\eesc} is 
reduced in the \teq{\zetae = 0} cases exhibited, signalling higher opacity; 
a reduction in opacity arises in \teq{\zetae = 180^{\circ}} cases.  Accordingly, one anticipates 
that in the upper hemisphere of Fig.~\ref{fig:splittosphere} at colatitudes \teq{\thetaE < \pi /2}, 
the zones of opacity will bulge somewhat.  In the lower, \teq{\pi /2 < \thetaE < \pi} 
hemisphere, the splittosphere will shrink or expand depending on the 
direction of emission relative to the local \teq{\Bvec}.  Generally, in such cases, photon 
trajectories with larger periastron radii will sample lower fields and lower opacities 
on average, and therefore yield smaller regions for opaque conditions.

The impact of general relativity on the determination of escape energies can be identified 
via plots like those in Figs.~\ref{fig:eesc_split_loops} and~\ref{fig:eesc_pairs_loops}.
To streamline this information, we now combine photon splitting and pair creation escape 
energies graphically in Fig.~\ref{fig:eesc_split_pairs_loops_GR}, where the \teq{y}-axis
represents the escape energy as perceived by an observer at infinity.  Solid curves represent 
escape energies for photon splitting.  Dashed curves correspond to a combination of 
pair creation and photon splitting, being clearly visible for
colatitudes \teq{\thetaE} where the pair creation escape energy is lower than that 
for splitting; hence the (+pc) designation in the legends.  Only results for the \teq{\perp} 
polarization are exhibited in Fig.~\ref{fig:eesc_split_pairs_loops_GR}, and similar 
character would appear for the \teq{\parallel} state, provided that all three CP-permitted 
modes of splitting operate.  The surface emission 
curves are depicted only for outward photon emission into the magnetosphere, i.e. truncated at the equator.

Flat spacetime curves are depicted for the surface emission and \teq{\rmax =2,5,10} 
examples as well as the GR ones to demonstrate how for most emission locales, general relativity reduces \teq{\eesc}, 
by a factor of up to two at \teq{\rmax < 2}, and by around \teq{10-20}\% for \teq{\rmax =10}.  
This \teq{\eesc} reduction is a direct consequence of the 
increase in field strength and the photon energy in the local inertial frame in the Schwarzschild metric.  
Accordingly, GR upgrades modify the constraints on the possible emission volumes only 
modestly, as depicted in Fig.~\ref{fig:splittosphere}.  Another feature of the plots is 
that the range of colatitudes for which pair creation is dominant is more restricted 
for the \teq{B_p=100} case.  This behavior follows from photon splitting being  
more effective at lower energies than pair creation.
Comparison of the footpoint colatitudes for the loop \teq{\rmax =2} examples in flat and 
curved spacetime in the northern hemisphere (\teq{\thetaE < \pi /2}) illustrates how,
when \teq{\rmax} is held fixed, 
general relativity moves the footpoints closer to the pole as the field there intensifies. 
A more subtle signature of the Schwarzschild metric
that appears in Fig.~\ref{fig:eesc_split_pairs_loops_GR}
is the fact that the endpoint colatitudes of the loop curves are not symmetrically spaced about the 
equator: the inward portions end at colatitudes more remote from the southern pole than 
the actual field line footpoint near the northern pole.  This is a consequence of the strong curvature of photon 
trajectories, as depicted in Fig.~\ref{fig:splittosphere}.  As the impact parameter \teq{b} declines, 
capture orbits can be realized, for which unattenuated photons cannot escape; this phenomenon 
prescribes the truncation colatitude of the curves.

\newpage

\section{Conclusions}

In this paper, opacities for photon splitting and pair creation have been calculated in the
magnetospheres of neutron stars, with an emphasis on magnetars. Attenuation
lengths and escape energies, the maximum energies for transparency from the emission point to
infinity, are computed for emission locales that are not exclusively tied to the stellar
surface or the magnetic poles. The presentation attaches special significance to loop emission
cases, which are more closely connected to physical emission models: in
slowly-rotating magnetars, motion of relativistic charges along field lines introduces strong
Doppler beaming of the radiation. Our calculations indicate that photon splitting dominates high
energy photon attenuation for supercritical field strengths, while in the slightly subcritical
field domain, pair creation is more often the leading attenuation mechanism.

For both persistent signals and flares in magnetars, photon splitting attenuation alone puts
constraints on the possible emission locales in their magnetospheres when accommodating the
maximum observed energies. For \teq{B_p=10} cases, magnetospheric regions within loops of
maximum altitudes \teq{\rmax\lesssim 2} (scaled in neutron star radii) are completely forbidden
by splitting attenuation opacity for persistent emission with energy above around 230 keV. For
the \teq{\rmax=5} loop case, emission regions with \teq{\thetaE>115^\circ} are excluded by
splitting. The inherent asymmetry comes from the restriction of leptons moving along field lines
from the northern (i.e., low \teq{\thetaE}) to the southern (\teq{\thetaE \sim 180^{\circ}})
magnetic footpoints when generating high-energy photons. The inward emission at high colatitudes
propagates into southern polar regions with strong fields and shorter radii of field curvature,
so that the opacities increase accordingly. Similarly, for \teq{B_p=100} cases, magnetospheric
regions with \teq{\rmax\lesssim 4} are completely forbidden for persistent emission with energy
above around 270 keV, and the \teq{\rmax=5} loop is permitted at low colatitudes with
\teq{\thetaE<85\degree}.  In Section 5, general relativistic modifications are addressed, 
and for most emission locales, it reduces \teq{\eesc}, 
by a factor of up to two at low altitudes \teq{\rmax \lesssim 2}, 
and by around \teq{10-20}\% for \teq{\rmax =10}.  

In Section~\ref{sec:resComp}, a model of attenuation by photon splitting combined with resonant inverse
Compton scattering was presented for the case where only photons with \teq{\perp} polarization
can split. This illustrated that the maximum upscattering (\teq{\efmax}) and escape (\teq{\eescsp}) energies both depend
strongly on the pulse phase of the rotating star even when toroidal field bundles are uniformly activated.
The asymmetry of the two hemispheres and the anti-correlation between \teq{\efmax} and
\teq{\eescsp} derive strong spectral and polarization signatures, highlighted in
Fig.~\ref{fig:modulation}. Allowing all CP-permitted polarization modes of splitting to proceed
will modify both the spectral and polarization signatures, yet still retaining a richness of
potential observational diagnostics; such will be the subject of future study.  These
signatures, if assessed using future phase-resolved observations from proposed MeV band telescopes
such as AMEGO and e-ASTROGAM \citep[e.g., see][]{Wadiasingh2019}, 
could identify which modes of photon splitting are operating, probe the emission region locale, 
and help determine the major geometric parameters of the star,
i.e. the inclination angle, \teq{\alpha}, between the magnetic and rotation axes, as well as an
observer's viewing angle, \teq{\zeta}.

\vspace{-10pt}
\section*{Acknowledgments}

The authors thank George Younes and the anonymous 
referee for suggestions helpful to the polishing of the manuscript.
M.~G.~B. acknowledges the generous support of the National Science Foundation
through grant AST-1517550, and NASA's {\it Fermi} Guest Investigator Program
through grant NNX16AR66G. Z.~W. is supported by the NASA postdoctoral 
fellowship program.

\vspace{-20pt}
\section*{Appendix}

\noindent {\bf Empirical Escape Energies for Loop Emission:}

The escape energy calculations for loop emission can be directly incorporated in various the emission models for magnetars.
The example addressed in Section \ref{sec:escape} is the resonant Compton upscattering scenario for hard X-ray tails in magnetars. 
It is expedient to employ compact analytic approximations for opacity in such models, 
and so here we assemble empirical fits to escape energies computed in Section \ref{sec:eesc_numerics}.
Well below pair threshold, the photon splitting attenuation coefficient is proportional to \teq{\varepsilon^5}, 
thus the optical depth for photons with a energy of \teq{\varepsilon} 
can be expressed as \teq{\tau=\exp{\left[-(\varepsilon/\eesc)^5\right]}}.
We developed an empirical approximation for the \teq{\perp\to\parallel\parallel} escape energy for photon emission along field loops, yielding the form
\begin{eqnarray}
   \varepsilon_{\rm esc}^{sp}&\approx&5.33\;B_p^{-1.2}\; \rmax^{3.4}\;\dover{\sin^{12.9}{\thetaE}}{\thetaE^{6.18}}\nonumber\\[-5.5pt]
 \label{eq:emp_split}\\[-5.5pt]
   &&+\;0.58\;B_p^{-0.5}\;\rmax\; \exp \bigl(-\thetaE^2 \bigr) + 0.1\;B_p^{-0.07}\quad.\nonumber
\end{eqnarray}
This was designed to best describe the shape of the loop curves exhibited in Figure 5. 
The first term in Eq.~(\ref{eq:emp_split}) represents the bell-shaped humps, 
while the second and the third terms deliver corrections for small and large colatitudes, respectively.
Eq.~(\ref{eq:emp_split}) works well for large \teq{\rmax} cases, 
but deviates slightly from numerical results for small \teq{\rmax} emission, i.e. \teq{\rmax<3}.
In addition, this formula is also approximately applicable to other splitting modes and the polarization-averaged mode, 
since their escape energies are similar (see Fig.~\ref{fig:eesc_split_loops}).
To apply this to the polarization-averaged case, intuition 
suggests multiplying the result in Eq.~(\ref{eq:emp_split}) by the factor 
\teq{\left[{{{{2\cal{M}}_1^2}/(3\cal{M}}_1^2+{\cal{M}}_2^2)}\right]^{1/5}}.  
Since the major portion of the loops is at altitudes where sub-critical fields 
exist, the \teq{{\cal M}_{\sigma}} in this factor can be replaced by constants, 
yielding a multiplicative factor \teq{(338/1083)^{1/5} \approx 0.792} for the 
polarization-averaged escape energy formula.

Similarly, we derived an empirical formula for \teq{\parallel \to e^+e^-} mode pair creation 
based on the results presented in Fig.~\ref{fig:eesc_pairs_loops} as well, which is expressed as
\begin{eqnarray}
   \varepsilon_{\rm esc}^{pp} & \approx &  \max \biggl\{ 2, \;\; 
       2.3\;B_p^{-1.05}\; \rmax^{3.1}\;\dover{\sin^{11.5}{\thetaE}}{\thetaE^{5.41}} \nonumber\\[-5.5pt]
 \label{eq:emp_pairs}\\[-5.5pt]
   &&  \qquad\qquad  + 1.26\;\thetaE^{-1.78/\rmax^{0.3}}\;\rmax^{0.5}\;e^{-\thetaE^2/2}  \biggr\} \quad .\nonumber
\end{eqnarray}
The piecewise form of the function originates from the horizontal absolute threshold tails in Fig.~\ref{fig:eesc_pairs_loops}, 
i.e., \teq{\eesc^{pp}} cannot be lower than the absolute threshold \teq{2m_ec^2}.
The empirical formula is not very accurate for low altitude cases where the 
escape energy curves decrease slightly with increasing emission colatitude.
Note this empirical formula is not applicable to the \teq{\perp} mode creation because of the 
subsequent dependence of the pair threshold on the field strength.

\end{document}